\documentclass[11pt]{article}

\usepackage[dvipsnames]{xcolor} 
\usepackage{comment,soul}
\usepackage[hidelinks]{hyperref}
\usepackage{enumitem}
\usepackage{amsmath,amssymb,epsf,cite,graphicx,subfigure}
\usepackage{amsmath,braket,tikzsymbols}
\usepackage[bbgreekl]{mathbbol}
\usepackage{comment}
\usepackage{mathtools}

\usepackage{blindtext}

\numberwithin{equation}{section}
\newcommand{\nn}{\nonumber}

\newcommand{\beq}{\begin{equation}}
\newcommand{\eeq}{\end{equation}}

\newcommand{\tr}{\text{tr}}

\newcommand{\cC}{\mathcal{C}}
\newcommand{\rmT}{\mathrm{T}}

\newcommand{\rvline}{\hspace*{-\arraycolsep}\vline\hspace*{-\arraycolsep}}

\newcommand\AJ[1]{{\color{RoyalBlue}\textbf{AJ:} #1}}

\usepackage[mathscr]{euscript}

\def\lie{\pounds}
\def\dow{\partial}
\def\Df{\mathrm{D}}
\def\df{\mathrm{d}}

\def\scX{\mathscr{X}}
\def\scB{\mathscr{B}}
\def\scK{\mathscr{K}}
\def\lb{\left(}
\def\rb{\right)}
\def\half{\frac12}

\def\eqb{\text{eqb}}
\def\ideal{\text{ideal}}
\def\hs{\text{hs}}
\def\nhs{\text{nhs}}

\def\sfA{\mathsf A}
\def\sfS{\mathsf S}

\def\tmu{\tilde\mu}
\def\tvarpi{\tilde\varpi}

\let\Im\relax
\let\Re\relax
\DeclareMathOperator{\Im}{Im}
\DeclareMathOperator{\Re}{Re}

\begin{document}
\def\arraystretch{1.3}
\baselineskip=15.5pt
%\pagestyle{plain}
%\setcounter{page}{1}
%--------+---------+---------+---------+---------+---------+---------+

\begin{center}
{\LARGE \bf Dipole superfluid hydrodynamics}
\vskip 1cm

\textbf{
Akash Jain$^{1,2,a}$, 
Kristan Jensen$^{3,b}$, Ruochuan Liu$^{3,c}$ and 
Eric Mefford$^{3,d}$}

\vspace{0.5cm}

{\small ${}^1$Institute for Theoretical Physics, University of Amsterdam, 1090
  GL Amsterdam, The Netherlands \vspace{.3cm}}

{\small ${}^2$Dutch Institute for Emergent Phenomena, 1090 GL Amsterdam, The Netherlands \vspace{.3cm}}

{\small ${}^3$Department of Physics and Astronomy, University of Victoria, Victoria, BC V8W 3P6, Canada\\}

\vspace{0.5cm}

{\tt \small ${}^a$a.jain@uva.nl,}
{\tt  \small ${}^b$kristanj@uvic.ca,}
{\tt \small ${}^c$liur@uvic.ca,}
{\tt \small ${}^d$ericmefford@uvic.ca\\}

\medskip

\end{center}
\thispagestyle{empty}

\vskip1cm

\begin{center} 
{\bf Abstract}
\end{center}

We construct a theory of hydrodynamic transport for systems with conserved dipole moment, U(1) charge, energy, and momentum. These models have been considered in the context of fractons, since their elementary and isolated charges are immobile by symmetry, and have two known translation-invariant gapless phases: a ``p-wave dipole superfluid'' phase where the dipole symmetry is spontaneously broken and a ``s-wave dipole superfluid'' phase where both the U(1) and dipole symmetries are spontaneously broken. We argue on grounds of symmetry and thermodynamics that there is no transitionally-invariant gapless fluid with unbroken dipole symmetry. In this work, we primarily focus on the hydrodynamic description of p-wave dipole superfluids, including leading dissipative corrections. That theory has, in a sense, a dynamical scaling exponent $z=2$, and its spectrum of fluctuations includes novel subdiffusive modes $\omega \sim -i k^4$ in the shear sector and magnon-like sound mode $\omega\sim \pm k^2 -i k^2$. By coupling the fluid to background fields, we find response functions of the various symmetry currents. We also present a preliminary generalization of our work to s-wave dipole superfluids, which resemble $z=1$ fluids and feature sound waves and diffusive shear modes, as in an ordinary fluid. However, the spectrum also contains a magnon-like second-sound mode $\omega\sim \pm k^2 \pm k^4 -i k^4$ with subdiffusive attenuation. 

\hspace{.3cm}

\newpage

\tableofcontents

%-----------------------------------------------
\section{Introduction}
\label{sec:intro}
%-----------------------------------------------

This work weaves together two threads of recent interest, namely fractons and hydrodynamics. Fractons arise in exotic lattice models of interest in both condensed matter and high energy physics~\cite{Chamon:2004lew, 2011AnPhy.326..839B, Haah:2011drr, Vijay:2015mka,Nandkishore:2018sel,Seiberg:2020bhn}. Generally speaking, these models do not possess a simple continuum effective field theory description at long distances and low energies. For example, they have excitations of restricted mobility, propagating only along sublattices of the microscopic lattice; sometimes have a non-topological, UV-sensitive ground state degeneracy, which, when present, is an example of UV/IR mixing; can have continuum limits that do not commute with the thermodynamic limit; and have exotic symmetries that are sensitive to the details of the underlying lattice. 

This last fact usefully organizes many of the unconventional features of these systems~\cite{Vijay:2016phm, Williamson:2016jiq, You:2018oai}. An instructive prototype is provided by the X-Cube model~\cite{Vijay:2016phm}, a soluble theory of quantum mechanical spins on a three-dimensional planar lattice. This model has a subsystem symmetry that act on spins along a sublattice, characterized by a large symmetry group $(\mathbb{Z}_2)^{2(L_x+L_y+L_z)-3}$ with $L_i$ being the number of lattice sites in the $i^{\rm th}$ direction. The ground states of this model are not invariant under this lattice-sensitive symmetry, leading to an exponentially large ground state degeneracy.\footnote{The X-Cube model actually possesses a dual subsystem symmetry as well, so that the ground state degeneracy is $\sim 4^{L_x+L_y+L_z}$ on a large lattice.} Despite how far these systems seem from an effective field theory (EFT) framework, it is sometimes possible to find a near-EFT description; see e.g.~\cite{Slagle:2017wrc,Seiberg:2020bhn,Seiberg:2020wsg}. Focusing on the X-Cube model, one can construct a field theory with exotic charge-2 bosonic matter featuring a U(1) subsystem symmetry and tune the potential so that the preferred ground state spontaneously breaks the U(1) symmetry to a $\mathbb{Z}_2$ subsystem symmetry. Upon coupling this gapless phase to a tensor gauge theory, thereby Higgsing the subsystem symmetry, one finds a massive phase with a description resembling that of a topological quantum field theory (TQFT). This TQFT-like description matches several features of the X-Cube model, including its global symmetry structure and ground state degeneracy.

Another, somewhat simpler, exotic spacetime symmetry considered in the context of fractons is dipole invariance, which, roughly speaking, concerns models where a U(1) charge and the associated dipole moment are conserved; see e.g.~\cite{Pretko:2018jbi}. We will give a precise definition in the next section. These models have previously been considered in the literature as a simpler alternative to those with subsystem symmetry, as their elementary and isolated charges are also immobile by symmetry. There are also good reasons to expect that systems with an approximate dipole symmetry are realized in nature. Elastic media in two spatial dimensions~\cite{Pretko:2017kvd}, vortices in superfluid helium~\cite{Doshi:2020jso}, and tilted optical lattices~\cite{tilted} are all expected to have dipole-invariant low-energy descriptions. We will focus on such systems in this work.

Like the subsystem symmetry, dipole symmetry also depends on the details of the underlying lattice. A complex scalar field $\Psi(t,\vec x)$ with U(1) charge $q$ transforms under U(1) dipole transformations along a phase $\Lambda$ and vector $\psi_i$ as $\Psi(t,\vec{x})\to \exp(iq\Lambda - iq\psi_ix^i)\Psi(t,\vec{x})$~\cite{Pretko:2018jbi}. In particular, dipole transformations are valued in the space of single-particle momenta. When the microscopic theory is defined on a lattice in finite volume, the dipole symmetry group is discrete and isomorphic to the dual lattice~\cite{Jensen:2022iww}. In the limit that the lattice spacing vanishes and the volume is infinite, the dipole symmetry becomes $\mathbb{R}^{d}$ with $d$ being the number of spatial dimensions.

Pretko has shown how to write down field theories of bosonic matter invariant under this exotic dipole symmetry~\cite{Pretko:2018jbi, Pretko:2020cko}; for similar models on a lattice see~\cite{Jensen:2022iww}. Unlike the case of subsystem symmetries, these models can be made rotationally invariant. In both cases, these bosonic theories are generically strongly correlated as terms with spatial derivatives in the effective action have four or more powers of the fundamental fields. This makes it a challenging task to map out their phase diagram. By tuning the tree-level potential to have a very deep well away from the origin, one might hope to spontaneously break the U(1) symmetry, in which case there is a candidate weakly coupled sigma model description of the phase. However, it is not a priori clear that interactions do not strongly renormalize the potential and restore the symmetry. More generally, one might wonder what other phases exist and with what symmetry breaking patterns.

The best answer to this question to date comes from an analysis of large $N$ versions of models with conserved dipole moment, where one can access the physics at finite interaction strength by studying models with large number of degrees of freedom~\cite{Jensen:2022iww}. Similar techniques may also be applied to study the large $N$ versions of models invariant under subsystem symmetries~\cite{subsystemWIP}. One considers models with an $N$-component charged scalar field $\Psi = (\Psi_1,\ldots,\Psi_N)$, for a large number of fields $N\gg 1$, with U($N$)-invariant interactions. The U($N$) transformations act on $\Psi$ in the usual way, $\Psi \to M\Psi$, where $M\in {\rm U}(N)$ is a constant matrix. The diagonal subgroup ${\rm U}(1)\subset {\rm U}(N)$, associated with $M = \exp(iq\Lambda)$ for some constant parameter $\Lambda$ and charge $q$, is the large $N$ version of the U(1) symmetry. For models with dipole symmetry, we further require invariance under $M = \exp(-iq\psi_ix^i)$. These models reveal three translation-invariant phases of matter with dipole and U(1) symmetries: 
\begin{enumerate}
    \item A gapped \emph{dipole-symmetric} phase in which both the dipole and U(1) symmetries are preserved. In the large $N$ models of~\cite{Jensen:2022iww}, this phase is found at lattice-scale temperatures or when the U(1) density vanishes. 
    
    \item A gapless \emph{p-wave dipole superfluid} phase in which the dipole symmetry is spontaneously 
    broken but the U(1) symmetry is preserved. This phase may be characterized by a neutral vector condensate $\langle i\Psi^\dagger\overset{\text{\tiny$\leftrightarrow$}}\dow_i\Psi \rangle$.
    
    \item A gapless \emph{s-wave dipole superfluid} phase in which the U(1) symmetry is spontaneously broken, which in turn leads to the spontaneous breaking of the dipole symmetry as well.\footnote{We will have more to say about this shortly, but, for now, note that the discrete dipole symmetry is continuous in infinite volume, leading to a conserved Noether charge, i.e. the dipole moment. That dipole moment $D^i$ does not commute with the momenta $P^i$ with a commutator $[P^i,D^j] = iQ\delta^{ij}$, where $Q$ is the U(1) charge~\cite{GromovMultipole}. This commutator simply expresses the fact that translating a charge $q$ by a displacement $\vec{\ell}$ changes the dipole moment by $q\,\vec{\ell}$. This commutator implies that if the U(1) symmetry is spontaneously broken but translations are unbroken, then dipole is spontaneously broken as well. \label{foot:both-breaking-algebra}} This phase is characterized by a charged scalar condensate $\langle\Psi\rangle$.
\end{enumerate}
In infinite volume, long-wavelength fluctuations may restore the dipole and U(1) symmetries, leading to a quasi-ordered rather than an ordered phase as in the two-dimensional XY model. At finite temperature (using arguments like that in~\cite{Distler:2021qzc}), this is expected to happen for the p-wave dipole superfluid phase in $d=1,2$ spatial dimensions, and in the s-wave dipole superfluid phase for $d\leq 4$. There may also be other phases in which the translation symmetry is spontaneously broken.

The gapped dipole-symmetric phase has immobile charged excitations resembling fractons, albeit with a trivial ground state degeneracy. The $n$-point functions of charged operators act as an order parameter for spontaneous breaking of dipole symmetry. Consider the two-point function $\langle\Psi^\dagger(t,\vec{x})\Psi(0,\vec 0)\rangle$, which is invariant under dipole transformations if and only if it vanishes when $x^i\neq 0$. More generally, an $n$-point function with charges $q_m$ inserted at positions $x^i_m$ is dipole-invariant only when it vanishes when $\sum_{m=1}^n q_m x^i_m \neq 0$.\footnote{A stronger statement holds at large $N$ thanks to large $N$ factorization. At leading order in large $N$ correlation functions of charged operators vanish unless the charges are coincident.} In this sense, the elementary and isolated charges cannot be transported in a dipole-symmetric phase. By contrast, dipole superfluid phases admit a ``dipole condensate,'' allowing the charges to move in pairs. One can also obtain a gapped phase with immobile excitations by coupling the $N=1$ version of the s-wave dipole superfluid phase to a tensor gauge theory~\cite{Gorantla:2022eem,Gorantla:2022ssr}. Without coupling to a tensor gauge theory, the dipole superfluid phases have a non-trivial ground state degeneracy equal to the number of lattice sites mandated by the spontaneous breaking of the dipole symmetry~\cite{Jensen:2022iww}. In the thermodynamic limit in infinite volume, where the dipole symmetry is continuous, there is a candidate weakly coupled sigma model description of the p-wave dipole superfluid phase in terms of a vector Goldstone $\phi_i$~\cite{Jensen:2022iww}, shifting under the dipole transformations as $\phi_i\to \phi_i -\psi_i$. There is also a candidate sigma model description of the s-wave dipole superfluid phase in terms of a scalar Goldstone $\phi$, shifting under the U(1) transformations as $\phi\to \phi-\Lambda$. The low-energy behaviours of these two superfluid phases are quite distinct from each other as we shall explore in this work.

In infinite volume and at low energies, many-body systems with conserved dipole moment can be universally described using the framework hydrodynamics, agnostic of the microscopic realizations of the model under consideration. Hydrodynamics is a relatively ancient subject, which has undergone a modern revolution thanks to insights from the AdS/CFT correspondence~\cite{Bhattacharyya:2007vjd, Banerjee:2008th} and effective field theory~\cite{Crossley:2015evo, Haehl:2018lcu, Jensen:2017kzi, Jain:2020vgc, Armas:2020mpr}. This modern perspective provides us with a rigorous prescription for building a dissipative hydrodynamic theory of fluids with spontaneously broken dipole symmetry. In this work, we will primarily focus on p-wave dipole superfluid hydrodynamics, where the U(1) symmetry is spontaneously unbroken. We will briefly discuss the generalization of our framework to s-wave dipole superfluid hydrodynamics, where both the U(1) and dipole symmetries are spontaneously broken, and will return with full details in an imminent publication~\cite{swavearticle}. Our main motivation in constructing these hydrodynamic models is to make predictions for future experiments. While there is good reason to expect that models with dipole symmetry can be realized in the lab, a conclusive discovery of these models in nature is yet lacking. Our theory of transport leads to model-independent predictions for the spectrum of fluctuations and the functional form of the response functions of conserved charges. We hope that these results will aid in the future discovery of fractonic matter.

Before we delve into the details of our formalism and results, we note that our work is not the first to tackle the hydrodynamic description for dipole-invariant systems; see~\cite{Gromov:2020yoc,Iaconis:2020zhc,Glorioso:2021bif,Grosvenor:2021rrt,Glodkowski:2022xje,Glorioso:2023chm}. Nonetheless, there are several distinctive features of our work. We introduce a consistent derivative counting appropriate for dipole superfluids that had not been identified before, allowing us to conclusively enumerate all transport coefficients at leading order in derivatives, as well as the first subleading derivative corrections. We discuss in detail how to couple these hydrodynamic theories to curved spacetime background, allowing us to give a complete account of hydrostatic equilibrium and so the thermodynamics of these phases. In particular, this allows us to transparently infer that for there to be non-trivial charge transport in a gapless transitionally-invariant phase of a system with dipole symmetry, the said dipole symmetry must be spontaneously broken. Relatedly, we can solve the hydrodynamic equations in the presence of background fields and thereby obtain the response functions of various conserved operators. Since we are in a superfluid phase, the system also admits equilibrium states with dissipationless dipole ``superflow'', characterized by a profile for the dipole Goldstones $\phi_i = x^{j}\xi_{ji }$ for some constant matrix $\xi_{ij}$. While we do not focus on these in this work, the framework we setup can be also used to describe transport in such more general states. Finally, while most of the previous work on the subject was focused on p-wave dipole superfluids, our work is the first to identify the qualitative features of s-wave dipole superfluids with conserved momentum. We provide a more detailed comparison to previous work in Appendix~\ref{app:comparison}.

The remainder of this manuscript is organized as follows. In Section~\ref{S:summary}, we give an overview of the rest of the paper, orienting the reader to our main results and approach. We set the stage for our analysis in Section~\ref{sec:preliminaries}, reviewing models with dipole symmetry, coupling to curved spacetime background, spontaneous breaking of dipole symmetry, and subtleties regarding the derivative counting scheme. We go on to construct the hydrostatic effective actions for p-wave dipole superfluids in Section~\ref{sec:hydrostatics} and the full theory of p-wave superfluid hydrodynamics in Section~\ref{modernhydro}. We use this theory to compute the spectrum of linearized fluctuations and response functions in Section~\ref{sec:response}. We sketch the construction of s-wave superfluid hydrodynamics in Section~\ref{S:swave}, and wrap up with a Discussion in Section~\ref{S:discussion}. Various technical results and a comparison with previous work are relegated to the Appendix.

\vspace{.05cm}
\noindent
\emph{Note:} While this work was near completion, we were made aware of the forthcoming work of~\cite{idealFractonHydro} that has significant overlap with our own and is set to appear on the same day.

\vspace{.05cm}
\noindent
\emph{Note added v2:} Fixed some typos. Changed the notation for the U(1) and dipole chemical potentials to $\tilde\mu$ and $\tilde\varpi$ respectively to highlight their dipole-invariant nature and implicit dependence on the dipole Goldstone.

%-----------------------------------------------
\section{Overview of hydrodynamics with dipole symmetry}
\label{S:summary}
%-----------------------------------------------

In this Section we provide a summary of the rest of the manuscript. We intend for it to serve as a guide for the technical results that follow.

The basic algorithm towards a hydrodynamic theory we follow in this work is determined almost entirely by symmetries, the symmetry-breaking pattern, and, crucially, on the coupling of conserved currents to slowly varying background fields. This algorithm is quite useful when it comes to constructing the hydrodynamic constitutive relations. It has the added advantage that we can thereby vary the constitutive relations with respect to the background fields and obtain the linear response functions of conserved currents. To avoid getting lost in technical details, here we summarize the overarching formalism and the main results of our construction. We shall focus on p-wave dipole superfluids for concreteness and return to s-wave case towards the end of our summary.

\vspace{1em}
\noindent
\textbf{Hydrodynamic fields and conservation laws.}---Our starting point is the Ward identities associated with the symmetries under consideration, together with the symmetry-breaking pattern, which in the present context refers to the spontaneous breaking of dipole and/or U(1) symmetry. We then postulate a hydrodynamic description in terms of a set of slowly varying hydrodynamic fields, one for each conserved current, and the Goldstone fields mandated by the symmetry-breaking pattern. Focusing on p-wave dipole superfluids for concreteness, the hydrodynamic fields are a local temperature $T$, chemical potential $\mu$, velocity $u^i$, and a dipole Goldstone $\phi_i$. The relevant symmetry currents are the energy density and flux $\epsilon^{t}$, $\epsilon^i$, momentum density $\pi_i$, symmetric spatial stress tensor $\tau^{ij}$, and U(1) density and flux $J^{t}$, $J^i$. The dipole symmetry implies a Ward identity, whereby the U(1) flux $J^i$ is expressible as the divergence of a symmetric dipole flux $J^{ij}$ through
\begin{equation}
    \label{E:dipoleWard}
    J^i = \partial_j J^{ij}\,.    
\end{equation}
Since the dipole symmetry is discrete in finite volume, one might expect there to be no associated Ward identity. However, as we discuss in Section~\ref{couplingtobackgrounds}, global dipole symmetry can be promoted to a continuous spurionic symmetry, so that dipole transformations imply the Ward identity \eqref{E:dipoleWard} in both finite and infinite volume and, suitably generalized, in curved spacetime as well.
The remaining Ward identities are the standard conservation equations for a non-relativistic system with translational and rotational symmetry, i.e.
\begin{align}\begin{split}
    \partial_t \epsilon^t + \partial_i \epsilon^i &=0, \\ 
    \partial_t \pi^i + \partial_j \tau^{ij} &=0, \\
    \partial_t J^t + \partial_i J^i &=0.
\label{eq:flat-cons}
\end{split}\end{align}
The curved space versions of these equations are slightly more complicated and will be outlined in Section \ref{sec:preliminaries}. We should emphasize that systems with dipole-invariance do not feature any boost symmetry, therefore the momentum density $\pi_i$ is not fixed in terms of the U(1) flux $J^i$ or the energy flux $\epsilon^i$, as it would be in Galilean and Lorentzian theories respectively. In any case, we impose these Ward identities as equations of motion in the hydrodynamic description. Note that there are as many equations as hydrodynamic variables, so that we have a well-defined differential system given that we specify the constitutive relations, i.e. how various conserved currents depend on the hydrodynamic fields. In particular, the dipole Ward identity can be understood as the Josephson condition for the dipole Goldstone.

\vspace{1em}
\noindent
\textbf{Dipole transformations and UV/IR mixing.}---The constitutive relations of our hydrodynamic theory must respect the symmetries under consideration. The conserved currents $\epsilon^t$, $\epsilon^i$, $\pi_i$, $\tau^{ij}$, $J^t$, $J^i$, and $J^{ij}$ transform under spacetime translations and spatial rotations as usual and are left invariant under global U(1) transformations. Global dipole transformations, however, do act non-trivially on $\pi_i$ and $\tau^{ij}$ as
\begin{align}
\begin{split}
  \pi_i
  &\to \pi_i - J^t \psi_i, \\
  \tau^{ij}
  &\to \tau^{ij} 
  - 2J^{(i} \psi^{j)}
  + \psi^k\partial_k J^{ij}\,.
  \label{eq:dipole-var-currents-flat}
\end{split}
\end{align}
One can check that this is a symmetry of the conservation equations \eqref{eq:flat-cons}. As we shall explain in our subsequent discussion, $\epsilon^t$, $\epsilon^i$ also transform under dipole transformations on a generic curved spacetime background, while $J^t$, $J^i$, $J^{ij}$ are left invariant; see~\cite{Jain:2021ibh} for more discussion. 

Note that due to these transformation properties, it is not possible to construct hydrodynamic constitutive relations for $\pi_i$ and $\tau^{ij}$ in the dipole-symmetric phase where the dipole symmetry is not spontaneously broken, provided that $J^t$, $J^i$, and $J^{ij}$ are non-vanishing. This is because in the absence of the dipole Goldstone $\phi_i$, all the hydrodynamic variables are dipole-invariant and so there are no low-energy fields available to reproduce the requisite transformation laws of $\pi_i$ and $\tau^{ij}$. This implies that $J^t$, $J^i$, and $J^{ij}$ must vanish in the dipole-symmetric phase, in which case $\pi_i$ and $\tau^{ij}$ are dipole-invariant themselves, and hence the low-energy description is effectively that of ordinary neutral hydrodynamics.

When the dipole symmetry is spontaneously broken, we can use the dipole Goldstone $\phi_i$ to construct the constitutive relations with correct transformation properties, e.g. $\pi_i \sim J^t\phi_i + \ldots$, where ellipsis characterize further dipole-invariant terms in the constitutive relations. Using suitable redefinitions involving the Goldstone field $\phi_i$, we can construct the dipole-invariant versions of various conserved currents and the respective Ward identities, denoted by ``tilde'' throughout this work. This allows us to formulate a manifestly dipole-invariant version of dipole superfluid dynamics, which can always be undone by performing a local dipole transformation with the parameter $\psi_i= \phi_i$.  For example, $\tilde\pi_i = \pi_i - J^t \phi_i$, etc.  This is analogous to constructing manifestly boost-invariant hydrodynamics in the Galilean setting~\cite{Jensen:2014ama}.

Since the dipole symmetry is lattice sensitive, our description exhibits UV/IR mixing characteristic of fracton models. However, the UV-sensitivity is relatively modest for dipole symmetries compared to subsystem symmetries.
Despite being a low-energy field, the dipole Goldstone field $\phi_i$ is UV-sensitive, in the sense that it is compact with periodicity $\sim 2\pi/a$, with $a$ being the lattice spacing~\cite{Jensen:2022iww}. This means that $n$-point functions of $\phi_i$ are UV-sensitive, but derivatives of $\phi_i$ are not. Thus, $\pi_i$ and $\tau^{ij}$ are UV-sensitive, even in the course-grained hydrodynamic limit, while the other conserved currents are not.

\vspace{1em}
\noindent
\textbf{Derivative counting.}---An important ingredient in the hydrodynamic description is the derivative counting scheme. Since hydrodynamics describes small fluctuations away from thermal equilibrium, we can organize the respective constitutive relations as a perturbative expansion in derivatives. However, for a consistent truncation of this expansion to any given order in derivatives, we must agree on the relative derivative ordering of various constituent fields. In textbook hydrodynamic treatments~\cite{landaubook,Kovtun:2012rj}, the local temperature $T$, chemical potential $\mu$, and velocity $u^i$ are treated as ${\cal O}(\partial^0)$ quantities. One then classifies terms in the constitutive relations according to how many space and time derivatives they contain, treating both space derivatives $\dow_i$ and time derivatives $\dow_t$ as ${\cal O}(\dow^1)$. The terms in the constitutive relations with no derivatives are part of ``ideal hydrodynamics,'' while those with a single derivative are the first-order corrections, and so on. 

The derivative counting scheme for dipole superfluids is considerably more intricate and is one of the central results of our work. Firstly, the dipole Goldstone field $\phi_i$ is counted as $O(\partial^{-1})$, as is conventional for a Goldstone field in superfluid hydrodynamics. The temperature $T$ and chemical potential $\mu$ are still counted as $O(\partial^0)$ quantities. However, the spatial velocity $u^i$ is counted as $O(\partial^1)$. The underlying physical reason is that the dipole symmetry forbids homogeneous equilibrium states with nonzero charge and nonzero velocity. Therefore, velocity is always a fluctuation in a state of nonzero density. A consistent counting scheme then requires that time derivatives $\dow_t$ scale as $O(\partial^2)$, while the spatial derivatives $\dow_i$ are still ${\cal O}(\dow^1)$. In other words, these systems are characterized by a dynamical scaling exponent $z=2$.

Due to the disparate counting of time and space derivatives, the density and flux components of conserved currents also have different power countings.
The energy density $\epsilon^t$, charge density $J^t$, spatial stress $\tau^{ij}$, and dipole flux $J^{ij}$ are $O(\partial^0)$; the energy flux $\epsilon^i$ and charge flux $J^i$ are $O(\partial^1)$; and the momentum density $\pi_i$ is $O(\partial^{-1})$. The unconventional scaling of $\pi_i$ is due to the fact that it transforms almost like a dipole Goldstone as in \eqref{eq:dipole-var-currents-flat}. With these scalings, the hydrodynamic equations are homogeneous in derivatives, and one can classify terms in the constitutive relations by how many derivatives they have relative to the leading scaling.

Note that not all allowed terms in the constitutive relations that can be written down at a particular derivative-order are physical. While the hydrodynamic variables have well-defined physical meaning in thermodynamic equilibrium, they do not have fixed definitions out-of-equilibrium in the hydrodynamic regime. One may use this freedom together with the leading order hydrodynamic equations to eliminate various unphysical terms from the constitutive relations. These steps are analogous to using field redefinitions and equations of motion to remove unphysical coupling constants in a field theory Lagrangian.

\vspace{1em}
\noindent
\textbf{Hydrostatic partition function.}---By coupling our hydrodynamic theory to a time-independent but slowly spatially-varying background spacetime, the fluid can be placed in hydrostatic equilibrium. These hydrostatic states can be simply described using Euclidean thermal field theory. When the state has finite correlation length, one can dimensionally reduce the theory on the thermal circle and obtain a gapped long wavelength description~\cite{Banerjee:2012iz,Jensen:2012jh}. These states are then described by a local partition function expressed in terms of the time-independent background fields and their derivatives. Varying the hydrostatic partition function with respect to background fields must reproduce the hydrostatic part of the conserved currents under consideration and thus imposes strong consistency constraints on the hydrodynamic constitutive relations. For systems with infinite correlation length, as in a superfluid, the hydrostatic partition function can be constructed using a Euclidean effective action for the Goldstone field coupled to time-independent background fields, which again yields strong hydrostatic constraints on the constitutive relations~\cite{Bhattacharyya:2012xi}.

Hydrostatic constraints provide us another argument for why transitionally-invariant  systems with conserved dipole moment must have their dipole symmetry spontaneously broken for there to be non-trivial charge transport at low-energies. 
In the dipole-symmetric phase, the correlation length is finite and the hydrostatic equilibrium states ought to be characterized by a local partition function constructed out of time-independent background fields. Let us consider the background sources $A_t$, $A_i$, and $a_{ij}$ coupled to $J^t$, $J^i$, and $J^{ij}$ respectively. If the dipole symmetry was absent, one could have a U(1)-invariant term in the partition function like $\half\chi(\mu_0+A_t)^2$, with $\mu_0$ the equilibrium chemical potential and $\chi$ the charge susceptibility, leading to nonzero equilibrium charge density $J^t \sim \chi (mu_0 + A_t)$. However, in the presence of a background frame velocity $v^i$ coupled to the momentum density $\pi_i$, the dipole transformations act on $A_t$ as $A_t \to A_t - \psi_i v^i$. Therefore, such a term is not compatible with dipole symmetry in the presence of momentum sources and forces the charge density to vanish.
More generally, it is an easy exercise to show that there are no terms compatible with dipole symmetry that depend on $A_t$, $A_i$, $a_{ij}$ and transport is effectively neutral. On the other hand, when the dipole symmetry is spontaneously broken, one can write down a dipole-invariant term in the Goldstone effective action as $\half\chi(\mu_0 + A_t - \phi_i v^i)^2$, resulting in non-trivial charge density in equilibrium.

\vspace{1em}
\noindent
\textbf{Second Law of thermodynamics.}---Out-of-equilibrium, the hydrodynamic constitutive relations must satisfy a local version of the Second Law of thermodynamics. That is, we require the existence of an entropy density and flux $s^{t}$, $s^i$, whose divergence is positive semi-definite everywhere when evaluated on any solution of the hydrodynamic equations.
In a homogeneous equilibrium state, these take the form $s^t = s$, $s^i = 0$, with $s$ being the thermodynamic entropy density. Historically, the local Second Law was imposed on hydrodynamic descriptions as a physically motivated constraint, but is now understood to follow from more fundamental considerations in Schwinger-Keldysh effective field theory~\cite{Glorioso:2016gsa, Haehl:2018uqv, Jensen:2018hhx}.

As it turns out, the Second Law requirement exactly reproduces all the ``equality-type'' constraints arising from hydrostatic consistency mentioned above. In addition, it also results in a set of ``inequality-type'' constraints, mandating certain dissipative transport coefficients at low derivative orders to be non-negative. In textbook hydrodynamics, these lead to the non-negativity of viscosity and conductivity. Fundamentally, these inequality constraints can be understood as arising from unitarity in a microscopic description viz. the non-negativity of spectral functions. The higher derivative corrections are, roughly speaking, always consistent with the positivity of entropy production provided that the leading corrections are, thanks to an argument by S.~Bhattacharyya~\cite{Bhattacharyya:2013lha}. 

\vspace{1em}
\noindent
\textbf{Constitutive relations.}---Let us briefly look at the constitutive relations allowed by the Second Law of thermodynamics in the absence of various background fields. 
Since the dipole Goldstone is counted as $O(\partial^{-1})$, there is a spatial tensor $\partial_i \phi_j$ at zeroth order in derivatives. This tensor is constant in a superflow state with $\phi^i=x^j\xi_{ji}$, and has a number of independent components that grows rapidly with the dimensionality of space. This leads to a dimension-dependent number of zeroth order scalars that can appear in the thermodynamic equation of state of a dipole superfluid. The complete theory of nonlinear hydrodynamics in general dimension is then quite complicated, analogous in some ways to nonlinear magnetohydrodynamics~\cite{Hernandez:2017mch, Armas:2018atq, Armas:2018zbe}. Motivated by the desire to predict future observations, we content ourselves to work at the level of linearized constitutive relations around a zero superflow state where the analysis is more tractable. However, we emphasize that our methods allow for a fully nonlinear hydrodynamic description of dipole superfluids.

Turning off the background fields and working to leading order in derivatives and up to linearized order in fluctuations around a zero superflow state, we find the constitutive relations for the conserved densities
\begin{align}
\begin{split}
    \epsilon^t & = \epsilon + {\cal O}(\partial^2)\,, \\
    \pi_i &= q\,\phi_i + {\cal O}(\partial)\,, \\ 
    J^t & = q + {\cal O}(\partial^2)\,,
\end{split}
\end{align}
where $\epsilon$ and $q$ are the thermodynamic energy and charge densities, respectively. The conventional kinetic term in the momentum density, $\pi_i \sim \rho\,u_i$ where $\rho$ is the kinetic mass density, appears as a leading derivative correction in a p-wave dipole superfluid.
The associated fluxes are given as
\begin{align}
\begin{split}
    \epsilon^i 
    & = (\epsilon+p) u^i 
    + p_d \dow_t\phi^i
    + \chi_{mn}\!\lb 
    \dow^i\partial_k\phi^k
    - \dow_k\dow^k\phi^i 
    \rb
    - \kappa\, \partial^iT 
    +  O(\partial^3)\,,
    \\
    \tau^{ij} 
    &= p\,\delta^{ij} 
    - p_d\,\delta^{ij}\partial_k\phi^k
    + O(\partial^2) \,,
    \\
    J^i 
    & = q\, u^i 
    - \chi_m\!\lb 
    \dow^i\partial_k\phi^k
    - \dow_k\dow^k\phi^i 
    \rb
    + O(\partial^3) \,,
    \\
    J^{ij} 
    & = p_d \delta^{ij} 
    - G_d\!\lb 
    \dow^i\phi^j + \dow^j\phi^i 
    - \frac{2}{d}\delta^{ij}\dow_k\phi^k 
    \rb
    + O(\partial^2) \,,
\end{split}
\label{eq:ideal_flat_constitutive}
\end{align}
where we have identified the thermodynamic pressure $p$, dipole pressure $p_d$, and susceptibilities $\chi_m$, $\chi_{mn}$, and $G_d$. Furthermore, $\kappa$ is the dissipative thermal conductivity transport coefficient, which, crucially, affects the constitutive relations at the same order as thermodynamics. This is a qualitatively novel feature of the dipole-invariant hydrodynamic description.

We have also enumerated the most general leading-derivative corrections to the constitutive relations. Accounting for the dependence on background fields, there are $28$ hydrostatic coefficients including the kinetic mass density $\rho$ at this order, 12 dissipative coefficients including the traditional shear viscosity $\eta$ and bulk viscosity $\zeta$, and 7 new non-dissipative non-equilibrium coefficients. More details can be found in Sections \ref{sec:hs-consti} and \ref{sec:nhs-consti}.

\vspace{1em}
\noindent
\textbf{Modes and correlators.}---We can then plug the constitutive relations above into the hydrodynamic equations and solve for the spectrum of linearized hydrodynamic fluctuations. In the longitudinal sector, the longitudinal velocity $u_\|$ is effectively gapped due to the dipole Ward identity, leaving the longitudinal component of the dipole Goldstone $\phi_\|$, temperature $T$, and chemical potential $\mu$ as the effective low-energy degrees of freedom. These mix to produce a pair of magnon-like modes with dispersion relations $\omega \sim  \pm k^2 - i k^2 + O(k^4)$ and a diffusive mode with dispersion relation $\omega \sim -i k^2 + O(k^4)$. The attenuation of the magnon-like modes and the diffusion constant are both controlled only by the thermal conductivity $\kappa$, meaning that if the system under consideration is a poor thermal conductor, these modes will become subdiffusive. This feature does not exist in ordinary fluids where other dissipative coefficients, e.g. the bulk viscosity $\zeta$, appear at the same order as $\kappa$. Similarly in the transverse sector, the transverse velocity $u_\perp$ is effectively gapped, while the transverse fluctuations of the dipole Goldstone $\phi_\perp$ give rise to a subdiffusive mode with dispersion relation $\omega \sim -i k^4$. The subdiffusion coefficient is fixed in terms of the shear viscosity $\eta$ of the dipole superfluid. 

We also compute the response functions of conserved densities and fluxes using our hydrodynamic theory in Section~\ref{S:responseFunctions}. The full suite of hydrodynamic correlation functions is quite complicated, therefore we focus on response functions in the transverse sector for illustrative purposes. We have included a supplementary Mathematica notebook with our arXiv submission that enables users to compute all longitudinal and transverse response functions depending on the particular application in mind. We also present the frequency-dependent optical responses at zero wavevector in Section~\ref{S:responseFunctions}. These allow us to find simple Kubo formulae for $\kappa$ and 8 out of 19 out-of-equilibrium transport coefficients at the order of leading-derivative corrections. The Kubo formulae for the remaining transport coefficients require response functions at nonzero wavevector and frequency and can be obtained using our supplementary Mathematica notebook.

\vspace{1em}
\noindent
\textbf{p-wave vs s-wave dipole superfluids.}---For concreteness and clarity of presentation, we primarily focus on p-wave dipole superfluids in this work, where the dipole symmetry is spontaneously broken but the U(1) symmetry is preserved. In Section~\ref{S:swave}, we briefly discuss the s-wave dipole superfluids, where the U(1) symmetry is also spontaneously broken, highlighting some important distinctions with the p-wave case. This phase is characterized by a scalar U(1) Goldstone $\phi$ and the vector Goldstone is no longer independent, $\phi_i = -\dow_i\phi - A_i$. Following a preliminary analysis, we find evidence that the mode spectrum of this phase has ordinary sound waves $\omega \sim \pm k - ik^2 + {\cal O}(k^3)$ in the longitudinal sector and ordinary shear diffusion $\omega \sim - ik^2 + {\cal O}(k^4)$ in the transverse sector. However, as a consequence of dipole symmetry, we do find additional magnon-like second sound waves $\omega \sim \pm k^2 \pm k^4 - ik^4 + {\cal O}(k^6)$ in the longitudinal sector with subdiffusive attenuation.
We leave a rigorous analysis of s-wave dipole superfluids to an imminent publication~\cite{swavearticle}. 

%-----------------------------------------------
\section{Preliminaries}
\label{sec:preliminaries}
%-----------------------------------------------

In this Section we review some background we require in order to construct a hydrodynamic theory for systems with conserved dipole moment. We start with a quick introduction to field theories with a dipole symmetry, followed by a review of how to couple such theories to a fixed curved spacetime background. We then discuss the spontaneous breaking of U(1) and dipole symmetry in these models. There we argue that the U(1) breaking leads to dipole breaking as well, and show how to use a dipole Goldstone to reformulate the currents and Ward identities in a dipole-invariant way. We end with an argument for the unconventional derivative counting in the hydrodynamic description of these models.

%-----------------------------------------------
\subsection{Dipole symmetry}
%-----------------------------------------------

We start optimistically with the goal of writing down a continuum quantum field theory of a unit-charge scalar field $\Psi(t,\vec{x})$ which, in infinite volume, is invariant under spacetime translations, rotations, U(1) phase rotation, and dipole transformations. Under a U(1) phase rotation $\Lambda$ and dipole transformation $\psi_i$, the scalar field transforms as
\beq
    \Psi(t,\vec{x}) \to 
    e^{i \Lambda - i\psi_ix^i}\,
    \Psi(t,\vec{x})\,.
\eeq
Pretko has shown how to write down theories of this sort~\cite{Pretko:2018jbi}. A representative Lagrangian is
\beq
\label{E:pretkoS}
    S = \int \df t\, \df^dx \left( 
    i \Psi^{\dagger} \partial_t \Psi 
    - \lambda \Df_{ij}(\Psi^{\dagger},\Psi^{\dagger}) 
    \Df^{ij}(\Psi,\Psi)
    - V(\Psi^*\Psi)\right)\,.
\eeq
The bi-local differential operator $\Df_{ij}$ involves two derivatives and acts on two fundamental fields of charges $q_1$, $q_2$ as
\beq
    \Df_{ij}(\Psi_1,\Psi_2) = \frac{1}{2}\left( \frac{q_1}{q_2}\Psi_1\partial_i \partial_j \Psi_2 - \partial_i \Psi_1 \partial_j \Psi_2 + (1 \leftrightarrow 2)\right)\,.
\eeq
By construction, this operator transforms covariantly under the aforementioned U(1) and dipole transformations, i.e. $\Df_{ij}(\Psi_1,\Psi_2) \to e^{i(q_1+q_2)(\Lambda-\psi_ix^i)} \Df_{ij}(\Psi_1,\Psi_2)$; see~\cite{Jain:2021ibh}. In particular, the usual spatial kinetic term $\dow_i\Psi^\dagger\dow^i\Psi$ is forbidden by dipole symmetries, and the simplest terms with spatial derivatives involve at least four powers of the charged fields written above. 

The dipole symmetry is simpler to understand in momentum space. Dipole transformations act on a momentum space field $\widetilde{\Psi}(t,\vec{k})$ of charge $q$ as
\begin{equation}
    \widetilde{\Psi}(t,\vec{k}) 
    \to \widetilde{\Psi}(t,
    \vec{k}-q \vec{\psi})\,,
\end{equation}
i.e. as translations in momentum space. Dipole-invariant interactions are those that are, in a sense, invariant under translations in momentum space. For example the interaction $\Df_{ij}(\Psi^{\dagger},\Psi^{\dagger})\Df^{ij}(\Psi,\Psi)$ can be expressed in momentum space as
\begin{equation}
\begin{split}
    \int \df t \df^dx \,
    &\Df_{ij}(\Psi^{\dagger},\Psi^{\dagger})
    \Df^{ij}(\Psi,\Psi)  \\ 
    &= \int \df t \frac{\df^dk_1}{(2\pi)^d} \ldots 
    \frac{\df^dk_4}{(2\pi)^d}\, 
    V_4(k_m)\,
    \widetilde{\Psi}^{\dagger}(t,-\vec{k}_1)\,
    \widetilde{\Psi}^{\dagger}(t,-\vec{k}_2)\,
    \widetilde{\Psi}(t,\vec{k}_3)\, 
    \widetilde{\Psi}(t,\vec{k}_4)\,,
\end{split}
\end{equation}
where the vertex is defined as
\beq
    V_4(k_m) = \frac{1}{4}
    \lb \vec{k}_{12}\cdot \vec{k}_{34}\rb^2 
    (2\pi)^d \delta^{(d)}(\vec k_{13} + \vec k_{24}),
\eeq
together with $\vec{k}_{mn} = \vec{k}_m - \vec{k}_n$.
Dipole invariance here is made manifest as the interaction only depends on the invariant differences of momenta. By going to momentum space, it is almost immediate to write down dipole-invariant theories on a hypercubic spatial lattice with lattice spacing $a$. The simplest lattice implementation of this interaction takes the same form as above, provided that we now interpret the integral over momenta as a sum over lattice momenta and the interaction is replaced with
\beq
    V_4(k_m) = \frac{4}{a^4} \left( \sum_{i=1}^d \sin\left( \frac{k_{12i}a}{2}\right)\sin\left( \frac{k_{34i}a}{2}\right)\right)^2 
    (2\pi)^d \delta^{(d)}(\vec k_{13} + \vec k_{24})\,,
\eeq
with the delta function understood as the appropriate one for lattice momenta.

What is the dipole symmetry group? As mentioned in the introduction, dipole transformations are valued in the space of single-particle momenta. This space depends on whether we are in finite or infinite volume, in the continuum or on a lattice. If by finite volume we mean a torus, then the four options are~\cite{Gorantla:2022eem,Jensen:2022iww,Gorantla:2022ssr}
\begin{enumerate}
    \item \emph{Infinite volume, continuum}: 
    Continuous non-compact dipole symmetry $\mathbb{R}^d$.
    \item \emph{Finite volume, continuum}: 
    Discrete non-compact dipole symmetry $\mathbb{Z}^d$.
    \item \emph{Infinite volume, lattice}: 
    Continuous compact dipole symmetry ${\rm U}(1)^d$.
    \item \emph{Finite volume, lattice}: 
    Discrete compact dipole symmetry 
    $\mathbb{Z}_{L_1} \times \ldots \times \mathbb{Z}_{L_d}$, where the lattice has $L_i$ sites in the $i^{\rm th}$ direction.
\end{enumerate}
In the lattice cases, the dipole transformations are valued in the Brillouin zone. For a lattice $\Gamma$, the dipole group is isomorphic to the dual lattice $\widetilde{\Gamma}$. Observe that lattice translations are valued in $\Gamma$, so that lattice translations and dipole transformations are dual to each other. Equivalently, as stated above, dipole transformations act as translations in momentum space.

In the continuum, there are conservation laws associated with spacetime translations and U(1) transformations given in \eqref{eq:flat-cons}, which become Ward identities in the quantum theory. Furthermore, from the action~\eqref{E:pretkoS}, one can compute the U(1) and dipole fluxes and check that they satisfy the relation given in \eqref{E:dipoleWard}. In the next subsection, we will see that one can define the current $J^{ij}$ by varying the action with respect to a suitable external field, and that this equation may also be understood as a Ward identity associated with a spurionic continuous dipole symmetry. 

The Hamiltonian $H$, total momenta $P_i$, and U(1) charge $Q$ can be constructed from the conserved densities appearing in \eqref{eq:flat-cons} as
\beq
    H = \int d^dx \,\epsilon^t\,, \qquad P_i = \int d^dx \,\pi_i\,, \qquad Q = \int d^dx\, J^t\,,
\eeq
and, in infinite volume, so that we have rotational symmetry, we have angular momenta $M_{ij} = - M_{ji}$ given by $M_{ij} = \int d^dx \left( \pi_j x_i - \pi_i x_j\right)$. Importantly, we do not require invariance under any kind of boost symmetry, Lorentzian or Galilean. Thus, we do not have a conserved boost generator. In infinite volume, so that the dipole symmetry is continuous, we also have the dipole moment
\beq
    D^i = \int d^dx \, x^i J^t\,,
\eeq
which is conserved thanks to~\eqref{E:dipoleWard} and \eqref{eq:flat-cons}, provided that $J^t$ falls off suitably fast at infinity. Note that this operator is only well-defined in infinite volume,\footnote{However, if space is a finite torus, note that $D^i$ is well-defined as an operator valued in the same torus thanks to charge quantization.} and that the flux associated with $D^i$ is almost but not quite $J^{ij}$. 
These charges generate an algebra with nonzero commutators~\cite{GromovMultipole}
\begin{align}
\begin{split}
    [M_{ij},D_k] &= i(\delta_{ik} D_j-\delta_{jk}D_i)\,,
    \\
    [M_{ij},P_k] & = i (\delta_{ik} P_j - \delta_{jk}P_i)\,, 
    \\
    [M_{ij},M_{kl}] &= i(\delta_{ik} M_{jl} - \delta_{il}M_{jk} - \delta_{jl}M_{ik} + \delta_{jk} M_{il})\,,
    \\
    [P_i,D_j] &= i\delta_{ij}Q\,.
    \label{DipolePoisson}
\end{split}
\end{align}
As a consequence of the symmetry algebra, the momentum operator shifts under dipole transformations, with $
\exp(-i\psi_k D^k) P_i \exp(i\psi_k D^k) = P_i - Q\psi_i$. 

The conserved currents transform as usual under spacetime symmetries and are invariant under U(1) transformations. Interestingly, there is a local version of the transformation law for momentum above, whereby the momentum density $\pi^i$ and spatial stress $\tau^{ij}$ are not invariant under dipole symmetry transformations~\cite{Jain:2021ibh}. The transformation laws for a constant dipole transformation in flat space are given in \eqref{eq:dipole-var-currents-flat}. As we will see in the next subsection, on a generic curved spacetime background, $\epsilon^t$, $\epsilon^i$ also transform under dipole transformations, while $J^t$, $J^i$, $J^{ij}$ are invariant.

%-----------------------------------------------
\subsection{Coupling to curved spacetime background}
\label{couplingtobackgrounds}
%-----------------------------------------------

In this subsection, we discuss how to couple field theories with dipole symmetries to curved spacetime background. This is a necessary ingredient to compute the response functions of conserved currents in the theory. It also makes manifest the transformation properties of these currents under global dipole transformations and will be indispensable for the construction of hydrostatic partition functions for dipole-invariant fluids. The discussion here is a digest of~\cite{Jain:2021ibh}; see also \cite{Bidussi:2021nmp}. Some of these matters were considered in~\cite{Slagle:2018kqf}.

%-----------------------------------------------
\subsubsection{Aristotelian backgrounds}
%-----------------------------------------------

We would like to couple dipole-symmetric field theories to a fixed curved spacetime. Systems of this sort do not respect any boost symmetry and generically couple to an ``Aristotelian'' spacetime~\cite{Novak:2019wqg, deBoer:2020xlc, Armas:2020mpr}, characterized by a nowhere-vanishing \emph{clock-form} $n_\mu$ and a degenerate \emph{spatial metric} tensor $h_{\mu\nu}$. Together, these background fields couple to the energy current, momentum density, and spatial stress tensor; we will return to the precise coupling structure in the following. In flat space, these sources take the trivial form $n_\mu = \delta_\mu^t$ and $h_{\mu\nu} = \delta_\mu^i\delta_\nu^j \delta_{ij}$. The zero eigenvector of $h_{\mu\nu}$ is called the \emph{frame velocity} $v^\mu$ and is normalized such that $n_\mu v^\mu = 1$. We can also define an inverse spatial metric $h^{\mu\nu}$ such that $h^{\mu\nu}n_\mu = 0$ and $h^{\mu\rho}h_{\rho\nu} = \delta^{\mu}_\nu - v^\mu n_\nu \equiv h^{\mu}_\nu$. Generally, we use $h^{\mu\nu}$ to raise indices. We can define a connection on this geometry 
\begin{align}
\begin{split}
\Gamma^{\lambda}{}_{\mu\nu} 
= v^\lambda\partial_\mu n_\nu 
+ \frac{1}{2}h^{\lambda\rho}\lb \partial_\mu h_{\nu\rho}
+\partial_\nu h_{\mu\rho} 
- \partial_\rho h_{\mu\nu} \rb,
\end{split}
\end{align}
where the covariant derivative $\nabla_{\mu}$ acts on tensors as
\beq
    \nabla_{\mu}\mathcal{T}^{\nu}{}_{\rho} = \partial_{\mu}\mathcal{T}^{\nu}{}_{\rho} + \Gamma^{\nu}{}_{\mu\sigma}\mathcal{T}^{\sigma}{}_{\rho} - \Gamma^{\sigma}{}_{\mu \rho}\mathcal{T}^{\nu}{}_{\sigma}\,.
\eeq
The covariant derivatives of the spacetime background reads
\begin{equation}
    \nabla_\lambda n_\nu = \nabla_\lambda h^{\mu\nu} = 0, \qquad 
    \nabla_\lambda h_{\mu\nu} = -n_{(\mu}\pounds_{v}h_{\nu)\lambda}, \qquad h_{\nu\mu}\nabla_\lambda v^\mu = \frac{1}{2}\pounds_v h_{\nu\lambda}.
\end{equation}
Here $\pounds_X$ denotes the Lie derivative along a vector field $X^\mu$. Note that this connection is torsional with $T^{\lambda}{}_{\mu\nu} \equiv 2\Gamma^\lambda{}_{[\mu\nu]} = v^\lambda F^n_{\mu\nu}$, where $F_{\mu\nu}^n = \partial_{\mu}n_{\nu}-\partial_{\nu}n_{\mu}$ is the vorticity of the clock-form. It is also helpful to define a volume measure $\gamma = \text{det}(n_\mu n_\nu+h_{\mu\nu})$. We can check that
\begin{align}
\frac{1}{\sqrt{\gamma}}\partial_\nu \sqrt{\gamma} 
= \Gamma^{\mu}{}_{\mu\nu} - v^\mu F^n_{\mu\nu}.
\end{align}
The curvature tensor associated with this connection is
\beq
    R^{\mu}{}_{\nu\rho\sigma} = \partial_{\rho}\Gamma^{\mu}{}_{\sigma\nu} - \partial_{\sigma}\Gamma^{\mu}{}_{\rho\nu} + \Gamma^{\mu}{}_{\rho \alpha}\Gamma^{\alpha}{}_{\sigma \nu} - \Gamma^{\mu}{}_{\sigma\alpha}\Gamma^{\alpha}{}_{\rho \nu}\,.
\eeq
In addition to the spacetime background, we introduce a background U(1) gauge field $A_\mu$ coupled to the conserved U(1) current. Finally, we need to introduce a symmetric tensor gauge field $a_{\mu\nu}$, satisfying $a_{\mu\nu}v^\nu = 0$, to couple to the symmetric dipole flux $J^{\mu\nu}$, satisfying $J^{\mu\nu}n_{\nu}=0$~\cite{Jain:2021ibh, Bidussi:2021nmp}. 

Due to the normalization conditions, the only independent components of the sources are $n_t$, $n_i$, $v^i$, $g_{ij}$, $A_t$, $A_i$, and $a_{ij}$. We can express them as
\begin{equation}
\begin{gathered}
    n_\mu =
    \begin{pmatrix}
         n_t \\ n_i
    \end{pmatrix}, \qquad 
    h_{\mu\nu} = 
    \begin{pmatrix}
       v^i v_i/(v^t)^2 & - v_j/v^t \\
       - v_i/v^t & g_{ij}
    \end{pmatrix}, \\ 
    v^\mu =
    \begin{pmatrix}
       (1 - v^i n_i)/n_t \\ v^i
    \end{pmatrix}, \qquad 
    h^{\mu\nu} =
    \begin{pmatrix}
        n^kn_k (v^t)^2 & -v^t(n^j-n^kn_k v^j) \\ 
        -v^t(n^i-n^kn_k v^i) & 
        n^k n_k v^i v^j - 2v^{(i}n^{j)}  + g^{ij}
    \end{pmatrix}, \\ 
    A_\mu =
    \begin{pmatrix}
         A_t \\ A_i
    \end{pmatrix}, \qquad 
    a_{\mu\nu} = 
    \begin{pmatrix}
        a_{ij}v^iv^j/(v^t)^2 & - a_{ij}v^i/v^t \\
       - a_{ij}v^j/v^t & a_{ij}
    \end{pmatrix}.
\end{gathered}
\label{eq:non-cov-sources}
\end{equation}
Here non-covariant indices are lowered and raised with $g_{ij}$ and its inverse $g^{ij}$ respectively. 

To define the dipole symmetry covariantly we promote it to a spurionic symmetry that can be imposed locally, in finite or infinite volume, and even in curved space. In flat space it acts by
\begin{align}
 \begin{split}
           A_t(t,\vec{x}) &\to A_t(t,\vec{x})\,, 
           \\
           A_i(t,\vec{x}) &\to A_i(t,\vec{x}) + \psi_i(t,\vec{x})\,, 
           \\
           a_{ij}(t,\vec{x}) &\to a_{ij}(t,\vec{x}) + \partial_i \psi_j(t,\vec{x}) + \partial_j \psi_i(t,\vec{x})\,,
\end{split}
\end{align}
with charged fields $\Psi$ inert. This spurionic symmetry implies the dipole Ward identity $J^i = \partial_j J^{ij}$ which must be accounted for in the low-energy hydrodynamic description we obtain later. 

In curved space, we impose invariance under diffeomorphisms, local U(1) transformations, and local dipole transformations. For an infinitesimal diffeomorphism $\chi^{\mu}$, U(1) transformation $\Lambda$, and dipole transformation $\psi_{\mu}$ satisfying $\psi_{\mu}v^{\mu}=0$, collectively denoted as $\hat\scX = (\chi^\mu,\Lambda,\psi_\mu)$, these act on the fields as
\begin{align}
\begin{split}
    \delta_{\hat\scX} n_\mu &= \lie_\chi n_\mu\,, 
    \\ 
    \delta_{\hat\scX} h_{\mu\nu} &= \lie_\chi h_{\mu\nu}\,, 
    \\
    \delta_{\hat\scX} A_\mu &= \lie_\chi A_\mu 
    + \partial_\mu\Lambda + \psi_\mu\,, 
    \\
    \delta_{\hat\scX} a_{\mu\nu} &= \lie_\chi a_{\mu\nu} 
    + h^{\rho}_\mu h^{\sigma}_\nu
    (\nabla_\rho \psi_\sigma + \nabla_\sigma \psi_\rho)\,.
\end{split}
\label{eq:back-var}
\end{align}
More details regarding these transformation properties can be found in~\cite{Jain:2021ibh}.

With the background sources in place, the curved space version of the scalar theory in~\eqref{E:pretkoS} reads
\beq
    S = \int \df^{d+1}x \sqrt{\gamma}\left( 
    i \Psi^{\dagger}v^{\mu}\Df_{\mu}\Psi 
    - \lambda \Df_{\mu\nu}(\Psi^\dagger,\Psi^\dagger)
    \Df^{\mu\nu}(\Psi,\Psi) 
    - V(\Psi^\dagger\Psi)\right)\,,
\eeq
with the derivative operators defined as
\begin{equation}
\begin{split}
    \Df_{\mu}\Psi 
    &= \nabla_{\mu}\Psi - i A_{\mu} \Psi\,, \\
    \Df_{\mu\nu}(\Psi,\Psi) 
    &= \frac{h_{\mu}^{\rho}h_{\nu}^{\sigma}}{2}\left( \Psi  \Df_{\rho}\Df_{\sigma} \Psi  
    + \Psi \Df_{\sigma}\Df_{\rho}\Psi
    - 2\Df_{\rho}\Psi \Df_{\sigma}\Psi\right) 
    + \frac{i}{2} a_{\mu\nu}\Psi^2\,.
\end{split}
\end{equation}
The definition of $\Df_\mu$ contains a Aristotelian covariant derivative when acting on tensor fields.
This action, so constructed, is invariant under finite diffeomorphisms, local U(1) transformations, and local dipole transformations, provided that we take the charged scalar to transform as 
\begin{align}
\begin{split}
    \delta_{\hat\scX} \Psi & = \lie_{\chi} \Psi + i \Lambda \Psi\,.
\end{split}
\end{align}

It is also helpful to define an effective dipole gauge field as
\begin{align}\begin{split}
A^{\lambda}_{\;\;\mu} 
    = n_\mu v^\rho F_{\rho\sigma}h^{\sigma\lambda}
    + \frac{1}{2}\left(h^{\rho}_\mu F_{\rho\sigma}h^{\sigma\lambda} 
    + a_{\mu\sigma}h^{\sigma\lambda}\right)\,,
\end{split}
\label{eq:Amixed-def}
\end{align}
satisfying $n_\lambda A^\lambda_{\;\;\mu} = 0$.
Here $F_{\mu\nu} = \partial_\mu A_\nu - \partial_\nu A_\mu$ is the U(1) field strength. This field contains information about both the U(1) gauge field $A_\mu$ and the symmetric tensor gauge field $a_{\mu\nu}$. It is a useful quantity because in the absence of Aristotelian sources, it transforms as a gauge field under dipole transformations, i.e. $A^\lambda_{\;\;\mu} \to A^\lambda_{\;\;\mu} + \partial_\mu \psi^\lambda$. More generally, its transformation is given as
\begin{align}\begin{split}
    A^{\lambda}_{\;\;\mu}
    \to A^\lambda_{\;\;\mu} 
    + \nabla_\mu\psi^{\lambda} 
    + n_\mu \psi^\nu\nabla_\nu v^\lambda.
    \label{eq:Amixed-trans}
\end{split}\end{align}
We can also construct the associated field strength 
\begin{align}\begin{split}
F^{\lambda}_{\;\;\mu\nu} 
= \nabla_\mu A^\lambda_{\;\;\nu} - \nabla_\nu A^{\lambda}_{\;\;\mu} + F^n_{\mu\nu}v^\rho A^\lambda_{\;\;\rho} + 2n_{[\mu}A^\rho_{\;\;\nu]}\nabla_\rho v^\lambda.
\end{split}\end{align}
This is invariant under dipole transformations in flat space, but transforms non-trivially when coupled to curved Aristotelian spacetime~\cite{Jain:2021ibh}.

%-----------------------------------------------
\subsubsection{Conservation equations in curved space}
%-----------------------------------------------

Having identified the appropriate background sources, we can define the conjugate symmetry currents. Given a field theory described by a generating functional $W=-i\ln{\cal Z}$, we define expectation values of the currents by
\begin{align}\begin{split}
\label{generatingfunctional1}
    \delta W
    = \int \df^{d+1}x\sqrt{\gamma} \left[
    - \epsilon^\mu \delta n_\mu 
    + \left(v^{(\mu}\pi^{\nu)}
    +\frac{1}{2}\tau^{\mu\nu}\right)\delta h_{\mu\nu} 
    + J^\mu \delta A_\mu
    + J^{\mu\nu}h_{\nu\lambda}\delta A^{\lambda}_{\;\;\mu}\right].
\end{split}\end{align}
Here $\epsilon^\mu$ is the energy current, $\pi^\mu$ (normalized as $\pi^\mu n_\mu = 0$) is the momentum density, $\tau^{\mu\nu}$ (normalized as $\tau^{\mu\nu} n_\nu = 0$ and $\tau^{\mu\nu} = \tau^{\nu\mu}$) is the stress tensor, $J^\mu$ is the charge current, and $J^{\mu\nu}$ (normalized as $J^{\mu\nu}n_\nu = 0$ and $J^{\mu\nu} = J^{\nu\mu}$)  is the dipole flux. In terms of the non-covariant currents appearing in \eqref{eq:flat-cons}, these are defined as
\begin{equation}
\begin{gathered}
    \epsilon^\mu = 
    \begin{pmatrix}
    \epsilon^t \\ \epsilon^i
    \end{pmatrix}, \qquad 
    \pi^\mu = 
    \begin{pmatrix}
       - \pi^i n_i/n_t \\ \pi^i
    \end{pmatrix}, \qquad 
    \tau^{\mu\nu} =
    \begin{pmatrix}
        \tau^{ij}n_in_j/n_t^2 & - \tau^{ij}n_i/n_t \\ 
        - \tau^{ij}n_j/n_t & \tau^{ij}
    \end{pmatrix} \\ 
    J^\mu = 
    \begin{pmatrix}
    J^t \\ J^i
    \end{pmatrix}, \qquad 
    J^{\mu\nu} =
    \begin{pmatrix}
        J^{ij}n_in_j/n_t^2 & - J^{ij}n_i/n_t \\ 
        - J^{ij}n_j/n_t & J^{ij}
    \end{pmatrix}.
\end{gathered}
\label{eq:non-cov-currents}
\end{equation}
Note that we have defined $J^{\mu\nu}$ using the coupling to the effective dipole gauge field $A^\lambda_{\;\;\mu}$. If we had instead coupled to the symmetric tensor gauge field $\delta a_{\mu\nu}$, various currents would need to be slightly adjusted on account of
\begin{align}
J^{\mu\nu}h_{\nu\lambda}\delta A^\lambda_{\;\;\mu} 
= J^{\mu\nu}v^\rho F_{\rho\nu}\delta n_\mu 
- J^{\mu\rho}A^\nu_{\;\;\rho}\delta h_{\mu\nu}
+ \frac{1}{2}J^{\mu\nu}\delta a_{\mu\nu}.
\end{align}
The benefit of using $A^\lambda_{\;\;\mu}$  instead of $a_{\mu\nu}$ as a source in (\ref{generatingfunctional1}) to define the various currents is that the transformations of the currents under the dipole shift symmetry, i.e.
\begin{align}
\begin{split}
  \epsilon^\mu
  &\to \epsilon^\mu
    + \lb 2J^{\mu(\rho}\psi^{\sigma)} - J^{\rho\sigma} \psi^\mu \rb
    \half\lie_v h_{\rho\sigma}\,,
    \\
  \pi^\mu
  &\to \pi^\mu - (J^\nu n_\nu) \psi^\mu
    + J^{\rho\mu} F^n_{\rho\sigma} \psi^\sigma\,,
    \\
  \tau^{\mu\nu}
  &\to \tau^{\mu\nu} 
  - 2\psi^{(\mu}\nabla'_\lambda J^{\nu)\lambda}
  + \nabla'_\lambda(\psi^\lambda J^{\mu\nu} )\,,
  \label{eq:dipole-var-currents}
\end{split}
\end{align}
are ``nicer'' in this case~\cite{Jain:2021ibh}. Note that we have introduced the notation $\nabla_\mu' \equiv \nabla_\mu + F^n_{\mu\nu} v^\nu$, defined such that we have $\int\df^{d+1}x\sqrt{\gamma}\,Y^{\mu\nu_1\ldots}\nabla_\mu X_{\nu_1\ldots} = -\int\df^{d+1}x\sqrt{\gamma}\,
\nabla'_\mu Y^{\mu\nu_1\ldots} X_{\nu_1\ldots}$ under integration-by-parts, plus a boundary term.

The currents $J^\mu$ and $J^{\mu\nu}$ are invariant under dipole shift transformations. Requiring the generating functional variation in \eqref{generatingfunctional1} to be invariant under diffeomorphisms, U(1) transformations, and dipole shift transformations in  \eqref{eq:back-var}, we are led to the respective conservation equations in the presence of background sources
\begin{align}\begin{split}
    \nabla_\mu' \epsilon^\mu 
    &= - v^\mu f_\mu 
    - (\tau^{\mu\nu}+\tau_d^{\mu\nu})
    h_{\lambda\nu}\nabla_\mu v^\lambda, \\
    \nabla_\mu'(v^\mu\pi^\nu + \tau^{\mu\nu}+\tau_d^{\mu\nu})
    &= h^{\nu\mu}f_\mu - \pi_\mu h^{\nu\lambda}\nabla_\lambda v^\mu, \\
    \nabla_\mu' J^\mu 
    &= 0, \\
    \nabla_\mu' J^{\mu\nu} 
    &= h_\mu^\nu J^\mu,
    \label{eq:cons-cov}
\end{split}\end{align}
where we have defined the effective Lorentz force $f_\mu$ and dipole stress tensor $\tau^{\mu\nu}_{d}$ as
\begin{align}\begin{split}
    f_\mu 
    &= -F_{\mu\nu}^n\epsilon^\nu 
    - h_{\mu\lambda} {A}^\lambda_{\;\;\nu}J^\nu 
    + F^{\lambda}_{~\mu\nu} J^{\nu}_{~\lambda}
    - n_\mu A^\lambda_{\;\;\rho} J^\rho_{\;\;\nu} \nabla_\lambda v^\nu, \\
    \tau_d^{\mu\nu} 
    &= -A^\mu_{~\rho} J^{\rho\nu}.
\end{split}\end{align}
These boil down to the conservation equations in \eqref{eq:flat-cons} in the absence of background sources. After some work, one can verify that the conservation equations are invariant under dipole shifts of the conserved currents given in \eqref{eq:dipole-var-currents}.

\iffalse:
\AJ{The energy and momentum equations can be combined into
\begin{align}
    \nabla_\mu'(
    - \epsilon^\mu n_\nu
    + v^\mu\pi_\nu 
    + \tau^{\mu}{}_\nu +\tau_d^{\mu}{}_\nu)
    &= f_\nu
    - \pi_\mu \nabla_\nu v^\mu
\end{align}
}
\fi

Were we to interpret $A^\lambda_{\;\;\mu}$ as an independent field transforming like \eqref{eq:Amixed-trans}, we could introduce an independent operator $J^{[\mu\nu]}\neq 0$ which would amount to adding intrinsic dipoles. However, in this work, we will always take symmetric dipole gauge field $a_{\mu\nu}$ to be fundamental, with $A^\lambda_{\;\;\mu}$ defined via \eqref{eq:Amixed-def}. Therefore, we don't have access to the operator $J^{[\mu\nu]}\neq 0$ in our description.

%-----------------------------------------------
\subsection{Spontaneous breaking of dipole symmetry}
\label{dipoleinvariantfields}
%-----------------------------------------------

We would like to obtain a thermodynamic and ultimately hydrodynamic description of many-body systems with dipole symmetry. A consequence of the dipole shift transformations \eqref{eq:dipole-var-currents-flat} is that the homogeneous equilibrium states of such a system must either have no charge or no momentum. A uniform slab of moving charge does not conserve dipole moment and hence is disallowed by dipole symmetry. A more formal argument using the representation theory of the dipole and translation symmetries leads to the same result~\cite{Jensen:2022iww}. Furthermore, as we explained in the Introduction, there is no transport of charge in a dipole-symmetric phase and the hydrodynamic description is effectively that of a neutral fluid.

Therefore, dipole-invariant systems with charge transport must have their dipole symmetry spontaneously broken.\footnote{A large $N$ model demonstrating this breaking was discovered in \cite{Jensen:2022iww}.} Spontaneous breaking of the dipole symmetry gives rise to a UV-sensitive degeneracy of ground states in dipole-invariant systems, as one finds in gapped phases of fracton order~\cite{Jensen:2022iww}. The spontaneous breaking of dipole symmetry is characterized by a vector Goldstone $\phi_i$ in the low-energy effective theory that transforms nonlinearly under dipole transformations $\psi_i$ as a shift, i.e.
\begin{equation}
    \phi_i \to \phi_i - \psi_i.
\end{equation}
More generally, we can also consider systems where the U(1) symmetry is spontaneously broken, giving rise to a scalar Goldstone $\phi$ that transforms as $\phi \to \phi - \Lambda$ under local U(1) transformations. Spontaneously breaking the U(1) symmetry automatically spontaneously breaks the associated dipole symmetry, but the would be dipole Goldstone $\phi_i$ is generally massive.\footnote{In a theory with a dipole Goldstone $\phi_i$ on top of the U(1) Goldstone $\phi$, we can create a dipole-invariant vector field $\phi_i + \dow_i\phi + A_i$. In this case, the free energy density of the system generically contains a mass term for $\phi_i$ like ${\cal F} \sim m^2 (\phi_i + \dow_i\phi + A_i)(\phi^i + \dow^i\phi + A^i)$. extremizing the free energy leads to $\phi_i = -\dow_i\phi - A_i + \ldots$ and can be used to algebraically integrate out $\phi_i$ from the low-energy description. The situation here is similar to how spontaneously breaking translational symmetry automatically spontaneously breaks the rotation symmetry as well. \label{foot:both-breaking-hs}} We will revisit this possibility in Section \ref{S:swave}. 

The identification of a Goldstone boson implementing the dipole shift symmetry in the hydrodynamics of dipole-invariant systems has also been described in \cite{Glorioso:2023chm,Glodkowski:2022xje}, where such a Goldstone boson is implicit in that the momentum density $\pi_i \propto J^t\phi_i$. This is a necessary consequence of the symmetry algebra \eqref{DipolePoisson}. However, this is only the case when $\langle \partial_i \phi_j \rangle = 0$, in the absence of external sources, and to leading order in gradients. A non-vanishing expectation value for this operator is analogous to a background superflow in a U(1) superfluid \cite{landaubook, Gouteraux:2022kpo}. In a forthcoming work \cite{EricDipoleSuperflow}, we investigate the consequences of such a superflow. Nevertheless, it is worthwhile to consider the simpler and perhaps more physical case of $\langle \partial_i \phi_j \rangle = 0$ first, which will be the concern of this manuscript. We emphasize that our construction differs somewhat from \cite{Glorioso:2023chm,Glodkowski:2022xje} in that we are able to explicitly identify the relationship between $\phi_i$ and fluctuating background sources and in particular are able to write down all response functions in this limit. Furthermore, we are able to include non-linear interactions among components of the dipole Goldstone. Our formalism also naturally extends to scenarios with nonzero $\langle \partial_i \phi_j \rangle$. 

We can introduce a curved space version of the dipole Goldstone $\phi_\mu$, normalized such that $\phi_\mu v^\mu = 0$, where the time-component is defined as $\phi_t = - \phi_iv^i/v^t$. The covariant dipole Goldstone field transforms under symmetry transformations parameterised by $\hat\scX = (\chi^\mu,\Lambda,\psi_\mu)$ as
\beq
    \delta_{\hat\scX}\phi_{\mu} = \lie_\chi\phi_\mu - \psi_\mu\,.
\eeq
The generating functional $W$ is given by a path integral over possible configurations of $\phi_\mu$, i.e. $W = -i\ln\int [\df\phi_\mu] \exp (iS[\phi])$. The off-shell variations of the effective action $S$ with respect to the background and dynamical fields can be parametrized similar to \eqref{generatingfunctional1} as 
\begin{align}
    \delta S 
    = \int \df^{d+1}x\sqrt{\gamma} \left[
    - \epsilon^\mu \delta n_\mu 
    + \left(v^{(\mu}\pi^{\nu)}
    +\frac{1}{2}\tau^{\mu\nu}\right)\delta h_{\mu\nu} 
    + J^\mu \delta A_\mu
    + J^{\mu\nu}h_{\nu\lambda}\delta A^{\lambda}_{~\mu}
    + X^\mu\delta\phi_\mu
    \right],
\label{eq:new-var-W}
\end{align}
where $X^\mu=0$, normalized as $X^\mu n_\mu = 0$, is the equation of motion for the dipole Goldstone. We note that the conservation equations for off-shell configurations of $\phi_\mu$ modify \eqref{eq:cons-cov} to 
\begin{align}\begin{split}
    \nabla_\mu' \epsilon^\mu 
    &= - v^\mu f_\mu 
    - (\tau^{\mu\nu}+\tau_d^{\mu\nu})
    h_{\lambda\nu}\nabla_\mu v^\lambda, \\
    \nabla_\mu'(v^\mu\pi^\nu + \tau^{\mu\nu}+\tau_d^{\mu\nu})
    &= h^{\nu\mu}f_\mu - \pi_\mu h^{\nu\lambda}\nabla_\lambda v^\mu, \\
    \nabla_\mu' J^\mu 
    &= 0, \\
    \nabla_\mu' J^{\mu\nu} 
    &= h_\mu^\nu J^\mu - X^\nu.
    \label{eq:cons-cov-off-shell}
\end{split}\end{align}
Note that the U(1) conservation equation remains the same; the energy and momentum conservation equations also remain the same in form, but the Lorentz force is now modified with
\begin{align}\begin{split}
    f_\mu 
    &= -F_{\mu\nu}^n\epsilon^\nu 
    - h_{\mu\lambda} {A}^\lambda_{\;\;\nu}J^\nu 
    + F^{\lambda}_{~\mu\nu} J^{\nu}_{~\lambda}
    - n_\mu A^\lambda_{\;\;\rho} J^\rho_{\;\;\nu} \nabla_\lambda v^\nu \\
    &\qquad 
    + \lb A^\lambda_{~\mu} + \nabla_\mu\phi^\lambda \rb X_\lambda
    - \nabla'_\nu (\phi_\mu X^\nu), \\
    \tau_d^{\mu\nu} 
    &= -A^\mu_{~\rho} J^{\rho\nu}.
    \label{eq:f-X}
\end{split}\end{align}
The dipole invariance of the action implies
\begin{align}
X^\mu = 0 \quad  \Longleftrightarrow \quad
\nabla_{\mu}'J^{\mu\nu} = h^\nu_\mu J^\mu,
\end{align}
so the equations of motion for the dipole Goldstone implement dipole conservation on-shell. 

We are welcome to work with the Goldstone as an independent field in the low-energy description, but we find that coupling to background sources is greatly simplified if we instead work with dipole-invariant combinations of the fields and currents. Using the transformation properties of the dipole Goldstone $\phi_\mu$, we can define dipole-invariant versions of the U(1) gauge field, symmetric tensor gauge field, and auxiliary dipole gauge field
\begin{align}\begin{split}
    \tilde{A}_\mu 
    &= A_\mu + \phi_\mu, \\
    \tilde{a}_{\mu\nu} 
    &= a_{\mu\nu} + h^\rho_\mu h^\sigma_\nu (\nabla_{\rho}\phi_\sigma + \nabla_\sigma \phi_\rho), \\ 
    \tilde{A}^\lambda_{\;\;\mu}
    &= A^\lambda_{\;\;\mu}
    + \nabla_\mu \phi^\lambda
    + n_\mu \phi^\nu \nabla_\nu v^\lambda \\ 
    &= n_\mu v^\rho\tilde{F}_{\rho\sigma}h^{\sigma\lambda} + \frac{1}{2}\left(h^\rho_\mu\tilde{F}_{\rho\sigma}h^{\sigma\lambda}+\tilde{a}_{\mu\sigma}h^{\sigma\lambda}\right),
\end{split}\end{align}
where $\tilde{F}_{\mu\nu}  = \partial_\mu \tilde{A}_\nu -\partial_\nu \tilde{A}_\mu$ and $\phi^{\mu}=h^{\mu\nu}\phi_{\nu}$. Similarly, we can define the field strength associated with the effective dipole gauge field
\begin{align}
\begin{split}
    \tilde{F}^{\lambda}_{\;\;\mu\nu} 
    = \nabla_\mu \tilde{A}^\lambda_{\;\;\nu} 
    - \nabla_\nu \tilde{A}^{\lambda}_{\;\;\mu} 
    + F^n_{\mu\nu}v^\rho \tilde{A}^\lambda_{\;\;\rho} 
    + 2n_{[\mu}\tilde{A}^\rho_{\;\;\nu]}\nabla_\rho v^\lambda.
\end{split}
\end{align}
We can also define dipole-invariant versions of the symmetry currents
\begin{align}
\begin{split}
\label{dipoleinvariantcurrents}
    \tilde\epsilon^\mu 
    &= \epsilon^\mu 
    + \lb 2J^{\mu(\rho}\phi^{\sigma)}
    - J^{\rho\sigma}\phi^\mu\rb
    \frac{1}{2}\pounds_v h_{\rho\sigma},\\
    \tilde\pi^\mu 
    &= \pi^\mu - (J^\nu n_\nu)\phi^\mu 
    + J^{\rho\mu}F^n_{\rho\sigma}\phi^\sigma, \\
    \tilde\tau^{\mu\nu}
    &= \tau^{\mu\nu} 
    - 2\phi^{(\mu}\nabla'_\lambda J^{\nu)\lambda}
    + \nabla_\lambda'\!\lb\phi^\lambda J^{\mu\nu}\rb,
\end{split}
\end{align} 
which are conjugate to variations of the dipole-invariant sources,
\begin{align}
\begin{split}
\label{deltaWdipoleinvariant}
    \delta S 
    &= \int \df^{d+1}x
    \sqrt{\gamma}\bigg[
    - \tilde\epsilon^\mu\delta n_\mu 
    + \lb v^{(\mu}\tilde\pi^{\nu)}
    + \frac{1}{2}\tilde\tau^{\mu\nu}\rb \delta h_{\mu\nu} 
    + J^\mu\delta \tilde{A}_\mu 
    + J^{\mu\nu}h_{\nu\lambda}\delta \tilde{A}^{\lambda}_{~\mu} \\
    &\hspace{22em}
    + \lb \nabla_\nu' J^{\nu\mu} 
    - h_\nu^\mu J^\nu 
    + X^\mu \rb \delta\phi_\mu
    \bigg].
\end{split}
\end{align}
All dependence on $\phi_\mu$ in this expression appears implicitly via the dipole-invariant combinations of various objects. The dipole-invariant currents satisfy an analogous set of conservation equations
\begin{align}
\begin{split}
    \nabla_\mu' \tilde\epsilon^\mu 
    &= -v^\mu \tilde f_\mu 
    - (\tilde\tau^{\mu\nu} + \tilde\tau^{\mu\nu}_d)
    h_{\lambda\nu}\nabla_\mu v^\lambda, 
    %\label{eq:inv-Wardidentity-1} 
    \\ 
    \nabla_\mu'\lb v^\mu \tilde\pi^\nu 
    + \tilde\tau^{\mu\nu} + \tilde\tau^{\mu\nu}_d \rb
    &= h^{\nu\mu}\tilde f_\mu 
    - \tilde\pi_\mu h^{\nu\lambda}\nabla_\lambda v^\mu,
    %\label{eq:inv-Wardidentity-2} 
    \\ 
    \nabla_\mu'J^\mu
    &= 0, %\label{eq:inv-Wardidentity-3} 
    \\
    \nabla_\mu' J^{\mu\nu} 
    &= h^\nu_\mu J^\mu - X^\nu,
    %\label{eq:inv-Wardidentity-4}
    \label{eq:dipole-invariant-Wardidentity}
\end{split}
\end{align}
where the dipole-invariant Lorentz force $\tilde f_\mu$ and dipole stress tensor $\tilde\tau_d^{\mu\nu}$ are now given as
\begin{align}
\begin{split}
    \tilde f_\mu 
    &= -F^n_{\mu\nu}\tilde\epsilon^\nu 
    - h_{\mu\lambda}\tilde{A}^\lambda_{~\nu}J^\nu 
    + \tilde{F}^\lambda_{\;\;\mu\nu}h_{\rho\lambda}J^{\nu\rho} 
    - n_\mu \tilde{A}^\lambda_{\;\;\rho}
    J^\rho_{\;\;\nu}\nabla_\lambda v^\nu
    + \tilde A^\lambda_{~\mu} X_\lambda, \\
    \tilde\tau_d^{\mu\nu} 
    &= - \tilde{A}^\mu_{\;\;\rho}J^{\rho\nu}.
    \label{eq:tf-X}
\end{split}
\end{align}
We should note that the dipole conservation equation in this interpretation is a consequence of the dipole Goldstone equation of motion, since all fields in \eqref{deltaWdipoleinvariant} are manifestly dipole-invariant.

In (\ref{dipoleinvariantcurrents}), we see that when $\langle \tilde\pi^\mu \rangle = 0$, the charge density $n_\nu \langle J^\nu \rangle = q$ is nonzero, and the clock-form torsion $F^n_{\mu\nu}$ vanishes, then the momentum density is proportional to the Goldstone boson,
\beq
    \langle \pi^\mu \rangle = q\langle \phi^\mu \rangle\,.
\eeq
It turns out that this identity holds in a homogeneous equilibrium state, but only under the assumptions mentioned above. A novel feature of our construction is that it allows us to discuss the momentum current and dipole Goldstone separately. This is also crucial when there is an equilibrium superflow, $\langle \partial_i\phi_j \rangle \neq 0$. For our construction of superfluid hydrodynamics to be consistent, the system must be linearly stable over some range of finite superflow, even in the absence of an underlying boost symmetry (an argument for standard U(1) superfluids is in \cite{Gouteraux:2022kpo}). We will explore this feature in an upcoming work \cite{EricDipoleSuperflow}. Now that we have discussed coupling the dipole Goldstone to a background spacetime, we can move ahead with our hydrodynamic construction.

%-----------------------------------------------
\subsection{Derivative counting}
\label{sec:derivative-counting}
%-----------------------------------------------

A distinctive feature of hydrodynamics for dipole-symmetric systems is the derivative counting of the hydrodynamic variables, symmetry currents, and background fields. For example,  p-wave dipole superfluids have a dynamical scaling exponent $z=2$, with time derivatives counting as second order in spatial derivatives $\dow_t \sim {\cal O}(\dow^2)$, $\dow_i\sim {\cal O}(\dow^1)$. This derivative counting will be crucial in the remainder of this manuscript, therefore we spend some time to arrive at it systematically.

To arrive a consistent derivative counting, we need to postulate the spectrum of hydrodynamic variables for these fluids. As we mentioned in the Introduction, we postulate a variable for each conserved quantity, a local temperature $T$, chemical potential $\mu$, and velocity $u^i$, in addition to a dipole Goldstone $\phi_i$. We are interested in describing thermodynamic systems with nonzero charge density, energy density, and pressure, so we enforce the leading order terms in $J^t$, $\epsilon^t$, and $\tau^{ij}$ to be ${\cal O}(\dow^0)$. We take the local temperature $T$ and chemical potential $\mu$ to similarly be $\mathcal{O}(\dow^0)$. In dipole-symmetric systems there are no homogeneous equilibrium states with nonzero charge density and finite velocity~\cite{Jensen:2022iww}. Therefore, velocity is always a fluctuation around an equilibrium state, and for this reason we take the spatial velocity $u^i$ to be $\mathcal{O}(\dow^1)$. As with the Goldstone mode of an ordinary superfluid, we take the dipole Goldstone to be of $\mathcal{O}(\dow^{-1})$.\footnote{One way to understand this scaling uses Euclidean thermal field theory. In dipole superfluid phases at finite temperature, the zero-frequency and long-wavelength physics is described by an effective theory of the Matsubara zero mode of the dipole Goldstone. In the absence of background fields, the effective theory has a leading order Lagrangian $\mathcal{L}_{\rm eff} = \half\lb B_d + \frac{d-2}{d}G_d \rb (\dow_k\phi^k)^2 + \half G_d\,\dow_i\phi_j\dow^i\phi^j + \hdots$, so that the zero-frequency small-wavevector two-point function of $\phi_i$ behaves as $\langle \phi_i(\vec{k})\phi_j (-\vec{k})\rangle \sim 1/k^2$. } This counting for the dipole Goldstone field is the same as in recent works~\cite{Glorioso:2023chm,Glodkowski:2022xje}.

The form of the dipole-invariant momentum density $\tilde{\pi}^{\mu}$ in~\eqref{dipoleinvariantcurrents} implies that $\pi^\mu$ contains a term proportional to $\phi^\mu$, meaning that $\pi^{\mu}$ scales as $\mathcal{O}(\dow^{-1})$. The source-free momentum conservation equation $\dow_t{\pi}^i + \partial_j \tau^{ij}=0$, together with the fact that $\tau^{ij} \sim \mathcal{O}(\dow^0)$, then implies that time derivatives count twice as much as spatial derivatives. The energy and U(1) conservation equations further imply that the fluxes $\epsilon^i$, $J^i$ are both of $\mathcal{O}(\partial^1)$. To summarize, we arrive at a counting for the leading order symmetry currents,
\begin{gather}
    J^t, \epsilon^t, \tau^{ij}, J^{ij} \sim {\cal O}(\dow^0) , \qquad 
    J^i, \epsilon^i \sim {\cal O}(\dow^1), \qquad 
    \pi_i \sim {\cal O}(\dow^{-1})\,.
\end{gather}
Given that the generating functional $W$ in \eqref{generatingfunctional1} should be counted as ${\cal O}(\dow^0)$, this in turn induces a counting prescription for the background fields conjugate to the currents,
\begin{gather}
    A_t, n_t, g_{ij}, a_{ij} \sim {\cal O}(\dow^0) , \qquad 
    A_i, n_i \sim {\cal O}(\dow^{-1}), \qquad 
    v^i \sim {\cal O}(\dow^{1}).
\end{gather}
This appropriate derivative counting of background spacetime fields is novel to our work.

With this in place, one can check that the dipole-invariant quantities $\tilde\epsilon^t$, $\tilde\epsilon^i$, $\tilde\pi^i$, $\tilde\tau^{ij}$, $\tilde A_t$, $\tilde A_i$, and $\tilde a_{ij}$ have the same derivative counting as their physical counterparts. It is also useful to note the derivative orders of different components of the connection
\begin{gather}
    \Gamma^t{}_{tt} \sim {\cal O}(\dow^2)\,, \qquad 
    \Gamma^t{}_{ti}, \Gamma^t{}_{it} \sim {\cal O}(\dow^1)\,, \qquad
    \Gamma^t{}_{ij} \sim {\cal O}(\dow^0)\,, \nn\\
    \Gamma^k{}_{tt} \sim {\cal O}(\dow^3)\,, \qquad 
    \Gamma^i{}_{tj}, \Gamma^i{}_{jt} \sim {\cal O}(\dow^2)\,, \qquad
    \Gamma^k{}_{ij} \sim {\cal O}(\dow^1)\,.
\end{gather}

Different derivative counting between space and time components of various objects makes this derivative counting scheme a bit unnatural to implement covariantly. Therefore, it is convenient to formally treat $\partial_\mu \sim {\cal O}(\dow^1)$ and $v^\mu \sim {\cal O}(\dow^1)$, so that the covariant time-derivative is $v^\mu\dow_\mu \sim {\cal O}(\dow^2)$. In the end, this covariant scheme gives rise to the same physics as the non-covariant scheme outlined above. For the conserved currents, this prescribes 
\begin{gather}
    \tau^{\mu\nu}, J^{\mu\nu} \sim {\cal O}(\dow^0) , \qquad 
    J^\mu, \epsilon^\mu \sim {\cal O}(\dow^1), \qquad 
    \pi_\mu \sim {\cal O}(\dow^{-1}),
\end{gather}
together with
\begin{gather}
    h_{\mu\nu}, a_{\mu\nu} \sim {\cal O}(\dow^0) , \qquad 
    A_\mu, n_\mu \sim {\cal O}(\dow^{-1}), \qquad 
    v^\mu \sim {\cal O}(\dow^1).
\end{gather}
Note that the covariant charge and energy density $n_\mu J^\mu$ and $n_\mu\epsilon^\mu$ are still counted as ${\cal O}(\dow^0)$ in this covariant scheme. For the Goldstone field and thermodynamic parameters, we have
\begin{equation}
    \phi_\mu \sim {\cal O}(\dow^{-1}), \qquad 
    \mu, T \sim {\cal O}(\dow^0), \qquad 
    u^\mu \sim {\cal O}(\dow^1),
\end{equation}
where $u^{\mu}$ is the covariant version of fluid velocity defined such that $u^\mu n_\mu = 1$. Note that under the covariant scheme, the connection has a uniform derivative order
\begin{equation}
    \Gamma^\lambda{}_{\mu\nu} \sim {\cal O}(\dow^1).
\end{equation}
The different countings of different time components can then be understood as arising from different contractions with $n_\mu$ and $v^\mu$, respectively.

%-----------------------------------------------
\section{Hot dipoles in thermal equilibrium}
\label{sec:hydrostatics}
%-----------------------------------------------

In this section, we discuss fluids with dipole symmetry in hydrostatic equilibrium. We start with a formal discussion of hydrostatic partition functions~\cite{Banerjee:2012iz,Jensen:2012jh,Bhattacharyya:2012xi,Jensen:2013kka} (see~\cite{Banerjee:2015uta, Jain:2020vgc, Armas:2020mpr} for discussion specialised to non-relativistic fluids), and again demonstrate the necessity of spontaneously breaking the dipole symmetry. We discuss ideal fluids with dipole symmetry in hydrostatic equilibrium and outline the roadmap to including derivative corrections.

%-----------------------------------------------
\subsection{Hydrostatic partition function}
%-----------------------------------------------

Let us warm up with a formal discussion of systems with conserved dipole moment at finite temperature. The primary object of interest in thermal equilibrium is the partition function(al) that can be used to obtain the thermodynamic charge densities and fluxes in equilibrium. In the grand canonical ensemble, in flat space, the equilibrium thermal state of a system can be described by the partition function
\begin{align}
  \exp(-\beta_0 W_\eqb)
  &= \tr\exp(-\beta_0\mathcal{H}_\scK),
\label{eq:W-def}
\end{align}
where $\beta_0 = 1/T_0$ is the inverse temperature of the thermal state and $\mathcal{H}_\scK$ is the Hamiltonian operator defined as
\begin{align}
    \mathcal{H}_\scK 
    = \int \df^dx \lb \epsilon^t 
    - u_0^i \pi_i - \mu_0 J^t \rb.
    \label{eq:H-K}
\end{align}
Here $\scK = (K^\mu,\Lambda_K)$ collectively labels the equilibrium thermodynamic temperature $T_0$, velocity $u^i_0$, and chemical potential $\mu_0$, via $K^\mu = \beta_0(1,u^i_0)$ and $\Lambda_K = \beta_0\mu_0$. In the context of field theory, the trace operation above can be understood as a path integral over all configurations of the dynamical fields on a Euclidean space where Euclidean time has periodicity $\beta_0$. Note that, while the definition of the
thermal partition function makes explicit reference to a preferred time-coordinate, both $\mathcal{H}_\scK$ and $W_\eqb$ are actually time-independent due to the conservation equations.

As we discussed in the previous subsection, for a system with conserved dipole moment, the momentum density $\pi_i$ shifts under dipole transformations as $\pi_i \to \pi_i - J^t\psi_i$. Consequently, the chemical potential $\mu_0$ must also shift as $\mu_0 \to \mu_0 + u^i_0 \psi_i$ while keeping $u^i_0$ unchanged, so as to keep the thermal state invariant. This non-invariance of the chemical potential has far reaching consequences for the thermodynamics of the system. In particular, in the ``ordinary'' phase where none of the symmetries are spontaneously broken, the grand canonical equation of state $p_0 = p(T_0,\mu_0,\vec u_0^2)$ is no longer allowed to have any dependence on $\mu_0$ as long as $u^i_0 \neq 0$. Consequently, the associated number density $n_0 = \dow p_0/\dow\mu_0$ is identically zero in equilibrium. This is to be expected of fractonic systems with conserved dipole moment, because a state with nonzero density cannot have nonzero momentum and vice-versa~\cite{Jensen:2022iww}. Therefore, to have non-trivial thermodynamic phases in a system with conserved dipole moment and momentum, the dipole symmetry must be spontaneously broken.

The definition of the thermal state can, in principle, be generalized on any time-independent background. The rationale is that, given enough time, any physical system coupled to time-independent sources will tend to relax back to its thermal equilibrium. To this end, consider coupling the physical system of interest to a set of background sources $n_\mu$, $h_{\mu\nu}$, $A_\mu$, and in the present case $A^\lambda_{~\mu}$, that admit a \emph{thermal isometry} $\scK = (K^\mu,\Lambda_K)$, i.e.
\begin{equation}
  \lie_K n_\mu = \lie_K h_{\mu\nu} = \lie_K A_\mu + \dow_\mu \Lambda_K
  = \lie_K A^\lambda_{~\mu} =  0.
\end{equation}
The isometry requirements also hold for $v^\mu$ and $h^{\mu\nu}$. The thermal isometry $\scK$ characterizes the observer with respect to which the system is in thermodynamic equilibrium, and the requirements above are the statement that background sources must be static in the reference frame of the equilibrium observer. Without loss of generality, we can take $K^\mu = \delta^\mu_t/T_0$ and $\Lambda_K = \mu_0/T_0$, however it is convenient to stick to the covariant notation. On a time-independent background, we can define a thermal state via the partition function~\eqref{eq:W-def} as before, but with the definition of
$\mathcal{H}_\scK$ generalized to include background sources as
\begin{align}\begin{split}
  \mathcal{H}_\scK
  &= T_0\int\df\Sigma_\mu
  \Big[ K^\nu n_\nu \epsilon^\mu
    - K^\nu h_{\nu\rho} \lb v^\mu \pi^{\rho}
    + \tau^{\mu\rho} + \tau^{\mu\rho}_{\text d} \rb
    - (\Lambda_K + K^\nu A_\nu) J^\mu
    - K^\nu A^\lambda_{~\nu} J^\mu_{~\lambda}
    \Big] \\ 
    &= \int\df^d x\,\sqrt{\gamma}
  \Bigg[ n_t \epsilon^t
    + \lb v_{i} + \frac{v_{k}v^k n_i}{1-v^kn_k} \rb \lb
    \pi^{i}
    - \frac{n_j}{1-v^kn_k}
    (\tau^{ji} + \tau^{ji}_{\text d}) \rb \\
    &\hspace{10em}
    - (\mu_0 + A_t) J^t
    - \lb A^i_{~t} + A^k_{~t} \frac{n_k v^i}{1-v^kn_k} \rb J^t_{~i}
    \Bigg].
\end{split}\end{align}
Here $\df\Sigma_\mu$ is the spatial volume element associated with a Cauchy slice transverse to the timelike vector $K^\mu$. In the second line, we have written the same expression in non-covariant notation for the benefit of the reader. It can be explicitly checked that $\mathcal{H}_\scK$ is invariant under ``time-independent'' symmetry transformations $\chi^\mu$, $\Lambda$, $\psi_\mu$, satisfying
\begin{equation}
  \lie_K \chi^\mu = \lie_K\Lambda - \lie_\chi \Lambda_K
  = \lie_K \psi_\mu = 0.
\end{equation}
Similarly, $\mathcal{H}_\scK$ is invariant under the choice of Cauchy slice used to define it, i.e. is time-independent, provided that the Ward identities are satisfied. This final property follows from the fact that the integrand in $\mathcal{H}_\scK$ above is conserved, i.e.
\begin{align}\begin{split}
  % - \epsilon^\mu \delta_\scB n_\mu
  % &+ \lb v^{(\mu} \pi^{\nu)} + \half \tau^{\mu\nu}
  % \rb \delta_\scB  h_{\mu\nu} + J^\mu \delta_\scB  A_\mu
  % + J^{\mu}{}_\lambda\delta_\scB 
  % A^\lambda{}_{\mu} \nn\\
  \nabla'_\mu
    &\Big( K^\nu n_\nu \epsilon^\mu
    - K^\lambda h_{\lambda\nu}\lb v^{\mu} \pi^{\nu} + \tau^{\mu\nu}
    + \tau^{\mu\nu}_{\text d} \rb 
    - \mu J^\mu
    - K^\nu A^\lambda_{~\nu} J^{\mu}_{~\lambda}
    \Big) \\
  &= 
    K^\nu n_\nu\lb \nabla'_\mu\epsilon^\mu
    + v^\mu f_\mu
    + (\tau^{\mu\rho} + \tau^{\mu\rho}_{\text{d}}) h_{\rho\nu} \nabla_\mu v^\nu \rb \\
  &\qquad
    - K^\lambda h_{\lambda\nu} \lb
    \nabla'_\mu \lb v^{\mu} \pi^{\nu}
    + \tau^{\mu\nu} + \tau^{\mu\nu}_{\text d} \rb
    - h^{\nu\mu}f_\mu
    + \pi^\mu \nabla_\mu v^\nu \rb \\
  &\qquad
    - \mu \nabla'_\mu J^\mu
    - K^\lambda A^\nu_{~\lambda}
    \lb \nabla'_\mu J^{\mu}{}_\nu - h_{\nu\mu}J^\mu \rb \\
  &= 0.
\end{split}\end{align}
The partition function $W_\eqb$ is now a functional of the background fields and can be used to compute the equilibrium expectation values and correlators of various currents. 

When the dipole symmetry is spontaneously broken, the hydrostatic generating functional~\cite{Jensen:2012jh, Banerjee:2012iz} can be expressed as a Euclidean path integral over the hydrostatic configurations of the Goldstone $\phi_\mu$ (as for an ordinary superfluid~\cite{Bhattacharyya:2012xi}), satisfying $\lie_K\phi_\mu = 0$, i.e.
\begin{equation}
  \exp(-\beta_0 W_\eqb)
  = \int [\df \phi_\mu] 
  \exp\lb -\int \df\Sigma_\mu K^\mu\,
  \mathcal{F}(n_\mu,h_{\mu\nu},
  \tilde A_\mu,\tilde a_{\mu\nu})
  \rb.
\end{equation}
Here, $\mathcal{F}$ is the free energy density, which is a scalar under diffeomorphisms and is required to be invariant under U(1) transformations. We have expressed the free energy in terms of dipole-invariant fields, so it is manifestly invariant under dipole shift symmetry. For later use, let us identify the local gauge- and dipole-invariant versions of thermodynamic parameters: local temperature $T$, velocity field $u^\mu$, and U(1) chemical potential $\mu$ as     
\begin{subequations}
\begin{align}\begin{split}
    T = \frac{1}{n_\nu K^\nu}, \qquad 
    u^{\mu} = \frac{K^\mu}{n_\nu K^\nu}, \qquad 
    \mu = \frac{\Lambda_K + K^\lambda {A}_\lambda}
    {n_\nu K^\nu}~.
\end{split}\end{align}
With these definitions, the velocity is normalized as $n_\mu u^\mu = 1$. Note that $T$, and $u^\mu$ are dipole-invariant but $\mu$ transforms as $\mu \to \mu + u^\mu\psi_\mu$. We can define a dipole-invariant version of the chemical potential by using the dipole-invariant gauge field $\tilde A_\mu$ as
\begin{equation}
    \tmu = \frac{\Lambda_K + K^\lambda \tilde{A}_\lambda}{n_\nu K^\nu}
    = \mu + u^\mu\phi_\mu~,
\end{equation}
\end{subequations}
has dependence on the dipole Goldstone $\phi_\mu$.
Formally, we take $K^\mu \sim {\cal O}(\dow^1)$, so that the thermodynamic variables have the right derivative counting as suggested in section \ref{sec:derivative-counting}. Note that, in non-covariant notation, this means $K^t\sim {\cal O}(\dow^0)$ and $K^i\sim{\cal O}(\dow^1)$.

%-----------------------------------------------
\subsection{Ideal p-wave dipole superfluids in equilibrium}
\label{sec:ideal-hydrostatics}
%-----------------------------------------------

As an example of the construction above, let us look at ideal p-wave dipole superfluids with a spontaneously broken dipole symmetry. These fluids are qualitatively distinct from ordinary ideal fluids due to the disparate derivative counting. In particular, the kinetic term proportional to $\vec u^2 = h_{\mu\nu}u^\mu u^\nu$ in the free energy gets suppressed to second order in derivatives, which is a nod to the fact that the kinetic energy of a fluid with conserved dipole moment is parametrically small. On the other hand, there are also terms like $F_n^2 = F_n^{\mu\nu}F^n_{\mu\nu}$ in the free energy that would ordinarily be counted as second order in derivatives, but get promoted to zeroth order in the presence of dipole symmetry.

%-----------------------------------------------
\subsubsection{Thermodynamics and constitutive relations}
%-----------------------------------------------

Ideal fluids are characterized by a free energy density ${\cal F}$ expressed in terms of zeroth order scalars made out of $n_\mu$, $h_{\mu\nu}$, $\tilde A_\mu$, and $\tilde A^\lambda_{~\mu}$. Because of the existence of zeroth order in derivative tensor fields $a_{\mu\nu}$, $F_{\mu\nu}$, and $F^n_{\mu\nu}$, there are in general many scalars one can write down by mutually contracting chains of these tensors. For simplicity, let us focus on terms in the free energy that are at most quadratic in fields. The ensuing constitutive currents will be at least linear in fields. These are the fields that will be relevant for linear response. The free energy density of an ideal fluid is characterized by the thermodynamic pressure
\begin{equation}
    {\cal F}_{\text{ideal}} = - p\!\lb
    T, \, \tmu, \, \tr\,\tilde a, \,
    \tilde a^2, \, \tilde F^2, \, F_n^2, \, \tilde F\cdot F_n\rb,
    \label{eq:ideal-free}
\end{equation}
where we have identified the zeroth order scalars
\begin{gather}
    T, \qquad \tmu, \qquad
    \tr\,\tilde a = h^{\mu\nu}\tilde a_{\mu\nu}, \qquad 
    \tilde a^2 = h^{\mu\nu}h^{\rho\sigma}
    \tilde a_{\mu\rho}\tilde a_{\nu\sigma}, \nn\\
    \tilde F^2 = h^{\mu\nu}h^{\rho\sigma}
    \tilde F_{\mu\rho}\tilde F_{\nu\sigma}, \qquad 
    F_n^2 = h^{\mu\nu}h^{\rho\sigma}
    F^n_{\mu\rho}F^n_{\nu\sigma}, \qquad 
    \tilde F\cdot F_n = h^{\mu\nu}h^{\rho\sigma}
    \tilde F_{\mu\rho}F^n_{\nu\sigma}.
    \label{eq:dow0-scalars}
\end{gather}
Note that there is no kinetic energy term proportional to $\vec u^2 = h_{\mu\nu}u^\mu u^\nu$ in the free energy at ideal order, as there would be in a non-boost-invariant fluid without dipole symmetry~\cite{Armas:2020mpr,Novak:2019wqg,deBoer:2020xlc}.\footnote{To be precise, the kinetic energy term is also implicitly present in relativistic or Galilean fluids, but is hidden inside the Lorentz- or Galilean-invariant definitions of $T$ and $\mu$.} This is because unlike ordinary fluids, the scalar $\vec u^2$ is counted as ${\cal O}(\dow^2)$ according to our derivative counting scheme outlined in subsection \ref{sec:derivative-counting}.

Using the thermodynamic derivatives of the pressure $p$ with respect to its arguments, we can define the thermodynamic relations 
\begin{equation}
\begin{split}
    \df p
    &= s\,\df T + q\,\df \tmu
    + \half p_d\,\df\,\tr\,\tilde a
    - \frac14 G_d\,\df\!\lb \tilde a^2 - \frac{1}{d} \tr^2\tilde a \rb
    - \frac{\chi_m}{4} \df \tilde F^2
    - \frac{\chi_n}{4} \df F_n^2
    - \frac{\chi_{mn}}{2} \df(\tilde F\cdot F_n), \\
    \epsilon
    &= - p + Ts + \tmu q. 
\end{split}
\end{equation}
We can identify the usual thermodynamic observables: the entropy density $s$, charge density $q$, and the energy density $\epsilon$. The coefficient $\chi_m$ can be understood as an inverse magnetic permeability, and we have similar coefficients $\chi_n$ and $\chi_{mn}$ related to the clock torsion $F^n_{\mu\nu}$. These coefficients normally show up at second derivative order in an ordinary fluid, but have been promoted to ideal order due to the derivative counting scheme of dipole fluids. In addition, we also have new thermodynamic coefficients specific to systems with conserved dipole moment: the dipole pressure $p_d$ and dipole ``shear modulus'' $G_d$.\footnote{An analogy can be drawn here with the theory of elasticity which also features a vector degree of freedom similar to $\delta\phi_i$, i.e. the displacement field. The analogue of the quantity $\tilde a_{ij} = a_{ij} + 2\dow_{(i}\phi_{j)}$ (in the absence of the spacetime sources) is the strain tensor. The coefficients $p_d$, $B_d$, and $G_d$ map to lattice pressure, bulk modulus, and shear modulus respectively; see e.g.~\cite{Armas:2019sbe, Armas:2020bmo}. Similarly, $\chi_{sd}/B_d$ and $\chi_{qd}/B_d$ are identified as thermal and charge expansion coefficients. However, unlike the theory of classical elasticity, we also have dependence on the analogue of antisymmetric strain tensor $\tilde F_{ij} = F_{ij} + 2\dow_{[i}\phi_{j]}$. It will be interesting to further explore this analogy in the context of fracton-elasticity duality~\cite{Pretko:2017kvd, Gromov:2017vir, Grosvenor:2021hkn}, which we leave for future work.} Eventually, we will be interested in linearized hydrodynamics which will be sensitive to at most quadratic fluctuations in the free energy. To this end, we can also define the thermodynamic derivatives of $s$, $q$, and $p_d$ leading to the susceptibilities
\begin{equation}
\begin{gathered}
    \chi_{ss} = \frac{\dow s}{\dow T}, \qquad 
    \chi_{sq} = \frac{\dow s}{\dow\tmu} = \frac{\dow q}{\dow T}, \qquad 
    \chi_{qq} = \frac{\dow q}{\dow\tmu}, \\ 
    B_d = -2\frac{\dow p_d}{\dow\, \tr\,\tilde a}, \qquad 
    \chi_{sd} = 2\frac{\dow s}{\dow\, \tr\,\tilde a}
    = \frac{\dow p_d}{\dow T}, \qquad 
    \chi_{qd} = 2\frac{\dow q}{\dow\, \tr\,\tilde a}
    = \frac{\dow p_d}{\dow \tmu}.
\end{gathered}
\label{eq:thermodynamicderivatives}
\end{equation}

Varying the partition function with respect to the dipole-invariant background sources, we can read out the equilibrium constitutive relations for the dipole-invariant quantities 
\begin{align}
\begin{split}
  \tilde\epsilon^\mu_\ideal 
  &= \epsilon\, u^\mu + p\, \vec u^\mu
  + p_d\, v^\rho \tilde F_{\rho}{}^\mu
  + \nabla'_\nu\!\lb \chi_{n} F_n^{\mu\nu}
  + \chi_{mn} \tilde F^{\mu\nu}
  \rb \\ 
  &\qquad 
  {\color{gray}+ \frac{1}{2}
  \lb \chi_n F_n^{\rho\sigma}F^n_{\rho\sigma} 
  + \chi_{mn} \tilde F^{\rho\sigma}F^n_{\rho\sigma} 
  \rb v^\mu} \\ 
  &\qquad 
    {\color{gray} -\lb 
  G_d\,\tilde a^{\langle\mu\nu\rangle}
  + \chi_m \tilde F^{\mu\nu}
    + \chi_{mn} F_n^{\mu\nu}
  \rb v^\rho \tilde F_{\rho\nu}
    - \lb \chi_n F_n^{\mu\nu} 
    + \chi_{mn} \tilde F^{\mu\nu}
    \rb v^\rho F^n_{\rho\nu}}
  + {\cal O}(\dow^3), \\
  \tilde\pi^\mu_\ideal
  &= {\cal O}(\dow), \\
  \tilde\tau^{\mu\nu}_\ideal
  &= 
  % \rho\,\vec u^\mu \vec u^\nu 
  p\, h^{\mu\nu}
  {\color{gray} \,+\,
  G_d\, \tilde a^{(\mu}_\lambda \tilde F^{\nu)\lambda}
  + \chi_m \tilde F^{\mu}_{~\lambda} \tilde F^{\nu\lambda}
  + \chi_n {F_n}^{\mu}_{~\lambda} F_n^{\nu\lambda}
  + 2\chi_{mn} \tilde F^{(\mu}_{~\lambda} F_n^{\nu)\lambda}}
  + {\cal O}(\dow^2), \\
  J^\mu_\ideal
  &= q\,u^\mu
  - \nabla'_\nu\!\lb \chi_m \tilde F^{\mu\nu}
  + \chi_{mn} F_n^{\mu\nu}
  \rb
  {\color{gray} \,-\, \frac{1}{2}
  \lb \chi_m \tilde F^{\rho\sigma}
  + \chi_{mn} F_n^{\rho\sigma}
  \rb F^n_{\rho\sigma}  v^\mu }
  + {\cal O}(\dow^3), \\
  J^{\mu\nu}_\ideal 
  &= p_d\, h^{\mu\nu} - G_d\,\tilde a^{\langle\mu\nu\rangle}
  + {\cal O}(\dow^2).
\end{split}
\label{eq:ideal-fluid}
\end{align}
Here, the indices on $\tilde a_{\mu\nu}$ and $\tilde F_{\mu\nu}$ are raised using $h^{\mu\nu}$, and angular brackets denote a symmetric-traceless combination $\tilde a^{\langle\mu\nu\rangle} = \tilde a^{\mu\nu} - \frac1d h^{\mu\nu} \tr\,\tilde a$. While deriving the above constitutive relations, we have used the fact that $\delta\phi_\mu = h_\mu^\nu\delta\phi_\nu + n_\mu \phi^\nu v^\rho \delta h_{\nu\rho}$ due to the normalization condition. The {\color{gray} grayed out} terms above are non-linear in fields. Since we decided to ignore cubic and higher non-linear terms in the free energy density \eqref{eq:ideal-free}, the non-linear contributions to the constitutive relations are not complete.

These constitutive relations generically depend on the equilibrium configurations of $\phi_\mu$ implicitly through $\tilde a_{\mu\nu}$ and $\tilde F_{\mu\nu}$. The respective classical configuration equations can be obtained by extremizing the free energy, leading to 
\begin{equation}
    q\,\vec u^\mu 
    = \nabla'_\nu\!\lb 
    p_d\, h^{\nu\mu} - G_d\,\tilde a^{\langle\nu\mu\rangle} 
    - \chi_m \tilde F^{\nu\mu}
    - \chi_{mn} F_n^{\nu\mu}
    \rb.
    \label{eq:phi-config-static}
\end{equation}
These relations validate our derivative counting scheme where it is observed that $\vec u^\mu$ is suppressed to ${\cal O}(\dow)$.

\begin{table}[t]
    \centering
    \begin{tabular}{c|ccc}
         & C & P & T  \\
         \hline 
         $\epsilon^t$, $n_t$, $T$ & $+$ & $+$ & $+$ \\ 
         $\epsilon^i$, $n_i$ & $+$ & $-$ & $-$ \\ 
         $\pi_i$, $v^i$, $u^i$ & $+$ & $-$ & $-$ \\ 
         $\tau^{ij}$, $h_{ij}$ & $+$ & $+$ & $+$ \\ 
         $J^t$, $A_t$, $\mu$ & $-$ & $+$ & $+$ \\ 
         $J^i$, $A_i$, $\phi_i$ & $-$ & $-$ & $-$ \\ 
         $J^{ij}$, $a_{ij}$ & $-$ & $+$ & $-$
    \end{tabular}
    \caption{Transformation properties of various quantities under charge conjugation C, parity P, and time-reversal T discrete transformations. We have only listed the independent components here; the transformation properties of remaining components can be obtained from here using normalization conditions. \label{tab:CPT}}
    \label{tab:my_label}
\end{table}

It is interesting to note the transformation properties of the constitutive relations under discrete symmetries that the system might possess; see table \ref{tab:CPT}. In particular, for systems respecting T symmetry, the coefficients $p_d$, $\chi_{sd}$, and $\chi_{qd}$ must vanish in equilibrium, while $B_d = -2\dow p_d/\dow\,\tr\,\tilde a$ can be non-vanishing. Another interesting case is CT symmetry. Since the equilibrium chemical potential $\mu_0$ flips sign under CT symmetry, this requires that $q$, $\chi_{sq}$, $\chi_{qd}$, and $\chi_{mn}$ must be odd functions of $\tmu$, while all the remaining objects must be even functions of $\tmu$. Note that unlike T symmetry, CT symmetry does not require any thermodynamic parameters in the theory to vanish. These discrete transformation properties will have important physical consequences in our subsequent hydrodynamic discussion.

%-----------------------------------------------
\subsubsection{Equilibrium configurations}
%-----------------------------------------------

To gain some intuition into the constitutive relations outlined above, let us turn off the background fields and look at the configuration equations for $\phi_i$. From \eqref{eq:phi-config-static} it follows that
\begin{equation}
    q\, u^i_0 = 
    \dow^i p_d
    - \dow_j \lb
    \lb G_d - \chi_m\rb \dow^i\phi^j 
    + \lb G_d + \chi_m \rb \dow^j \phi^i 
    - \frac{2}{d}G_d \delta^{ij} \dow_k\phi^k
    \rb.
    \label{eq:equilibrium_u0}
\end{equation}
The scalar coefficients in this equation are further functions of
\begin{equation}
\begin{gathered}
    T=T_0, \qquad 
    \tmu = \mu_0 + u_0^i\phi_i, \qquad 
    \vec u^2 = u^i_0 u^0_i, \\
    \tr\,\tilde a = 2\dow_i\phi^i, \qquad 
    \tilde a^2 = 
    2\dow_i\phi_j\dow^i\phi^j
    + 2\dow_i\phi_j\dow^j\phi^i, \qquad 
    \tilde F^2 
    = 
    2\dow_i\phi_j\dow^i\phi^j
    - 2\dow_i\phi_j\dow^j\phi^i.
\end{gathered}
\end{equation}
We are interested in equilibrium states with vanishing equilibrium velocity $u^i_0 = 0$ and non-vanishing charge density $q = q_0$. In this case, the generic ground state solution for $\phi_i$ takes a linear form
\begin{equation}
\begin{split}
    \langle\phi_i\rangle 
    &= \frac{1}{q_0}\pi^0_i
    + x^k\xi_{ki}, \\
    \langle\dow_i\phi_j\rangle 
    &= \xi_{ij},
    \label{eq:classical-config}
\end{split}
\end{equation}
for arbitrary constant $\xi_{ij}$ and $\pi^i_0$. The respective constitutive relations are
\begin{equation}
\begin{split}
    \langle\tilde\epsilon^t_\ideal \rangle
    &= \epsilon, \\
    \langle\tilde\tau^{ij}_\ideal\rangle 
    &= p\,\delta^{ij}
    + \lb G_d\, \xi_\sfS^{k(i} \xi_\sfA^{j)l}
    + \chi_m \xi_\sfA^{ik} \xi_\sfA^{jl} \rb \delta_{kl}, \\
    \langle J^t_\ideal\rangle 
    &= q, \\
    \langle J^{ij}_\ideal\rangle 
    &= p_d\, \delta^{ij} - G_d\xi_\sfS^{\langle ij\rangle}, \\ 
    \langle\tilde\epsilon^i_\ideal \rangle
    &= \langle\tilde\pi^i_\ideal \rangle
    = \langle J^i_\ideal\rangle = 0,
    \label{eq:eqb-state}
\end{split}
\end{equation}
where $\xi_{ij}^\sfS = 2\xi_{(ij)}$ and $\xi^\sfA_{ij} = 2\xi_{[ij]}$.
The coefficients appearing here are understood to be evaluated on the solution for $\phi_i$. Note that even though the dipole-invariant momentum density is zero, the physical momentum density is nonzero due to nonzero expectation value of $\phi_i$. We find the physical momentum density and stress tensor 
\begin{equation}
\begin{split}
    \langle\pi^i_\ideal\rangle 
    &= q\langle\phi^i \rangle \\ 
    &= \pi^0_i
    + \half q\, x^k\! \lb 
    \xi_{ki}^{\sfS}
    + \xi^{\sfA}_{ki}
    \rb, \\
    \langle\tau^{ij}_\ideal \rangle
    &= \langle\tilde\tau^{ij}_\ideal \rangle
    - \langle J^{ij}_\ideal \rangle 
    \langle \dow_k\phi^k\rangle
    \\ 
    &= \lb p
    - \half p_d\, \delta^{kl}\xi^\sfS_{kl} \rb \delta^{ij}
    + \lb G_d\, \xi_\sfS^{k(i} \xi_\sfA^{j)l}
    + G_d\,\xi_\sfS^{\langle ij\rangle} \xi_\sfS^{kl}
    + \chi_m \xi_\sfA^{ik} \xi_\sfA^{jl} \rb \delta_{kl}.
\end{split}
\end{equation}
The energy density/flux remain the same as their invariant counterparts in flat space. It is also possible to find more general solutions, but we will not explore these in this work. 

Looking at the explicit equilibrium solutions allows us a further opportunity to verify our derivative counting scheme. Since the generic solution for $\phi_i$ in \eqref{eq:classical-config} involves a linear term in the spatial coordinates $x^i$, it follows that $\phi_i\sim{\cal O}(\dow^{-1})$.

Lastly, we note that not all equilibrium configurations in \eqref{eq:classical-config} are thermodynamically equivalent. The thermodynamically favored solution is one that minimizes the free energy. Throughout this work, we will mostly be interested in the zero superflow state $\langle\phi_i\rangle = \pi^0_i/q$, $\langle\dow_i\phi_j\rangle = 0$. This state is thermodynamically stable if
\begin{equation}
    p_d = 0, \qquad 
    B_d, G_d, \chi_m, \chi_n > 0,
\end{equation}
when evaluated on the solution $\langle\phi_i\rangle = \pi^0_i/q$.
When these stability conditions are not satisfied, the system will prefer to transition to a nonzero superflow state $\langle\dow_i\phi_j\rangle \neq 0$.

%-----------------------------------------------
\subsection{Derivative counting and higher-derivative corrections}
%-----------------------------------------------

The ideal fluid free energy can generically admit higher-derivative corrections. The most physically relevant among these is the kinetic term that appears at second order in derivatives, i.e.
\begin{equation}
    {\cal F} \sim - p - \half \rho\, \vec u^2 + \ldots
    + {\cal O}(\dow^3),
\end{equation}
where $\rho$ is the kinetic mass density. In a Galilean fluid, $\rho$ is proportional to the charge density $q$, while in a relativistic fluid, it is proportional to $\epsilon+p$. Since a fluid with conserved dipole moment does not have any boost symmetries, $\rho$ is generically an independent thermodynamic coefficient. Since we are only interested in the free energy up to quadratic order in fluctuations, we can take $\rho$ to be a constant. Including the kinetic energy term modifies the momentum density to take more conventional form
\begin{equation}
    \tilde\pi^\mu_{\text{hs}} = \rho\,\vec u^\mu + \ldots + {\cal O}(\dow^2).
\end{equation}
The contributions to all other currents is at least quadratic in fluctuations. Interestingly, the kinetic energy term is not the only hydrostatic scalar we can introduce in the free energy density at second order in derivatives. There are many more hydrostatic coefficients at this order which are as important as $\rho$ that must be taken into account for consistent truncation of the derivative expansion. These contributions are represented by ellipses in the expression above.

More generally, we can use the derivative counting scheme outlined in section \ref{sec:derivative-counting} to arrange the free energy density ${\cal F}$ in a perturbative series. Schematically, we have
\begin{align}
    {\cal F} = - p 
    - \sum_I \lambda^{(1)}_I {\cal S}^{(1)}_I
    - \sum_I \lambda^{(2)}_I {\cal S}^{(2)}_I + \ldots,
\end{align}
where $S^{(n)}_I$ denotes the collection of all independent hydrostatic scalars at ${\cal O}(\dow^n)$, while the coefficients $p$ and $\lambda^{(n)}_I$ are arbitrary functions of ${\cal O}(\dow^0)$ scalars given in \eqref{eq:dow0-scalars}. As it turns out, there are no hydrostatic scalars at ${\cal O}(\dow^1)$ for a parity-preserving fluid. There are a total of 27 independent hydrostatic scalars at ${\cal O}(\dow^2)$ in addition to $\vec u^2$ that can affect the linearized constitutive relations; we have worked these out in Appendix \ref{app:HS-scalars}. These lead to corrections to the ideal fluid constitutive relations, which enter at the same order as those coming from the kinetic term $\half\rho\,\vec u^2$. 

As an illustration, let us look at all the hydrostatic scalars that affect the momentum density. There are seven such scalars in addition to $\vec u^2$, leading to
\begin{align}
\begin{split}
    {\cal F} 
    &= - p - \half \rho\, \vec u^2 
    - \lambda_2\vec u^\mu\dow_\mu T
    - \lambda_3\vec u^\mu\dow_\mu\tmu 
    - \half\lambda_4\vec u^\mu\dow_\mu\tr\,\tilde a
    \\
    & \quad
    - \lambda_5\, \tilde a^{\mu\nu}\nabla_\mu \vec u_\nu 
    - \lambda_6\, \tilde F^{\mu\nu}\nabla_\mu \vec u_\nu 
    - \lambda_7\, F_n^{\mu\nu}\nabla_\mu \vec u_\nu - \lambda_{27}\,R\,.
\end{split}
\end{align}
Here, $R = R^{\lambda}_{\;\;\mu\lambda\nu}h^{\mu\nu}$ is the Ricci scalar and we have dropped the superscript ``$(2)$'' from the transport coefficients for ease of reading. This results in the full hydrostatic momentum density up to second order in derivatives and linear order in fluctuations
\begin{align}
\begin{split}
    \tilde{\pi}^\mu_{\hs} 
    &= \rho\, \vec{u}^\mu 
    + \lambda_2 \nabla^\mu T
    + \lambda_3 \nabla^\mu\tmu
    + \half \lambda_4 \nabla^\mu\tr\,\tilde a 
    \\
    &\qquad 
    - \nabla'_\nu\lb 
    \lambda_5 \tilde{a}^{\nu\mu}
    + \lambda_6 \tilde{F}^{\nu\mu}
    + (\lambda_7-\lambda_{27}) F_n^{\nu\mu}
    \rb
    + {\cal O}(\dow^2)\,.
\end{split}
\end{align}
These and the 21 other hydrostatic coefficients also contribute to the remaining conserved currents. However, we will not report these expressions here. Instead, in the next section we will depart from hydrostatic equilibrium and construct a hydrodynamic theory with conserved dipole moment.

%-----------------------------------------------
\section{Hydrodynamics of p-wave dipole superfluids}
\label{modernhydro}
%-----------------------------------------------

Using the ingredients outlined in sections \ref{sec:preliminaries} and \ref{sec:hydrostatics}, we will now develop a theory of hydrodynamics for p-wave dipole superfluids using the techniques of~\cite{Banerjee:2012iz, Jensen:2012jh, Crossley:2015evo,Haehl:2015uoc,Jensen:2017kzi,Jensen:2014aia,Geracie:2015xfa,Novak:2019wqg,deBoer:2020xlc,Armas:2020mpr}. We will explicitly write down the constitutive relations for our hydrodynamic theory up to subleading orders in derivatives and up to linearized order in fields. We will use this theory to obtain the dispersion relations and response in section \ref{sec:response}.

%----------------------------------
\subsection{The setup and the adiabaticity equation}
\label{sec:adiabaticity}
%----------------------------------

Thermal fluctuations of a many-body system around thermal equilibrium are described by the framework of hydrodynamics. The starting point of hydrodynamics are the conservation equations associated with the global symmetries of the system under consideration, which can be used to obtain the dynamical evolution of various thermodynamic observables. For our case of interest, the conservation equations are given by \eqref{eq:cons-cov} and the relevant hydrodynamic fields are given by a set of symmetry parameters: $\hat\scB = (\beta^\mu,\Lambda_\beta,\psi_\mu^\beta)$. In thermal equilibrium, the hydrodynamic fields are aligned along the thermal isometry as $\hat\scK = (K^\mu,\Lambda_K,0)$, meaning that the fluid flow is stationary with respect to the thermal observer. However, out-of-equilibrium, the hydrodynamic fields take a meaning of their own. Note that we have introduced one hydrodynamic field per conservation equation, so the system of equations is closed provided that we specify the \emph{hydrodynamic constitutive relations}, i.e how the conserved currents of the theory are fixed in terms of the dynamical and background field content. The hydrodynamic fields are valued in the Lie algebra of the symmetry group and transform under an infinitesimal symmetry transformation $\hat\scX = (\chi^\mu,\Lambda,\psi_\mu)$ as
\begin{equation}
\begin{split}
    \delta_{\hat\scX}\beta^\mu 
    &= \lie_\chi\beta^\mu, \\ 
    \delta_{\hat\scX}\Lambda_\beta 
    &= \lie_\chi\Lambda_\beta - \lie_\beta\Lambda, \\ 
    \delta_{\hat\scX}\psi_\mu^\beta 
    &= \lie_\chi \psi_\mu^\beta 
    - h_\mu^\nu\lie_\beta\psi_\nu.
\end{split}
\end{equation}
More details regarding these symmetry transformations can be found in~\cite{Jain:2021ibh}. Since the dipole symmetry is spontaneously broken in a p-wave dipole superfluid, we also have the associated Goldstone $\phi_\mu$ in the hydrodynamic theory. This will need to be supplied with its own equations of motion, known as the \emph{Josephson equation}. For now, we take this to have a schematic form 
\begin{equation}
    X^\mu = 0,
    \label{eq:Joseph-general}
\end{equation}
where $X^\mu$ is some vector field satisfying $X^\mu n_\mu = 0$. In hydrostatic equilibrium, $X^\mu$ could be obtained by varying the free energy with respect to $\phi_\mu$. We do not have access to a free energy description out-of-equilibrium, so we will need to implement a different way to find $X^\mu$. We will get to it by the end of this subsection; let us keep $X^\mu$ as an arbitrary operator for now. Together, the dynamical fields can be used to construct dipole and gauge-invariant hydrodynamic fields
\begin{align}\begin{gathered}
    T = \frac{1}{n_\nu\beta^\nu}, \qquad 
    u^{\mu} = \frac{\beta^\mu}{n_\nu\beta^\nu}, \qquad 
    \tmu = \frac{\Lambda_\beta + \beta^\lambda \tilde{A}_\lambda}
    {n_\nu\beta^\nu}, \\
    \tvarpi_\mu = \frac{
    \psi_\mu^\beta 
    - h_\mu^\rho\lie_\beta \phi_\rho 
    + h_{\mu\lambda}\beta^\rho 
    \tilde A^\lambda_{~\rho}}{n_\nu\beta^\nu}.
\end{gathered}
\label{eq:hydrofields}
\end{align}
identified as fluid temperature $T$, velocity $u^\mu$, U(1) chemical potential $\tmu$, and dipole chemical potential $\tvarpi_\mu$ respectively. We will shortly see that $\tvarpi_\mu$ can be integrated out of the hydrodynamic theory using the Josephson equation for $\phi_\mu$. The situation here is similar to ordinary superfluids where the U(1) chemical potential $\mu$ can be integrated out using the Josephson equation for the U(1) Goldstone $\phi$. 

To complete the hydrodynamic set of equations, we need to provide a set of constitutive relations for how $\epsilon^\mu$, $\pi^\mu$, $\tau^{\mu\nu}$, $J^\mu$, $J^{\mu\nu}$, and $X^\mu$ are fixed in terms of the dynamical fields $\beta^\mu$, $\Lambda_\beta$, $\psi^\beta_\mu$, $\phi_\mu$ and the background fields $n_\mu$, $h_{\mu\nu}$, $A_\mu$, $A^\lambda_{~\mu}$. The constitutive relations are required to satisfy the local version of the Second Law of thermodynamics, i.e. there must exist an entropy current $s^\mu$ whose divergence is positive semi-definite, i.e.
\begin{align}\begin{split}
\nabla_\mu' s^\mu = \Delta \geq 0,
\label{eq:second-law}
\end{split}\end{align}
for \emph{all} solutions of the conservation equations \eqref{eq:cons-cov-off-shell}, when the dipole Goldstone field $\phi_\mu$ is kept off-shell. Note that the Second Law is just an inequality statement, however the fact that it is imposed for every possible hydrodynamic configuration results in strong constraints on the functional form of the constitutive relations. We emphasize that the Second Law is imposed for off-shell configurations of the Goldstone $\phi_\mu$, which appears to be a stronger requirement than the conventional statement of the Second Law of thermodynamics. Ultimately, this generality serves to fix the Josephson equation \eqref{eq:Joseph-general}. See~\cite{Jain:2016rlz} for more discussion in this regard.

In conventional hydrodynamics, one writes down the most generic expressions for the constitutive relations and entropy current allowed by symmetries, truncated to some desired order in derivatives, and investigates the constraints resulting from imposing \eqref{eq:second-law}; see e.g.~\cite{landaubook}. However, an alternative and simpler route to allowed constitutive relations involves the construction of an off-shell formulation of the Second Law. The details of this procedure for relativistic and Galilean hydrodynamics can be found in~\cite{Liu1972MethodOL,Loganayagam:2011mu,Haehl:2015pja,Jensen:2014ama,Jain:2018jxj}, and for the boost-agnostic hydrodynamics relevant to this work in~\cite{Armas:2020mpr}. Here we find an analogue of this off-shell constraint for systems with dipole symmetry, whose final form is given in \eqref{eq:adiabaticity}. To derive this, consider an off-shell generalization of the Second Law statement by adding combinations of conservation equations 
\begin{align}
\begin{split}
    \nabla'_\mu s^\mu 
    &- \beta^\rho n_\rho \Big[\nabla'_\mu\epsilon^\mu + \ldots \Big]
    + \beta^\rho h_{\rho\nu} \Big[ 
    \nabla'_\mu (v^\mu\pi^\nu + \tau^{\mu\nu} 
    + \tau^{\mu\nu}_d) + \ldots \Big] \\
    &\qquad 
    + (\Lambda_\beta  + \beta^\lambda A_\lambda) \nabla'_\mu J^\mu 
    + (\psi_\nu^\beta + h_{\nu\lambda}\beta^\rho A^\lambda_{~\rho}) 
    \Big[ \nabla'_\mu J^{\mu\nu} - h^\nu_\mu J^\mu + X^\nu \Big]
    = \Delta \geq 0.
    \label{eq:off-shell-2nd}
\end{split}
\end{align}
A priori, the coefficients multiplying the conservation equations in this expression can be arbitrary. Earlier, we introduced $\beta^\mu$, $\Lambda_\beta$, and $\psi^\beta_\mu$ as a means to solve the conservation equations out-of-equilibrium, though we emphasize that they have no intrinsic physical meaning. However, the off-shell Second Law can now be seen as a particular definition for the hydrodynamic fields, which turns out to be equivalent to the requirement that the hydrodynamic fields are aligned with the thermal isometry $\scK$ in hydrostatic equilibrium, i.e. in hydrostatic equilibrium $\beta^\mu = K^\mu$ and $\Lambda_\beta = \Lambda_K$, and $\psi^\beta_\mu = 0$. We will return to the redefinition freedom of hydrodynamic fields in section \ref{sec:nhs-consti}. Having fixed these definitions, the off-shell statement \eqref{eq:off-shell-2nd} is entirely equivalent to the on-shell statement of the Second Law of thermodynamics in \eqref{eq:second-law}. We can also massage the Second Law into a more useful form by defining the free energy current
\begin{align}\begin{split}
    N^\mu 
    &= s^\mu 
    - \beta^\nu n_\nu \epsilon^\mu 
    + \beta^\rho h_{\rho\nu}\lb v^\mu\pi^\nu + \tau^{\mu\nu}+\tau_d^{\mu\nu} \rb
    + \lb \Lambda_\beta + \beta^\rho A_\rho \rb J^\mu  \\
    &\qquad 
    + \lb \psi_\nu^\beta 
    + \beta^\rho h_{\lambda\nu}A^{\lambda}_{~\rho} \rb 
    J^{\mu\nu}
    + \beta^\rho \phi_\rho X^\mu,
    \label{eq:free-E-current}
\end{split}\end{align}
which converts the Second Law statement into the so-called \emph{adiabaticity equation}
\begin{align}\begin{split}
    \nabla_\mu' N^\mu &=
    - \epsilon^\mu\delta_{\hat\scB} n_\mu 
    + \lb v^\mu \pi^\nu+\frac{1}{2}\tau^{\mu\nu}\rb 
    \delta_{\hat\scB} h_{\mu\nu}
    + J^\mu \delta_{\hat\scB} A_\mu
    + J^{\mu\nu}h_{\nu\lambda} \delta_{\hat\scB} A^{\lambda}_{~\mu}
    + X^\mu \delta_{\hat\scB}\phi_\mu
    + \Delta.
    \label{eq:adiabaticity-phys}
\end{split}\end{align}
Here $\delta_{\hat\scB}$ denotes a diffeomorphism, gauge transformation, and dipole transformation along the hydrodynamic fields $\scB = (\beta^\mu,\Lambda_\beta,\psi^\beta_\mu)$. Note that there is no entropy production $\Delta$ in hydrostatic equilibrium and $\delta_{\hat\scB} = \delta_\scK$ acting on background fields vanishes. Consequently, the right-hand-side of \eqref{eq:adiabaticity-phys} is trivially zero in equilibrium and the free energy current $N^\mu$ is identically conserved. In fact, the equilibrium value of $N^\mu$ is precisely $-{\cal F}K^\mu$, where ${\cal F}$ is the hydrostatic free energy density introduced in section \ref{sec:hydrostatics}.

For simplicity and convenience, it is useful to recast above equations in terms of dipole-invariant objects. First, the off-shell Second Law of thermodynamics in \eqref{eq:off-shell-2nd} can be re-expressed as
\begin{align}\begin{split}
    \nabla'_\mu \tilde s^\mu 
    &- \frac1T \Big[\nabla'_\mu \tilde\epsilon^\mu + \ldots \Big]
    + \frac{\vec u_{\nu}}{T} \Big[ 
    \nabla'_\mu (v^\mu\tilde\pi^\nu + \tilde\tau^{\mu\nu} 
    + \tilde\tau^{\mu\nu}_d) + \ldots \Big] \\
    &\qquad\qquad\qquad 
    + \frac{\tmu}{T} \nabla'_\mu J^\mu 
    + \frac{\tvarpi_\nu}{T}
    \Big[ \nabla'_\mu J^{\mu\nu} - h^\nu_\mu J^\mu + X^\nu \Big]
    = \Delta \geq 0.
    \label{eq:off-shell-2nd-inv}
\end{split}\end{align}
where the dipole-invariant entropy current is defined as
\begin{equation}
    \tilde s^\mu = s^\mu 
    + \beta^\rho \phi_\rho
    \lb \nabla'_\nu J^{\nu\mu} 
    - h^\mu_\nu J^\nu + X^\mu \rb.
\end{equation}
The transformation vanishes on-shell, so the two entropy currents are physically indistinguishable. Similarly, we define the dipole-invariant free energy current
\begin{equation}
\begin{split}
    \tilde N^\mu
    &= N^\mu 
    - \lb 2\phi^{(\rho} J^{\sigma)\mu} 
    - \phi^{\mu} J^{\rho\sigma}
    \rb \half \lie_\beta h_{\rho\sigma}
    - J^{\mu\nu} \tilde\psi_\nu^\beta
    + \frac{1}{\sqrt{\gamma}}
    \dow_\nu\!\lb\sqrt{\gamma}\,
    2\phi^{[\nu} J^{\mu]}_{~\rho}\beta^\rho\rb \\ 
    &=
    \tilde s^\mu 
    - \frac1T \tilde\epsilon^\mu 
    + \frac{\vec u_\nu}{T}
    \lb v^\mu\tilde\pi^\nu 
    + \tilde\tau^{\mu\nu} 
    + \tilde\tau_d^{\mu\nu} \rb
    + \frac{\tmu}{T} J^\mu
    + \frac{u^\rho}{T} 
    h_{\lambda\nu} \tilde A^{\lambda}_{~\rho} 
    J^{\mu\nu},
\end{split}
\end{equation}
where $\tilde\psi^\beta_\mu = \psi^\beta_\mu - h_\mu^\nu\lie_\beta \phi_\nu$.
Using this, the adiabaticity equation becomes 
\begin{align}\begin{split}
    \nabla_\mu' \tilde N^\mu &=
    - \tilde\epsilon^\mu\delta_{\scB} n_\mu 
    + \lb v^\mu \tilde\pi^\nu+\frac{1}{2}\tilde\tau^{\mu\nu}\rb 
    \delta_{\scB} h_{\mu\nu}
    + J^\mu \delta_{\scB} \tilde A_\mu 
    + J^{\mu\nu}h_{\nu\lambda} \delta_{\scB} \tilde A^{\lambda}_{~\mu}  \\
    &\hspace{15em}
    + \lb \nabla'_\nu J^{\mu\nu} 
    - h^\mu_\nu J^\nu + X^\mu \rb 
    \lb \delta_\scB\phi_\mu - \psi^\beta_\mu \rb
    + \Delta,
    \label{eq:adiabaticity-inv}
\end{split}\end{align}
where $\delta_\scB$ denotes a Lie derivative along $\beta^\mu$ together with a gauge transformation along $\Lambda_\beta$. This equation encodes the same information as the original adiabaticity equation in \eqref{eq:adiabaticity-phys}, only expressed in terms of dipole-invariant observables.

Let us return to the Josephson equation for the dipole Goldstone field $\phi_\mu$. There is a trivial solution of the adiabaticity equation \eqref{eq:adiabaticity-inv} given by 
\begin{equation}
\begin{split}
    X^\mu 
    &= - \nabla'_\nu J^{\mu\nu} 
    + h^\mu_\nu J^\nu - T\sigma_\phi h^{\mu\nu}\lb \delta_\scB\phi_\nu - \psi^\beta_\nu \rb, \\
    \Delta 
    &= T\sigma_\phi h^{\mu\nu} \lb \delta_\scB\phi_\mu - \psi^\beta_\mu \rb
    \lb \delta_\scB\phi_\nu - \psi^\beta_\nu \rb,
\end{split}
\end{equation}
for some positive coefficient $\sigma_\phi$. Upon imposing $X^\mu = 0$, this implies that
\begin{equation}
\begin{gathered}
    h^\mu_\nu\lie_\beta\phi_\mu = \psi^\beta_\nu 
    - \frac{1}{T\sigma_\phi} h_{\nu\mu} \lb \nabla'_\rho J^{\mu\rho} 
    - h_{\mu\nu} J^\mu \rb
    + \ldots, \\ 
    \implies 
    u^\rho \tilde A^\mu_{~\rho}
    = \tvarpi^\mu
    - \frac{1}{\sigma_\phi} \lb \nabla'_\rho J^{\mu\rho} 
    - h^\mu_\nu J^\nu \rb
    + \ldots.
\end{gathered}
\label{eq:joseph}
\end{equation}
In principle, this equation can get further derivative corrections compatible with the adiabaticity equation, denoted by ellipses. This equation is supposed to determine the dynamics of $\phi_\mu$. In practice, however, we can view this as defining $\tvarpi_\mu$ or $\psi_\mu^\beta$ in terms of $\phi_\mu$ and other hydrodynamic fields. In turn, the dipole Ward identity can be used to obtain the dynamics of $\phi_\mu$. Upon taking this view, the final term in the adiabaticity equation \eqref{eq:adiabaticity-inv} drops out, and we are left with a simpler equation
\begin{align}\begin{split}
    \nabla_\mu' \tilde N^\mu &=
    - \tilde\epsilon^\mu\delta_{\scB} n_\mu 
    + \lb v^\mu \tilde\pi^\nu+\frac{1}{2}\tilde\tau^{\mu\nu}\rb 
    \delta_{\scB} h_{\mu\nu}
    + J^\mu \delta_{\scB} \tilde A_\mu 
    + J^{\mu\nu}h_{\nu\lambda} \delta_{\scB} \tilde A^{\lambda}_{~\mu}
    + \Delta.
    \label{eq:adiabaticity}
\end{split}\end{align}
Thus, to obtain the hydrodynamic constitutive relations allowed by the Second Law of thermodynamics, we need to find the most general expressions for the dipole-invariant currents $\tilde\epsilon^\mu$, $\tilde\pi^\mu$, $\tilde\tau^{\mu\nu}$, $J^\mu$, and $J^{\mu\nu}$ in terms of the hydrodynamic fields $\beta^\mu$, $\Lambda_\beta$, $\phi_\mu$ and the background fields $n_\mu$, $h_{\mu\nu}$, $A_\mu$, $A^\lambda_{~\mu}$, for some free energy current $\tilde N^\mu$ that satisfy the \eqref{eq:adiabaticity} for some positive semi-definite quadratic form $\Delta \geq 0$. 

%-----------------------------------------------
\subsection{Constitutive relations: hydrostatic sector}
\label{sec:hs-consti}
%-----------------------------------------------

The constitutive relations allowed by the adiabaticity equation \eqref{eq:adiabaticity} can be broadly classified into hydrostatic and non-hydrostatic sectors. The hydrostatic sector concerns the part of the constitutive relations that remains non-trivial in equilibrium after the hydrostatic constraints have been implemented. The hydrostatic constitutive relations can be characterized by the free energy density ${\cal F}$ introduced in Section \ref{sec:hydrostatics} and does not cause entropy production. On the other hand, the non-hydrostatic sector concerns part of the constitutive relations that vanish in equilibrium. They are further classified into the dissipative and non-hydrostatic non-dissipative sectors, depending on whether they contribute to entropy production or not.

Ideal fluids are part of the hydrostatic sector. It can be checked that the ideal fluid constitutive relations presented in \eqref{eq:ideal-fluid} satisfy the adiabaticity equation with the free energy current
\begin{align}
    \tilde N^\mu_\ideal &= p\,\beta^\mu + 
    \lb \chi_m \tilde F^{\mu\nu} + \chi_{mn} F_n^{\mu\nu} \rb \delta_\scB \tilde A_\nu 
    + \lb \chi_n F_n^{\mu\nu} + \chi_{mn} \tilde F^{\mu\nu} \rb \delta_\scB \tilde n_\nu 
    \nn\\
    &\qquad
    {\color{gray} -\, \frac{1}{\sqrt{\gamma}}\dow_\nu\!\bigg[ 
    \sqrt{\gamma}\,\frac{\mu}{T}
    \lb \chi_m \tilde F^{\mu\nu} + \chi_{mn} F_n^{\mu\nu} \rb 
    + \sqrt{\gamma}\,\frac{1}{T} \lb \chi_n F_n^{\mu\nu} + \chi_{mn} \tilde F^{\mu\nu} \rb
    \bigg]},
\end{align}
and $\Delta_\ideal = 0$.
Note that the additional terms in the first line vanish in equilibrium. The terms in the second line have been added for convenience; they are a total derivative and trivially drop out of the adiabaticity equation. The corresponding entropy current is given by
\begin{equation}
\begin{split}
    \tilde s^\mu_\ideal 
    &= s\, u^\mu.
\end{split}
\end{equation}
Therefore entropy flow in an ideal fluid is purely in the direction of the fluid velocity.

To derive the hydrostatic constitutive relations associated with more general corrections to the hydrostatic free energy density ${\cal F}$, we note the identity
\begin{equation}
\begin{split}
    \nabla'_\mu ({\cal F}\beta^\mu)
    &= {\cal F} v^\mu \delta_\scB n_\mu 
    + \half {\cal F} h^{\mu\nu}\delta_\scB h_{\mu\nu} 
    + \delta_\scB{\cal F} \\ 
    &= \lb \frac{\delta{\cal F}}{\delta n_\mu} 
    + {\cal F} v^\mu \rb \delta_\scB n_\mu 
    + \lb \frac{\delta{\cal F}}{\delta h_{\mu\nu}} 
    + \half {\cal F} h^{\mu\nu} \rb \delta_\scB h_{\mu\nu} 
    + \frac{\delta{\cal F}}{\delta \tilde A_\mu} \delta_\scB A_\mu \\
    &\qquad
    + \frac{\delta{\cal F}}{\delta \tilde a_{\mu\nu}} \delta_\scB a_{\mu\nu}
    - \nabla'_\mu\Theta_{\cal F}^\mu.
\end{split}
\end{equation}
Here we have utilised the Euler-Lagrange derivatives of ${\cal F}$ and $\nabla'_\mu\Theta_{\cal F}^\mu$ denotes the leftover total derivative term. This identically solves the adiabaticity equation, if we identify the hydrostatic constitutive relations
\begin{align}
\begin{split}
    \tilde\epsilon^\mu_\hs 
    &= \frac{\delta{\cal F}}{\delta n_\mu} 
    - h^\mu_\rho \frac{\delta{\cal F}}{\delta \tilde a_{\rho\nu}}
    v^\sigma\tilde F_{\sigma\nu}
    + {\cal F} v^\mu, \\
    \tilde\pi^{\mu}_\hs 
    &= - n_\nu\frac{\delta{\cal F}}{\delta h_{\mu\nu}}, \\
    \tilde\tau^{\mu\nu}_\hs 
    &= - 2 h^{\mu}_\rho h^\nu_\sigma \lb
    \frac{\delta{\cal F}}{\delta h_{\rho\sigma}} 
    + 2\frac{\delta{\cal F}}{\delta \tilde a_{\lambda(\rho}}
    \tilde A^{\sigma)}_{~\lambda} \rb
    - {\cal F} h^{\mu\nu}
    , \\ 
    J^\mu_\hs 
    &= - \frac{\delta{\cal F}}{\delta \tilde A_\mu}, \\ 
    J^{\mu\nu}_\hs 
    &= - 2h^\mu_\rho h^\nu_\sigma 
    \frac{\delta{\cal F}}{\delta \tilde a_{\rho\sigma}},
\end{split}
\end{align}
together with the free energy current 
\begin{equation}
    \tilde N^\mu_\hs = - {\cal F}\beta^\mu - \Theta^\mu_{\cal F},
\end{equation}
and $\Delta_\hs = 0$. Therefore, transport in the hydrostatic sector is non-dissipative. 

%-----------------------------------------------
\subsection{Constitutive relations: non-hydrostatic sector}
\label{sec:nhs-consti}
%-----------------------------------------------

As discussed earlier, equilibrium hydrodynamics is subject to the hydrostatic constraint $\delta_\scK(\ldots)=0$. Out-of-equilibrium, these constraints do not apply and the constitutive relations pick up new contributions, called non-hydrostatic constitutive relations. Due to the construction of the adiabaticity equation, it is natural to write the corrections in terms of $\delta_\scB$ acting on background sources, since they identically drop out in hydrostatic equilibrium when $\delta_\scB = \delta_\scK$ is promoted to an isometry.

Before we get to the non-hydrostatic constitutive relations, we need to address the issue of hydrodynamic frames. As we have stated, $\beta^\mu$, $\Lambda_\beta$ are arbitrary fields that we chose to describe the hydrodynamic system and we can always redefine them to suit our needs. These induce redefinitions of temperature, chemical potential, and fluid velocity
\begin{align}
    T \to T + \delta T, \qquad 
    \tmu \to \tmu + \delta\tmu, \qquad 
    u^\mu \to u^\mu + \delta u^\mu,
\end{align}
where $n_\mu\delta u^\mu = 0$. Incidentally, in writing the off-shell Second Law of thermodynamics \eqref{eq:off-shell-2nd}, or equivalently the adiabaticity equation \eqref{eq:adiabaticity}, and choosing the hydrodynamic fields to be the Lagrange multipliers for equations of motion, we further fixed part of the redefinition freedom available to us. Such partial fixing is known as the ``thermodynamic frame'' of hydrodynamics. However, this still leaves some redefinition freedom where we only shift the hydrodynamic fields with non-hydrostatic variables; see~\cite{Armas:2020mpr}. Heuristically, since the energy, momentum, and charge conservation equations become trivial in equilibrium, they can be seen as relating certain linear combinations of non-hydrostatic data out-of-equilibrium. Therefore, we can eliminate certain non-hydrostatic data from the constitutive relations by suitably redefining the hydrodynamic variables. A suitable choice involves disallowing any non-hydrostatic corrections to the energy, momentum, and charge densities, i.e. choosing
\begin{align}
    n_\mu{\tilde\epsilon}_\nhs^\mu 
    = \tilde\pi^\mu_\nhs 
    = n_\mu J^\mu_\nhs  = 0.
    \label{eq:thermo-frame}
\end{align}
This is the so-called ``thermodynamic density frame''. Conveniently, this frame is known not to exhibit the unphysical instabilities in the linearized spectrum that can appear in conventional choices like the Landau and Eckart frames \cite{Armas:2020mpr, Kovtun:2019hdm, Hiscock:1983zz, Bemfica:2020zjp}. In addition, we have a similar redefinition freedom in the dipole Goldstone field
\begin{equation}
    \phi_\mu \to \phi_\mu + \delta\phi_\mu,
\end{equation}
where $v^\mu\delta\phi_\mu = 0$. This will allows us to additionally fix
\begin{equation}
    h_{\mu\nu} J^\nu_\nhs = 0.
\end{equation}

Having fixed the hydrodynamic frame, non-hydrostatic corrections can only appear in the energy flux $h^\mu_\nu\tilde\epsilon^\nu$, stress tensor $\tau^{\mu\nu}$, and the dipole flux $J^{\mu\nu}$.
The non-hydrostatic constitutive relations satisfy a simpler adiabaticity equation 
\begin{equation}
    \nabla'_\mu \tilde N^\mu_\nhs
    = - \tilde\epsilon^\mu_\nhs \delta_\scB n_\mu 
    + \frac{1}{2}\tilde\tau^{\mu\nu}_\nhs
    \delta_\scB h_{\mu\nu}
    + J^\mu_\nhs \delta_\scB \tilde{A}_\mu
    + J^{\mu\nu}_\nhs h_{\nu\lambda} \delta_\scB \tilde{A}^{\lambda}_{\;\;\mu}
    + \Delta_\nhs.
    \label{eq:adiabaticity-nhs}
\end{equation}
There are two kinds of solutions to this equation. First, we have non-hydrostatic non-dissipative constitutive relations, where contributions to the adiabaticity equation cancel among themselves and lead to no entropy production. They are naturally left unconstrained by the Second Law requirement. Second, we have dissipative constitutive relations, where terms arrange themselves into a quadratic form that is required to be sign constrained. Note that the Second Law does not disallow any non-hydrostatic transport coefficients; at most it imposes inequality constraints on them.

Let us note the derivative order of the non-hydrostatic data appearing in \eqref{eq:adiabaticity-nhs}. We find 
\begin{equation}
    h^{\mu\nu} \delta_\scB n_\mu,~~
    h^{\mu\nu} \delta_\scB \tilde A_\mu \sim {\cal O}(\dow^1), \qquad 
    h^{\mu\rho}h^{\nu\sigma}\delta_\scB h_{\rho\sigma},~~
    h_\mu^\rho h_{\nu\lambda} \delta_\scB \tilde{A}^{\lambda}_{\;\;\rho}
    \sim {\cal O}(\dow^2).
    \label{eq:nhs-derivative-counting}
\end{equation}
This disparate derivative counting between vector and other sectors has interesting consequences for non-hydrostatic transport as we discuss below.

%-----------------------------------------------
\subsubsection{Zero-derivative order}
\label{sec:zero-order-diss}
%-----------------------------------------------

An interesting consequence of the derivative ordering \eqref{eq:nhs-derivative-counting} is that we obtain a dissipative transport coefficient at the same order as ideal fluids discussed in section \ref{sec:ideal-hydrostatics}. This is given by
\begin{align}
\begin{split}
    \tilde\epsilon_{\nhs}^\mu
    = T^2 \kappa\, h^{\mu\nu}\delta_\scB n_\nu
    + {\cal O}(\dow^2),
\end{split}
\label{eq:nonhs_leading}
\end{align}
where $\kappa$ is thermal conductivity. The entropy production quadratic form and the associated free energy current are given as
\begin{equation}
    \Delta 
    = T^2\kappa\,h^{\mu\nu} \delta_\scB n_\mu \delta_\scB n_\nu
    + {\cal O}(\dow^3), \qquad 
    N^\mu_\nhs = {\cal O}(\dow^2).
\end{equation}
The positivity of entropy production requires that the thermal conductivity is non-negative
\begin{equation}
    \kappa \geq 0.
\end{equation}
All other non-hydrostatic transport is at least second order in derivatives.

%-----------------------------------------------
\subsubsection{Two-derivative order}
%-----------------------------------------------

The non-hydrostatic sector is more interesting at subleading orders in derivative expansion. For systems that respect parity invariance, there are no possible contributions at one derivative order. However, at two  derivative order, keeping only terms that contribute to linearized hydrodynamics around a state with $\langle u^i\rangle = 0$  and $\langle\phi^i \rangle = \pi^i_0/q$, the full set of non-hydrostatic constitutive relations are given by
\begin{align}
\begin{split}
    \begin{pmatrix}
    \tilde\epsilon_{\nhs}^\mu \\
    \tilde\tau_{\nhs}^{\mu\nu} \\ 
    J_{\nhs}^{\mu\nu}
    \end{pmatrix} 
    = - T
    \begin{pmatrix}
    {\mathfrak{D}}_{\epsilon\epsilon}^{\mu\rho}  
    & {\mathfrak{D}}_{\epsilon\tau}^{\mu(\rho\sigma)}
    & {\mathfrak{D}}_{\epsilon d}^{\mu(\rho\sigma)} \\
    {\mathfrak{D}}'^{\mu(\rho\sigma)}_{\epsilon\tau} 
    & {\mathfrak{D}}_{\tau\tau}^{(\mu\nu)(\rho\sigma)}
    & {\mathfrak{D}}_{\tau d}^{(\mu\nu)(\rho\sigma)}\\
    {\mathfrak{D}}'^{\mu(\rho\sigma)}_{\epsilon d}  
    & {\mathfrak{D}}_{\tau d}'^{(\rho\sigma)(\mu\nu)} 
    & {\mathfrak{D}}_{dd}^{(\mu\nu)(\rho\sigma)}
    \end{pmatrix}
    \begin{pmatrix} 
        - h_\rho^\lambda\delta_\scB n_{\lambda} \\
        \frac{1}{2} h_\rho^\lambda h_\sigma^\tau
        \delta_\scB h_{\lambda\tau} \\ 
        h_{\lambda(\rho}h_{\sigma)}^\tau
        \delta_\scB \tilde{A}^{\lambda}_{~\tau}
    \end{pmatrix},
\end{split}
\label{eq:nonhs_subleading1}
\end{align}
where the objects appearing above are differential operators
\begin{align}\begin{split}
    {\mathfrak{D}}_{\epsilon\epsilon}^{\mu\nu}
    &= T\kappa\,h^{\mu\nu} 
    + \bar\Omega_\epsilon\, h^{\mu\nu} \delta_\scB
    - \sigma_\epsilon\nabla^{\mu}\nabla^{\nu}
    - \gamma_\epsilon\, h^{\mu\nu}\nabla_\rho\nabla^\rho
    + {\cal O}(\dow^3), \\ 
    {\mathfrak{D}}_{\epsilon\tau}^{\mu(\rho\sigma)}
    &= (\bar\sigma_{\epsilon\tau} + \sigma_{\epsilon\tau})\,
    h^{\rho\sigma} \nabla^\mu
    + 2(\bar\gamma_{\epsilon\tau} + \gamma_{\epsilon\tau})\,
    h^{\mu\langle\rho}\nabla^{\sigma\rangle}
    + {\cal O}(\dow^2), \\
    {\mathfrak{D}}_{\epsilon\tau}'^{\mu(\rho\sigma)}
    &= (\bar\sigma_{\epsilon\tau} - \sigma_{\epsilon\tau})\, 
    h^{\rho\sigma}\nabla^\mu
    + 2(\bar\gamma_{\epsilon\tau} - \gamma_{\epsilon\tau})\,
    h^{\mu\langle\rho}\nabla^{\sigma\rangle}
    + {\cal O}(\dow^2), \\
    {\mathfrak{D}}_{\epsilon d}^{\mu(\rho\sigma)}
    &= (\bar\sigma_{\epsilon d} + \sigma_{\epsilon d})\,
    h^{\rho\sigma}\nabla^\mu
    + 2(\bar\gamma_{\epsilon d} + \gamma_{\epsilon d})\,
    h^{\mu\langle\rho}\nabla^{\sigma\rangle}
    + {\cal O}(\dow^2), \\
    {\mathfrak{D}}_{\epsilon d}'^{\mu(\rho\sigma)}
    &= (\bar\sigma_{\epsilon d} - \sigma_{\epsilon d})\, 
    h^{\rho\sigma}\nabla^\mu
    + 2(\bar\gamma_{\epsilon d} - \gamma_{\epsilon d})\,
    h^{\mu\langle\rho}\nabla^{\sigma\rangle}
    + {\cal O}(\dow^2), \\
    {\mathfrak{D}}_{\tau\tau}^{(\mu\nu)(\rho\sigma)}
    &= \zeta\, h^{\mu\nu}h^{\rho\sigma}
    + 2\eta\,h^{\mu\langle\rho}h^{\sigma\rangle\nu}
    + {\cal O}(\dow), \\
    {\mathfrak{D}}_{\tau d}^{(\mu\nu)(\rho\sigma)}
    &= (\zeta_{\tau d} + \bar\zeta_{\tau d})\, 
    h^{\mu\nu}h^{\rho\sigma}
    + 2(\eta_{\tau d} + \bar\eta_{\tau d})\,
    h^{\mu\langle\rho}h^{\sigma\rangle\nu}
    + {\cal O}(\dow), \\ 
    {\mathfrak{D}}_{\tau d}'^{(\mu\nu)(\rho\sigma)}
    &= (\zeta_{\tau d} - \bar\zeta_{\tau d})\, 
    h^{\mu\nu}h^{\rho\sigma}
    + 2(\eta_{\tau d} - \bar\eta_{\tau d})\,
    h^{\mu\langle\rho}h^{\sigma\rangle\nu}
    + {\cal O}(\dow), \\ 
    {\mathfrak{D}}_{dd}^{(\mu\lambda)(\rho\sigma)}
    &= \zeta_d\, h^{\rho\sigma}h^{\mu\lambda}
    + 2\eta_d\,h^{\mu\langle\rho}h^{\sigma\rangle\nu}
    + {\cal O}(\dow),
\end{split}
\label{eq:nonhs_subleading2}
\end{align}
together with $h^{\mu\langle\rho}h^{\sigma\rangle\lambda} = h^{\mu(\rho}h^{\sigma)\lambda} - \frac{1}{d}h^{\mu\lambda}h^{\rho\sigma}$ and $\nabla^\mu = h^{\mu\nu}\nabla_\nu$. There are 19 new coefficients at second order. The coefficients $\zeta$ and $\eta$ are bulk and shear viscosities respectively, which have now been suppressed to second order due to the derivative suppression of fluid velocity. Five coefficients $\bar\Omega_\epsilon$, $\sigma_{\epsilon\tau}$, $\bar\sigma_{\epsilon\tau}$, $\gamma_{\epsilon\tau}$, and $\bar\gamma_{\epsilon\tau}$ appear at second order in ordinary hydrodynamics and still show up at second order in the presence of dipole conservation.\footnote{In flat spacetime, the $\bar\Omega_\epsilon$ term in the energy flux goes as $\propto \dow^i\dow_t T$. Using equations of motion, this will likely be replaced with terms involving just spatial derivatives and might get absorbed into a redefinition of other transport coefficients at this order. At the time of writing, we are not able to conclusively conclude that the same is true in the presence of spacetime sources as well.} On the other hand, there are two coefficients $\sigma_\epsilon$, $\gamma_\epsilon$ that would ordinarily show up at third order, but which are now promoted to second order in a dipole-conserving fluid. The remaining ten coefficients $\sigma_{\epsilon d}$, $\bar\sigma_{\epsilon d}$, $\gamma_{\epsilon d}$, $\bar\gamma_{\epsilon d}$, $\zeta_{\tau d}$, $\bar\zeta_{\tau d}$ $\eta_{\tau d}$, $\bar\eta_{\tau d}$, $\zeta_d$, and $\eta_d$ are specific to the conserved dipole moment. We emphasize that when expanding around an equilibrium state with $\langle\vec{u}^i\rangle\neq 0$ or $\langle \dow_i\phi_j\rangle \neq 0$, there would be many more transport coefficients.

After some computation, we can check that these constitutive relations satisfy the adibaticity equation with the entropy production quadratic form
\begin{align}\begin{split}
    \Delta &= \kappa\,h^{\mu\nu} {\cal V}_\mu {\cal V}_\nu
    \\
    &\qquad 
    + T
    \begin{pmatrix}
        h^{\mu\nu} \half\delta_\scB h_{\mu\nu} \\ 
        h^\mu_\nu \delta_\scB\tilde A^\nu_{~\mu}
    \end{pmatrix}^\rmT
    \begin{pmatrix}
        \zeta & \zeta_{\tau d} \\ 
        \zeta_{\tau d} & \zeta_d 
    \end{pmatrix}
    \begin{pmatrix}
        h^{\rho\sigma} \half\delta_\scB h_{\rho\sigma} \\ 
        h^\rho_\sigma \delta_\scB\tilde A^\sigma_{~\rho}
    \end{pmatrix} \\
    &\qquad 
    + 2T h^{\rho\langle\mu} h^{\nu\rangle\sigma}
    \begin{pmatrix}
        \half\delta_\scB h_{\mu\nu} \\ 
        h_{\mu\lambda} \delta_\scB\tilde A^\lambda_{~\nu}
    \end{pmatrix}^\rmT
    \begin{pmatrix}
        \eta & \eta_{\tau d} \\ 
        \eta_{\tau d} & \eta_d
    \end{pmatrix}
    \begin{pmatrix} 
        \half\delta_\scB h_{\rho\sigma} \\ 
        h_{\rho\lambda} \delta_\scB\tilde A^\lambda_{~\sigma}
    \end{pmatrix}
    + {\cal O}(\dow^5),
    \label{eq:Delta-2}
\end{split}
\end{align}
where we have defined 
\begin{align}
\begin{split}
    {\cal V}_\mu
    &=
    T\delta_\scB n_\mu 
    - \frac{\sigma_\epsilon}{2\kappa} \nabla_\mu \nabla^\rho\delta_\scB n_\rho 
    - \frac{\sigma_{\epsilon\tau}
    - \frac2d \gamma_{\epsilon\tau}
    }{\kappa}
    \nabla_\mu \lb \half h^{\rho\sigma} \delta_\scB h_{\rho\sigma} \rb
    - \frac{\sigma_{\epsilon d}
    - \frac2d \gamma_{\epsilon d}}{\kappa}
    \nabla_\mu \lb h_\lambda^\rho \delta_\scB A^\lambda_{~\rho} \rb \\
    &\qquad
    - \frac{\gamma_\epsilon}{2\kappa} \nabla_\rho \nabla^\rho
    \delta_\scB n_\mu 
    - \frac{\gamma_{\epsilon\tau}}{\kappa} 
    \nabla^\rho \delta_\scB h_{\rho\mu}
    - \frac{\gamma_{\epsilon d}}{\kappa}
    \nabla^\rho \lb 2h_{\lambda(\mu}\delta_\scB A^\lambda_{~\rho)}
    \rb + O(\partial^4)\,,
\end{split}
\end{align}
together with the free energy current
\begin{align}
\begin{split}
    N^\mu 
    &= - u^\mu \frac{1}{2}\bar\Omega_\epsilon
    h^{\rho\sigma} \delta_\scB n_\rho \delta_\scB n_\sigma 
    \\
    &\qquad 
    + T \half 
    \lb \bar\sigma_{\epsilon\tau} - \sigma_{\epsilon\tau} \rb
    h^{\mu\nu}h^{\rho\sigma} \delta_\scB n_\nu \delta_\scB h_{\rho\sigma}
    + T (\bar\gamma_{\epsilon\tau} - \gamma_{\epsilon\tau})\, 
    \delta_\scB n_\nu 
     h^{\rho\langle\mu}h^{\nu\rangle\sigma}
    \delta_\scB h_{\rho\sigma} \\ 
    &\qquad 
    + T
    \lb \bar\sigma_{\epsilon d} - \sigma_{\epsilon d} \rb
    h^{\mu\nu}h^{\sigma}_\rho
    \delta_\scB n_\nu \delta_\scB \tilde A^\rho_{~\sigma}
    + 2T 
    (\bar\gamma_{\epsilon d} - \gamma_{\epsilon d})\,
    \delta_\scB n_\nu 
     h^{\langle\mu}_\rho h^{\nu\rangle\sigma}
    \delta_\scB \tilde A^\rho_{~\sigma}
    + {\cal O}(\dow^4)\,.
\end{split}
\end{align}
We have ignored the terms that are cubic or higher order in fields while deriving these expressions. Furthermore, to manifest the quadratic form structure of the entropy production $\Delta$, we have included some ${\cal O}(\dow^6)$ terms in $\Delta$, arising when ${\cal O}(\dow^3)$ terms in ${\cal V}_\mu$ multiply each other. Formally, we can think of these as some contributions to $\Delta$ coming from four-derivative order constitutive relations that are not relevant at our derivative order of interest. See~\cite{Bhattacharyya:2013lha, Bhattacharyya:2014bha} for more details on this procedure. The first line of~\eqref{eq:Delta-2} refers to dissipation in the vector sector of the constitutive relations, the second to dissipation in the scalar sector, and the third to dissipation in the tensor sector.

We see that all 7 barred coefficients drop out from entropy production and can be identified as non-hydrostatic non-dissipative. The remaining 12 second order coefficients, together with the zeroth order thermal conductivity $\kappa$, are dissipative and lead to entropy production. The positivity of $\Delta$ requires that\footnote{If $\kappa = 0$, then the higher-order vector sector coefficients $\sigma_\epsilon, \sigma_{\epsilon\tau}$ etc. are constrained. However we expect that limit to be unphysical, i.e. that $\kappa$ never vanishes, analogous to the empirical lower bounds on viscosity.} 
\begin{gather}
    \kappa, \zeta, \eta \geq 0, \qquad 
    \zeta_d \geq \zeta_{\tau d}^2/\zeta, \qquad 
    \eta_d \geq \eta_{\tau d}^2/\eta.
    \label{eq:nonhs_inequalities}
\end{gather}

The non-hydrostatic constitutive relations are subject to additional constraints in cases where the microscopic system obeys microscopic time-reversal invariance. In the presence of dissipation, the macroscopic constitutive relations are typically not expected to be time-reversal invariant. Nonetheless, provided the microscopic theory under consideration is time-reversal invariant, the two-point response functions obtained from the dissipative hydrodynamic construction must obey certain discrete symmetry properties, known as the Onsager relations. We will discuss these in detail in the next section. For now, let us record the implications of the Onsager's relations. Under T invariance, the following non-hydrostatic transport coefficients must vanish\footnote{Ordinarily, imposing T symmetry switches off all non-hydrostatic non-dissipative transport coefficients in a parity-preserving hydrodynamic theory. This is because, ordinarily, all odd/even rank tensors (in spatial indices) in a parity-preserving theory are odd/even under T. In the presence of dipole symmetry, however, we have even rank objects $J^{ij}$, $a_{ij}$ that are odd under T. Consequently, Onsager's relation end up switching off certain dissipative coefficients appearing in $J^{ij}$.}
\begin{gather}
    \bar\Omega_\epsilon, \qquad 
    \sigma_{\epsilon\tau}, \qquad 
    \gamma_{\epsilon\tau}, \qquad 
    \sigma_{\epsilon d}, \qquad 
    \gamma_{\epsilon d}, \qquad 
    \zeta_{\tau d}, \qquad 
    \eta_{\tau d}.
    \label{eq:onsager-T-diss}
\end{gather}
On the other hand, under CT symmetry the following non-hydrostatic coefficients must be odd functions of the chemical potential $\tmu$, i.e.
\begin{gather}
    \bar\Omega_\epsilon, \qquad 
    \sigma_{\epsilon\tau}, \qquad 
    \gamma_{\epsilon\tau}, \qquad 
    \bar\sigma_{\epsilon d}, \qquad 
    \bar\gamma_{\epsilon d}, \qquad 
    \bar\zeta_{\tau d}, \qquad 
    \bar\eta_{\tau d}.
\end{gather}

This concludes our discussion of hydrodynamics with spontaneously broken dipole symmetry up to second order in derivatives and linear order in fluctuations. In the next section, we will use this theory to derive the hydrodynamic predictions for mode spectrum and linearized response functions.

%-----------------------------------------------
\section{Dispersion relations and linear response functions}
\label{sec:response}
%-----------------------------------------------

In this section, we solve the linearized hydrodynamic equations and use these to obtain the mode spectrum and linear response functions of various hydrodynamic operators. In the longitudinal sector, we find a thermal diffusion mode $\omega = -iD_{\|}k^2$ and pair of ``magnon-like'' propagating sound modes $\omega \sim \pm v_{s}k^2 -i \Gamma_s k^2$. On the other hand in the transverse sector, we find a ``subdiffusive'' shear mode $\omega \sim -i D_\perp k^4$. Subdiffusion is a common feature of kinematically constrained systems as has been emphasized in \cite{Iaconis:2019hab, Iaconis:2020zhc, Glodkowski:2022xje, Glorioso:2023chm}.

%-----------------------------------------------
\subsection{Response functions from hydrodynamics}
%-----------------------------------------------

Let us start with a quick overview of response functions in hydrodynamics; for more details see~\cite{Kovtun:2012rj}.
Within linear response, turning on a small external source $\delta s(t')$ for an operator ${O}(t')$ at some time $t'$, leads to a small deformation at later times of all operators ${O}(t)$ which do not commute with ${O}(t')$, i.e. $[{O}(t),{O}(t')]\neq 0$. The deformation can be expressed as
\begin{align}\begin{split}
    \delta \langle {O} (t,{\vec x})\rangle_s
    = -\int\df t'\,\df^d x' 
    G^{R}_{OO}(t-t',{\vec x}-{\vec x}')
    \delta s(t',{\vec x}')
    + {\cal O}(\delta s^2),
\end{split}\end{align}
where $\langle \ldots \rangle_s$ denotes expectation values in the presence of background sources and the object
\begin{align}\begin{split}
    G^{R}_{OO}(t-t',{\vec x}-{\vec x'}) 
    &\equiv 
    - i\theta(t-t')
    \langle[O(t,{\vec x}),O(t',{\vec x'})]\rangle \\ 
    &= - \frac{\delta}{\delta s(t',\vec x')} 
    \langle O(t,\vec x) \rangle_s \bigg|_{\delta s = 0}.
\end{split}\end{align}
is known as the response function or a retarded correlation function. It is often convenient to talk about the Fourier transform of the response function instead
\begin{align}\begin{split}
    G^{R}_{OO}(\omega,\vec k) 
    &= \int \df t\,\df^d x\,\df t'\,\df^d x'
    {\rm e}^{i\omega (t-t') - ik_i (x^i-x'^i)}
    G^{R}_{OO}(t-t',{\vec x}-{\vec x'}) 
    \\ 
    &= - \frac{\delta}{\delta s(\omega,\vec k)} 
    \langle O(\omega,\vec k) \rangle_s \bigg|_{\delta s = 0}.
    \label{eq:corr-def}
\end{split}\end{align}
The generalization to multiple operators is straightforward. The poles of the Fourier transformed response function can be used to read off the dispersion relations of the hydrodynamic system. The fact that the response function is only nonzero for $t>t'$, means that poles always lie in the lower-half complex $\omega$ plane.  Given a set of hydrodynamic equations coupled to background sources, we can use these to solve for hydrodynamic observables in terms of the respective background sources and use these to explicitly compute the response functions~\cite{Kovtun:2012rj}. This method goes beyond traditional Kadanoff-Martin approach~\cite{1963AnPhy..24..419K}, as it allows us to compute response functions of not just conserved densities but also their respective fluxes, while keeping track of the necessary contact terms required to satisfy the Ward identities.

We are interested in response functions for a dipole-invariant system in flat spacetime in the absence of external sources. To this end, we introduce plane wave fluctuations of the hydrodynamic fields
\begin{align}
\begin{gathered}
    T = T_0 + \delta T\, e^{-i\omega t + ik_i x^i}, \qquad 
    u^i = \delta u^i\, e^{-i\omega t + ik_i x^i}, \qquad 
    \tmu = \mu_0 + \delta\tmu\, e^{-i\omega t + ik_i x^i}.
\end{gathered}
\end{align}
Furthermore, we are interested in states with zero superflow, so we introduce fluctuations of the Goldstone field as\footnote{The states with nonzero superflow are spatially inhomogeneous, so a simple linearized analysis in terms of Fourier modes is not accessible. Note that the case for spontaneously broken dipole symmetry is qualitatively distinct from the states with nonzero superflow in usual superfluids with spontaneously broken U(1) symmetry, because fluctuations in the U(1)-invariant operators are still homogeneous. In the present case, however, due to covariant derivatives in the definition, the fluctuations of U(1) and dipole-invariant operator $\tilde a_{ij}$ is non-homogeneous.}
\begin{equation}
    \phi_i = \frac{1}{q} \pi_i^0 
    + \delta\phi_i\, e^{-i\omega t + ik_i x^i}.
\end{equation}
We also introduce fluctuations of the background sources on top of the flat background, i.e.
\begin{align}
\begin{gathered}
    n_t = 1 + \delta n_t\, e^{-i\omega t + ik_i x^i}, \qquad 
    n_i = \delta n_i\, e^{-i\omega t + ik_i x^i}, \\
    g_{ij} = \delta_{ij} + \delta g_{ij}\,
    e^{-i\omega t + ik_i x^i}, \qquad 
    v^{i} = \delta v^i\, e^{-i\omega t + ik_i x^i}, \\
    A_{t}
    = \delta A_t\, e^{-i\omega t + ik_i x^i}, \qquad 
    A_{i}
    = \delta A_i\,
    e^{-i\omega t + ik_i x^i}, \qquad 
    a_{ij}
    = \delta a_{ij}\, e^{-i\omega t + ik_i x^i}.
 \end{gathered}
 \end{align}
The set of sources and operators for our hydrodynamic theory can be read off using \eqref{generatingfunctional1} together with \eqref{eq:non-cov-sources} and \eqref{eq:non-cov-currents} as
\begin{equation}
    \begin{split}
        \delta s
        &= \begin{pmatrix}
            {- \delta n_t} \\
            {- \delta n_i} \\
            {- \delta v^i} \\
            \half \delta g_{ij} \\
            \delta A_t \\
            \delta A_i \\ 
            \half \delta a_{ij}
        \end{pmatrix}, \qquad
        \langle {O} \rangle_s
        = \sqrt{\gamma}\,
        \begin{pmatrix}
            \epsilon^t + \pi_k v^k \\
            \epsilon^i - J^{ik} F_{tk} \\
            \pi_i - \tau_{ik}n^k \\
            \tau^{ij} - 2A^{(i}_{~k}J^{j)k} \\
            J^t \\
            J^i \\
            J^{ij}
        \end{pmatrix}.
    \end{split}
    \label{eq:physical-operators}
\end{equation}
where the operators are evaluated on the solutions of the equations of motion. The operators presented here are only correct up to linear order in background fluctuations; these are sufficient to compute two-point response functions. Our goal is to solve the hydrodynamic equations to determine $\delta T$, $\delta u^i$, $\delta\tmu$, and $\delta\phi_i$ in terms of the background field variations $\delta n_t$, $\delta n_i$, $\delta g_{ij}$, $\delta v^i$, $\delta A_t$, $\delta A_i$, and $\delta a_{ij}$. Having done that, we can substitute these to find operator expectation values $\langle O\rangle_s$ in terms of background fields and use these to obtain the response functions and dispersion relations of our hydrodynamic model.

If we require the underlying microscopic description of the theory to feature certain discrete time-reversal symmetry, e.g. T or CT, the response functions satisfy the Onsager's reciprocity relations
\begin{equation}
    G^{R}_{OO}(\omega,\vec k) 
    = \Theta \cdot \Big[G^{R}_{OO}(\omega,-\vec k)\Big]^\rmT \cdot \Theta,
    \label{eq:onsager}
\end{equation}
where $\Theta$ is a diagonal matrix denoting the eigenvalues of operators under the said discrete transformation given in table \ref{tab:CPT}; see \cite{Kovtun:2012rj} for more details. Note that the overall sign of $\pi^i_0$ and $\mu_0$ must be flipped under T and CT respectively. 

%-----------------------------------------------
\subsection{Dispersion relations}
%-----------------------------------------------

Let us first obtain the dispersion relations for our hydrodynamic theory. To get some physical intuition of the subsequent results, it is useful to first ignore the dipole Ward identity and leave the dipole Goldstone $\phi_\mu$ off-shell. The ensuing mode spectrum will be that of a boost-agnostic fluid without dipole symmetry; see~\cite{Armas:2020mpr}. Under this assumption, we can solve the energy-momentum and charge conservation equations for the hydrodynamic fields $\beta^\mu$ and $\Lambda_\beta$ in terms of the dipole-invariant background sources $\delta\tilde s$. The equations can be written schematically as
\begin{align}
   \begin{pmatrix}
        {\tilde{\mathcal{M}}}_{\|}(\omega,k)_{3\times 3}
        & \rvline &
        {\mathbb 0}\\
        \hline 
        {\mathbb 0}& \rvline & {\tilde{\mathcal{M}}}_{\perp}(\omega,k)
    \end{pmatrix}
    \begin{pmatrix} 
        \delta\beta^{t}\\ 
        \delta \beta_{\|}\\ 
        \delta \Lambda_\beta\\
        \hline 
        \delta \beta_\perp
    \end{pmatrix} 
    = \tilde{\mathcal{S}}(\omega, k)\,\delta \tilde{s}.
\end{align}
Here we have denoted $\delta\beta_{\|} = \hat{k}_i\delta\beta^i$ and $\delta \beta_\perp^i = (\delta^{i}_j -\hat{k}^i\hat{k}_j)\delta\beta^j$ to be the longitudinal and transverse projections of $\beta^i$ in the Fourier basis, where $\hat k^i = k^i/|k|$. Furthermore, $\delta\tilde{s}$ denotes the dipole-invariant version of background fields $\delta s$, combined with the dipole Goldstone $\phi_\mu$. Here $\tilde{\cal S}$ and $\tilde{\cal M}$ are matrices of suitable dimensionality. 
Setting $\delta \tilde{s} = 0$, the $\phi_\mu$-off-shell modes are found by solving $\det \tilde{\cal M}_{\|} = \det\tilde{\cal M}_{\perp} = 0$. We find a sound and thermal diffusion mode in the longitudinal sector 
\begin{align}
\begin{split}
    \tilde\omega 
    &= \pm \sqrt{\frac{q^2\chi_{ss}-2qs\chi_{sq}
    +s^2\chi_{qq}}{\rho\, \Xi}} 
    - \frac{i}{2}\left(\frac{\zeta+\eta}{\rho} + \frac{(q\chi_{sq}-s\chi_{qq})^2\kappa}{T\Xi(q^2\chi_{ss}-2qs\chi_{sq}+s^2\chi_{qq})}\right)k^2+ \mathcal{O}(k^3), \\ 
    \tilde\omega 
    &= -i \frac{q^2\kappa}{T(s^2\chi_{qq}-2qs\chi_{sq}+q^2\chi_{ss})}k^2 + \mathcal{O}(k^3),
\end{split}
\end{align}
where $\Xi = \chi_{ss}\chi_{qq}- \chi_{sq}^2$. The notation $\tilde\omega$ above is to remind us that these are \emph{not} the physical modes of our theory. Similarly, we find a shear mode in the transverse sector
\begin{align}
    \tilde\omega = -i \frac{\eta}{\rho}k^2+ O(k^3).
    \label{eq:tran-mode-inv}
\end{align}
This mode spectrum should be qualitatively familiar from ordinary hydrodynamics.

Physical dispersion relations in our model are obtained when we put the Goldstone $\phi_\mu$ on-shell by solving the dipole Ward identity. In other words, we solve for $\delta \beta_\mu, \delta \Lambda_\beta, \delta \phi_\mu$ in terms of the physical background sources $\delta s$. The equations of motion can be schematically written as
\begin{align}
   \begin{pmatrix}
        {\mathcal{M}}_{\|}(\omega,k)_{4\times 4} & \rvline &
        {\mathbb 0}\\
        \hline 
        {\mathbb 0}& \rvline & {\mathcal{M}}_{\perp}(\omega,k)_{2\times 2}
    \end{pmatrix}
    \begin{pmatrix}
        \delta\beta^{t}\\ 
        \delta \beta_{\|}\\ 
        \delta \Lambda_\beta\\ 
        \delta \phi_{\|}\\ 
        \hline 
        \delta \beta_\perp\\ 
        \delta \phi_{\perp}
    \end{pmatrix} 
    = \mathcal{S}(\omega, k)\,\delta s.
\end{align}
where we have introduced the decomposition for $\phi_i$ similar to the one introduced for $\beta^i$ above. Setting $\delta s = 0$, the physical mode spectrum can be obtained by setting $\det\mathcal{M}_{\|} = \det \mathcal{M}_\perp = 0$. In the longitudinal sector, we find the linearly propagating sound modes are suppressed to magnon-like modes and the thermal diffusion mode stays intact, i.e.
\begin{equation}
\begin{split}
    \omega 
    &= \pm v_{s}k^2 
    - i\Gamma_{s} k^2 + {\cal O}(k^4), \\ 
    \omega 
    &= -iD_\| k^2 + {\cal O}(k^4).
\end{split}
\end{equation}
In the transverse sector, we find that the shear diffusion mode is suppressed to become sub-diffusive, i.e.
\begin{equation}
    \omega = - i D_{\perp} k^4.
    \label{eq:subdiff-shear}
\end{equation}
Here the magnon velocity and attenuation, and the diffusion constants are given by
\begin{equation}
\begin{split}
    2\Gamma_{s} + D_\|
    &= \frac{\chi_{qq}}{T\Xi} \kappa, \\ 
    v_{s}^2 + \Gamma_{s}^2 + 2D_\|\Gamma_s
    &= \frac{\chi_{ss}B'_d + \chi_{sd}^2}{\Xi}
    - \frac{2s}{q} 
    \frac{\chi_{sq}B'_d + \chi_{qd}\chi_{sd}}{\Xi}
    + \frac{s^2}{q^2} 
    \frac{\chi_{qq} B'_d
    + \chi_{qd}^2}{\Xi}, \\ 
    (v_{s}^2 + \Gamma_{s}^2) D_\|
    &= \frac{B'_d}{T\Xi} \kappa, \\ 
    D_{\perp}
    &= \frac{\eta}{q^2} G'_d.
\end{split}
\end{equation}
where $B'_d = B_d + 2\frac{d-1}{d}G_d$ and $G'_d = G_d + \chi_m$. 

Note that we do not find any real $k^4$ contribution to the subdiffusive shear mode in the transverse sector. This fact can be understood as a generic consequence of rotational invariance and reality of the hydrodynamic equations of motion. Hydrodynamic response functions can schematically be expressed as $G^R(\omega, k) \sim P(\omega, k)/Q(\omega, k)$, where $P$ and $Q$ are polynomials in $\omega$ and $k$. We can always rescale these polynomials in a way that the reality and rotation-invariance of hydrodynamic equations implies $P^*(\omega, k) = P(-\omega,-k) = P(-\omega,k)$ and $Q^*(\omega,k) = Q(-\omega,-k) = Q(-\omega,k)$. In particular, this implies that both $P$ and $Q$ are real polynomials of $i\omega$ and $k^2$, and thus their roots for $i\omega$ are either real or arise in conjugate complex pairs. In particular, for the roots of $Q$ of the form $\omega = \omega_n(k)$ representing the mode spectrum of the theory, this means that either $\Re\omega_n(k) = 0$ or that there exists another mode $\omega = \omega_m(k)$ such that $\Re\omega_n(k)= -\Re\omega_m(k)$. The transverse sector of our theory has precisely one mode $\omega = \omega_\perp(k)$ and, hence, it follows that $\Re\omega_\perp(k) = 0$, explaining the absence of the real $k^4$ contribution. It also explains why the pair of magnon-like sound modes in the longitudinal sector have oppositely signed real parts and the thermal diffusion mode is purely imaginary. 

The fact that dissipation in the longitudinal sector appears at ${\cal O}(k^2)$, the same order as the thermodynamic contributions, emphasizes the importance of thermal fluctuations in a dipole superfluid. As we have seen, dissipation at this order is controlled by a single transport coefficient $\kappa$ from the vector sector. Other dissipative transport coefficients from the vector sector, whose signs are not fixed by the positivity of entropy production when $\kappa \neq 0$, will appear at ${\cal O}(k^4)$ in the dispersion relations and contribute to fracton-like subdiffusive dissipation. Setting $\kappa = 0$ for illustrative purposes, we indeed find that thermal diffusion mode becomes subdiffusive
\begin{align}
    \omega \big|_{\kappa = 0} 
    = - i k^4
    \frac{B_d'(\sigma_\epsilon + \gamma_\epsilon)}{\Xi T^2 v_{s}^2} +\mathcal{O}(k^6).
\end{align}
When $\kappa = 0$, the coefficients $\gamma_\epsilon$, $\sigma_\epsilon$ are constrained by entropy production to be positive.

%-----------------------------------------------
\subsection{Response functions}
\label{S:responseFunctions}
%-----------------------------------------------

In this section, we look at the hydrodynamic response functions in the transverse sector, involving the operators $\pi_\perp$, $J_\perp$, and $\epsilon_\perp$. The response functions for other operators in the transverse sector, namely $\tau_{\| \perp}$ and $J_{\|\perp}$, can be obtained from these using Ward identities. The response functions in the longitudinal sector can also be easily obtained using our formalism, but the respective expressions are too involved to report here. We have provided a supplementary Mathematica notebook with this submission for the readers who might want to give it a go. We will also present the results for optical response functions obtained in the limit $k\to 0$, in which limit the theory becomes isotropic and the transverse and longitudinal sectors coincide. 

Our formalism can predict accurate results for response functions up to first non-trivial correction in the gradient expansion, which are $k^2$ corrections on top the leading order result. Further taking into account the isotropy of the hydrodynamic model, this means that we can infer results for $G^R_{\pi_\perp\pi_\perp}$ that are accurate up to $\mathcal{O}(k^2)$ corrections, those for  $G^R_{J_\perp \pi_\perp}$, $G^R_{\epsilon_\perp\pi_\perp}$ up to $\mathcal{O}(k^4)$ corrections, and finally $G^R_{J_\perp J_\perp}$, $G^R_{\epsilon_\perp\epsilon_\perp}$, $G^R_{J_\perp\epsilon_\perp}$ up to $\mathcal{O}(k^6)$ corrections. However, due to the sheer number of transport coefficients, the general results are quite complicated to present here even in the transverse sector. These have been included in the supplementary Mathematica notebook.

Let us first look at the response function of dipole-invariant momenta $\tilde\pi_\perp$, obtained by keeping the dipole Goldstone $\phi_\mu$ off-shell by ignoring the dipole Ward identity. We find
\begin{align}
    \tilde G^R_{\tilde{\pi}_\perp\tilde{\pi}_\perp}
    &= \frac{\eta k^2}
    {i\omega - \eta/\rho\, k^2}
    + {\cal O}(k^2).
\end{align}
This resembles the respective result from ordinary hydrodynamics without dipole symmetries. We notice that the pole of this correlator is given by the transverse $\phi_\mu$-off-shell shear mode in \eqref{eq:tran-mode-inv}. Instead, if we take $\phi_\mu$ on-shell by solving the dipole Ward identity, we find the response function of physical momenta $\pi_\perp$, leading to
\begin{align}
    G^R_{\pi_\perp\pi_\perp}
    &= \frac{\eta k^2}{Q_\perp}
    \left(1 + \frac{k^2}{q^2}
    \lb 2q\lambda_\perp + \rho G_d'\rb 
    \right)
    + {\cal O}(k^2),
\end{align}
where $\lambda_\perp = \lambda_5+\lambda_6$ and we have defined the denominator
\begin{align}
    Q_\perp
    = i\omega\lb 
    1+ \frac{2k^2}{q}
    \lb \bar{\eta}_{\tau d} + \lambda_\perp \rb
    \rb 
    + \frac{k^2}{q^2} G_d'
    \lb i\omega \rho - \eta k^2 \rb
    + \mathcal{O}(k^6).
\end{align}
There are two poles of this correlator: one given by the subdiffusive hydrodynamic shear mode in \eqref{eq:subdiff-shear} and another non-hydrodynamic mode at $\omega \sim -i/k^2 + {\cal O}(k^0)$ that is not relevant for the low-energy spectrum. The dictionary between the dipole-invariant and physical response functions is explained in appendix \ref{app:responsefunctions}. Notice that, in the limit $q\to 0$, the physical response function reduces to its dipole-invariant value, meaning that dipole symmetry has no consequence for response functions in chargeless states. It is also interesting to note that as $\omega\to 0$, we recover the hydrostatic response function
\begin{align}
    \lim_{\omega\to 0}  G^R_{\pi_\perp \pi_\perp} 
    = -\frac{q^2}{G'_d k^2} + \mathcal{O}(k^0),
\end{align} 
justifying the role of $\pi_i$ as a dipole Goldstone.

For the remaining response functions, we only look at the respective physical versions where the Goldstone $\phi_\mu$ has been taken on-shell. We find the diagonal response functions
\begin{align}
\begin{split}
    G^R_{\epsilon_\perp \epsilon_\perp} 
    &= \frac{1}{Q_\perp}
    \left( \omega^2 T\kappa
    + i\omega k^2\lb
    \frac{(\epsilon+p)^2}{q^2} G_d'
    + \frac{\epsilon+p}{q}2\chi_{mn}
    + \chi_n
    \rb \right)
    + {\cal O}(k^4), \\
    G^R_{J_\perp J_\perp} 
    &= \frac{i\omega k^2 G_d}{Q_\perp} 
    + {\cal O}(k^4).
\end{split}
\end{align}
Our formalism can also predict the ${\cal O}(k^4)$ corrections to these correlators, but these expressions are quite involved and not illuminating. We can also consider the off-diagonal response functions. For the off-diagonal response involving the momentum density, we find
\begin{align}
    G^{R\pm}_{J_\perp\pi_\perp}
    &= \frac{1}{Q_\perp}
    \lb i\omega q 
    + \frac{i\omega k^2}{q} \lb 
    \frac{\rho}{q} G_d'
    + \bar{\eta}_{\tau d}
    \pm \eta_{\tau d}
    + 2\lambda_\perp \rb  
    - \frac{\chi_{m}}{q} \eta k^4
    \rb 
    + {\cal O}(k^4)
    , \nn\\ 
    G^{R\pm }_{\epsilon_\perp \pi_\perp } 
    &= \frac{1}{Q_\perp}\Bigg(
    i\omega \epsilon 
    + i\omega k^2
    \lb 
    \bar\gamma_{\epsilon\tau} 
    \pm \gamma_{\epsilon\tau} 
    \mp \frac{p_d}{q}\eta
    + \frac{\epsilon+p}{q}
    \lb 
    \frac{\rho}{q} G'_d
    \pm \eta_{\tau d} 
    \rb
    + \frac{\epsilon-p}{q} \bar\eta_{\tau d}
    + \frac{\epsilon}{q}2\lambda_\perp
    - \lambda_{27}
    \rb
    \nn\\
    &\hspace{40pt}
    - \frac{k^2}{q^2} p G_d' (i\omega\rho - \eta k^2)
    + \frac{k^4}{q} \eta\chi_{mn}
    \Bigg)
    + {\cal O}(k^4).
    % \nn\\
    % G^{R\pm}_{\tau_{\| \perp} \pi_\perp} 
    % &=  \frac{\omega k\eta}{Q_\perp}
    % + {\cal O}(k^3), \nn\\ 
    % %
    % G^{R\pm}_{\tau_{\| \perp}J_\perp} \nn\\
    % %
    % G^{R\pm}_{\tau_{\| \perp}\epsilon_\perp} 
    % &= 
    % \frac{1}{Q_\perp }
    % \lb \omega k^3 \frac{G_d}{q}\eta 
    % - i\omega^2 k
    % (\bar\eta_{\tau d} \pm \eta_{\tau d}) \rb 
    % + {\cal O}(k^5).
\end{align}
We have used the notation $G^{R+}_{AB} = G^R_{AB}$ and $G^{R-}_{AB} = G^R_{BA}$. Here we see that the off-diagonal terms are symmetric up to three signs of $p_d$ and certain dissipative coefficients. Imposing Onsager reciprocity relations as in \eqref{eq:onsager-T-diss} sets these coefficients to zero, leading to symmetric off-diagonal response. Similarly, for the energy-charge off-diagonal correlator we find
\begin{equation}
    G^{R\pm}_{\epsilon_\perp J_\perp} 
    = \frac{1}{Q_\perp}
    \lb \pm p_d \omega^2
    + i\omega k^2\frac{\epsilon+p}{q}G_d \rb
    + {\cal O}(k^4).
\end{equation}
Again, we have dropped the ${\cal O}(k^4)$ corrections to this correlators for clarity.

Now, let us look at the frequency-dependent correlators at zero wavevector. In this limit, all correlators involving the conserved densities $\epsilon^t$, $J^t$, and $\pi_i$ and the U(1) flux $J^i$ are zero due to the Ward identities (up to contact terms). The optical correlators then read
\begin{align}
\begin{split}
    G^R_{\epsilon^i\epsilon^j}(\omega) 
    &= - \lb \frac{\rho\,p_d^2}{q^2} \omega^2
    + i\omega T\kappa \rb \delta^{ij}, \\
    G^R_{\tau^{ij}\tau^{kl}}(\omega) 
    &= \left(\frac{q^2\chi_{ss} - 2qs\chi_{sq} + s^2\chi_{qq}}{\Xi }
    - i\omega \zeta \right) \delta^{ij}\delta^{kl}
    + (p - i\omega\eta) 2\delta^{k\langle i}\delta^{j\rangle l}, \\
    G^R_{J^{ij}J^{kl}}(\omega) 
    &= \left(\frac{\chi_{qd}^2\chi_{ss} 
    - 2\chi_{qd}\chi_{sd}\chi_{sq} + \chi_{sd}^2\chi_{qq}}{\Xi }
    + B_d - i\omega \zeta_d \right) \delta^{ij}\delta^{kl}
    + (G_d - i\omega\eta_d) 2\delta^{k\langle i}\delta^{j\rangle l}, \\
    G^R_{\tau^{ij}J^{kl}}(\omega) 
    &= \left(\frac{q\chi_{qd}\chi_{ss} 
    - (q\chi_{sd}+s\chi_{qd})\chi_{sq} + s\chi_{sd}\chi_{qq}}{\Xi }
    - i\omega (\zeta_{\tau d}+\bar\zeta_{\tau d}) \right) \delta^{ij}\delta^{kl}\\
    &\qquad 
    + \lb p_d - i\omega(\eta_{\tau d}+\bar\eta_{\tau d})\rb
    2\delta^{k\langle i}\delta^{j\rangle l}, \\
    G^R_{J^{ij}\tau^{kl}}(\omega) 
    &= \left(\frac{q\chi_{qd}\chi_{ss} 
    - (q\chi_{sd}+s\chi_{qd})\chi_{sq} + s\chi_{sd}\chi_{qq}}{\Xi }
    - i\omega (\zeta_{\tau d}-\bar\zeta_{\tau d}) \right) \delta^{ij}\delta^{kl}\\
    &\qquad 
    + \lb p_d - i\omega(\eta_{\tau d}-\bar\eta_{\tau d})\rb
    2\delta^{k\langle i}\delta^{j\rangle l}.
    \label{eq:optical_response}
\end{split}
\end{align}
We can use these to find the Kubo formulae
\begin{align}
\begin{split}
    \kappa 
    &=
    \lim_{\omega\to 0}\lim_{k\to 0}
    \frac{-1}{T\omega} \Im G^R_{\epsilon_\perp\epsilon_\perp}, \\
    \eta
    &=
    \lim_{\omega\to 0}\lim_{k\to 0}
    \frac{-1}{\omega} \Im G^R_{\tau_{\perp\|}\tau_{\perp\|}}, \\
    \zeta + 2\frac{d-1}{d}\eta
    &=
    \lim_{\omega\to 0}\lim_{k\to 0}
    \frac{-1}{\omega} \Im G^R_{\tau_{\|\|}\tau_{\|\|}}, \\ 
    \eta_d
    &=
    \lim_{\omega\to 0}\lim_{k\to 0}
    \frac{-1}{\omega} \Im G^R_{J_{\perp\|}J_{\perp\|}},  \\
    \zeta_d + 2\frac{d-1}{d}\eta_d
    &=
    \lim_{\omega\to 0}\lim_{k\to 0}
    \frac{-1}{\omega} \Im G^R_{J_{\|\|}J_{\|\|}}, \\
    \eta_{\tau d}
    &=
    \lim_{\omega\to 0}\lim_{k\to 0}
    \frac{-1}{2\omega} \Im \lb G^R_{\tau_{\perp\|}J_{\perp\|}}
    + G^R_{J_{\perp\|}\tau_{\perp\|}}
    \rb , \\
    \zeta_{\tau d} + 2\frac{d-1}{d}\eta_{\tau d}
    &=
    \lim_{\omega\to 0}\lim_{k\to 0}
    \frac{-1}{2\omega} \Im \lb G^R_{\tau_{\|\|}J_{\|\|}}
    + G^R_{J_{\|\|}\tau_{\|\|}}
    \rb, \\
    \bar\eta_{\tau d}
    &=
    \lim_{\omega\to 0}\lim_{k\to 0}
    \frac{-1}{2\omega} \Im \lb G^R_{\tau_{\perp\|}J_{\perp\|}}
    - G^R_{J_{\perp\|}\tau_{\perp\|}}
    \rb , \\
    \bar\zeta_{\tau d} + 2\frac{d-1}{d}\bar\eta_{\tau d}
    &=
    \lim_{\omega\to 0}\lim_{k\to 0}
    \frac{-1}{2\omega} \Im \lb G^R_{\tau_{\|\|}J_{\|\|}}
    - G^R_{J_{\|\|}\tau_{\|\|}}
    \rb.
    \label{eq:Kubo_formulae}
\end{split}
\end{align}
The Kubo formulae for other transport coefficients require us to probe the response functions at nonzero wavevector. We shall not report these in the manuscript.

%-----------------------------------------------
\section{Towards s-wave dipole superfluids}
\label{S:swave}
%-----------------------------------------------
 
In our discussion thus far, we have presented a detailed analysis of p-wave dipole superfluids where only the dipole symmetry is spontaneously broken, but the U(1) symmetry is left intact. In this section, we briefly outline how our results can be generalized to s-wave dipole superfluids where the U(1) symmetry is spontaneously broken as well. At nonzero temperature these phases are only expected to exist in spatial dimension $d>4$, below which infrared fluctuations of the Goldstone restore the U(1) symmetry at long distance.

We leave a comprehensive analysis to a forthcoming manuscript \cite{swavearticle}, but record the most salient features here. In particular, we find that s-wave dipole superfluids admit ordinary sound modes and so resemble a model with a dynamical scaling exponent $z=1$ instead of $z=2$ as for p-wave dipole superfluids. This raises the prospect that the natural derivative counting scheme for this phase is different than in the p-wave phase, which we comment on below and will address in~\cite{swavearticle}.

\vspace{1em}
\noindent 
\textbf{Hydrodynamic equations.}---Instead of an independent vector Goldstone $\phi_\mu$, s-wave dipole superfluids are expressed in terms of a scalar U(1) Goldstone $\phi$ transforming according to
\begin{equation}
    \delta_{\hat\scX}\phi = \lie_\chi\phi - \Lambda.
\end{equation}
The dipole symmetry is automatically spontaneously broken as a result of the spontaneously broken U(1) symmetry, and the associated vector Goldstone is given by 
\begin{equation}
    \phi_\mu = - h_\mu^\nu \lb \dow_\nu \phi + A_\nu \rb.
    \label{eq:swave_vector_constraint}
\end{equation}
Introducing independent Goldstones for U(1) and dipole symmetries is redundant from the perspective of low-energy effective field theory; see footnotes \ref{foot:both-breaking-algebra} and \ref{foot:both-breaking-hs}. 

Most of the discussion of p-wave dipole superfluids can be extended to s-wave by simply replacing $\phi_\mu$ with its definition above in terms of $\phi$. In particular, after substituting for $\phi_\mu$, the U(1)- and dipole-invariant versions of the background gauge fields become 
\begin{equation}
\begin{split}
    \tilde A_\mu 
    &= n_\mu \Phi - \dow_\mu\phi,\\ 
    \tilde{a}_{\mu\nu} 
    &= a_{\mu\nu} 
    - 2h_{(\mu}^\rho h_{\nu)}^\sigma    
    \nabla_\rho\nabla_\sigma \phi 
    - 2h_{(\mu}^\rho h_{\nu)}^\sigma    
    \nabla_\rho A_\sigma,
\end{split}
\end{equation}
where we have defined $\Phi = v^\mu(A_\mu + \partial_\mu \phi)$. We note that, unlike the p-wave case, $\tilde a_{\mu\nu}$ here mixes both $a_{\mu\nu}$ and $A_\mu$. Furthermore, $\tilde A_\mu$ is only sensitive to the temporal part of the U(1) gauge field $v^\mu A_\mu$. The resulting dipole-invariant chemical potential is given as
\begin{equation}
    \tmu 
    = \frac{\Lambda_\beta + \beta^\mu\tilde A_\mu}{\beta^\lambda n_\lambda}
    = \Phi 
    - T\delta_\scB \phi,
\end{equation}
meaning that, in hydrostatic equilibrium when $\delta_\scB\phi = 0$, the chemical potential is just $\Phi$. 

Taking these observations into account, one can construct a theory of hydrodynamics for s-wave dipole superfluids similar to our discussion in Section \ref{sec:adiabaticity}. The hydrodynamical variables can be taken to be a local temperature $T$, chemical potential $\tmu$, velocity $u^{\mu}$ (satisfying $u^{\mu}n_{\mu}=1$), and scalar Goldstone $\phi$. The hydrodynamical equations are the Ward identities for energy, momentum, and U(1) current conservation, together with a Josephson condition. At the level of constitutive relations the vector current $h^{\mu}_{\nu}J^{\nu}$ is exactly fixed to be the gradient of the dipole current, $h^{\mu}_{\nu}J^{\nu} = \nabla_{\nu}'J^{\mu\nu}$, and we supply constitutive relations for the energy current, momentum current, spatial stress density, charge density $n_{\mu}J^{\mu}$, and dipole current $J^{\mu\nu}$.

The derivation of the adiabaticity equation also follows exactly the same as before in Section \ref{sec:adiabaticity}, culminating into its final form in terms of dipole-invariant objects
\begin{align}\begin{split}
    \nabla_\mu' \tilde N^\mu &=
    - \tilde\epsilon^\mu
    \delta_{\scB} n_\mu 
    + \lb v^\mu \tilde\pi^\nu+\frac{1}{2}\tilde\tau^{\mu\nu}\rb 
    \delta_{\scB} h_{\mu\nu}
    + J^\mu \delta_{\scB}\lb\Phi\, n_\mu\rb 
    + J^{\mu\nu}h_{\nu\lambda} \delta_{\scB} \tilde A^{\lambda}_{~\mu} \\
    &\hspace{15em}
    + \lb  X + \nabla'_\mu J^\mu \rb \delta_\scB\phi 
    + \Delta.
    \label{eq:adiabaticity-swave}
\end{split}\end{align}
Recall from above that the spatial part of $\tilde A_\mu$ is not an independent field in the s-wave phase and thus has been substituted in terms of $\phi$ in this expression. 

An argument similar to \eqref{eq:joseph} gives a trivial solution of the adiabaticity equation
\begin{equation}
    X = - \nabla'_\mu J^{\mu} 
    - T\sigma_\phi\delta_\scB\phi
    + \ldots, \qquad 
    \Delta = T\sigma_\phi (\delta_\scB\phi)^2,
    \label{eq:Delta-joseph-swave}
\end{equation}
for some positive coefficient $\sigma_\phi$. Upon imposing the Josephson equation $X=0$, we find 
\begin{equation}
    \delta_\scB\phi = 
    - \frac{1}{T\sigma_\phi} \nabla'_\mu J^\mu
    + \ldots
    \quad\implies\quad 
    \Phi = \tmu 
    - \frac{1}{\sigma_\phi} \nabla'_\mu J^\mu
    + \ldots,
    \label{eq:joseph-swave}
\end{equation}
where ellipses represent further corrections that can be added consistent with the adiabaticity equation. Noting that $\Phi = v^\mu\dow_\mu\phi$ in the absence of background fields, this equation is reminiscent of the Josephson equation from ordinary U(1) superfluids~\cite{Bhattacharya:2011eea}. We could proceed as we did following \eqref{eq:joseph}, by treating the Josephson equation as the defining equation for $\tmu$ and drop the $\delta_\scB\phi$ in adiabaticity equation \eqref{eq:adiabaticity-swave}. However, it is customary to fix the definition of $\tmu$ by forbidding the U(1) density $J^\mu n_\mu$ to obtain any non-hydrostatic corrections. In this interpretation, \eqref{eq:joseph-swave} should be seen as determining the time-derivatives of the scalar Goldstone $v^\mu\dow_\mu\phi$ in terms of $\tmu - v^\mu A_\mu$ and further derivative corrections.

\vspace{1em}
\noindent 
\textbf{Derivative counting and constitutive relations.}---With the hydrodynamic variables and equations out of the way, we require a derivative counting scheme to construct constitutive relations. Given that our derivative counting for p-wave dipole superfluids required that $\phi_\mu\sim{\cal O}(\dow^{-1})$, we anticipate $\phi$  to be ${\cal O}(\dow^{-2})$ along with the p-wave scalings for the fields and currents. However, as we shall see below, the linearized mode spectrum of the theory suggests another potential derivative counting scheme where we take $\phi \sim {\cal O}(\dow^{-1})$ and $\phi_\mu\sim{\cal O}(\dow^0)$, together with dynamical scaling exponent $z=1$, meaning that $v^\mu \sim {\cal O}(\dow^0)$ and $v^\mu\dow_\mu\sim{\cal O}(\dow^1)$. In this counting, all the hydrodynamic fields $u^\mu$, $T$, $\tmu$ are ${\cal O}(\dow^0)$, similar to ordinary hydrodynamics. Following through the consistency arguments similar to those in Section \ref{sec:derivative-counting}, we find that, under this alternative counting, the conserved currents $\epsilon^\mu$, $\pi_\mu$, $\tau^{\mu\nu}$, $J^\mu$ are all ${\cal O}(\dow^0)$, except $J^{\mu\nu}$  that is formally ${\cal O}(\dow^{-1})$. Correspondingly, the background fields $n_\mu$, $v^\mu$, $h_{\mu\nu}$, $A_\mu$ are ${\cal O}(\dow^0)$, while $a_{\mu\nu}$ is ${\cal O}(\dow^1)$. The same countings would apply for the dipole-invariant versions of these quantities as well. 

At this time it is unclear which of these counting schemes is the one that organizes the s-wave hydrodynamics. We will address this question in~\cite{swavearticle}; for now we proceed agnostically in order to draw out some of the basic physical features of these fluids.

To illustrate the qualitative features of s-wave dipole superfluids, let us write down a representative set of constitutive relations consistent with the adiabaticity equation \eqref{eq:adiabaticity-swave}. We leave a more thorough analysis for future work~\cite{swavearticle}. The hydrostatic sector of the constitutive relations is characterized by the free energy density ${\cal F}$. Let us include a pressure term, $p_0(T,\mu)$, a ``kinetic energy'' term $\propto \vec{u}^2$, a term $\propto \text{tr}\,\tilde{a}$ that contributes to the dipole current one-point function, and terms $\sim \tilde{a}^2$ that, in hydrostatics, serve as kinetic terms for the Goldstone $\phi$: 
\begin{equation}
    {\cal F}
    = - p_0(T,\tmu) - \frac{1}{2}\rho \vec{u}^2
    - \frac12 p_d\, \tr\,\tilde a
    + \frac18 B_d\, \tr^2\tilde a
    + \frac14 G_d\! \lb \tilde a^2 
    - \frac1d \tr^2\tilde a \rb
    + \ldots .
\end{equation}
We emphasize that this does not characterize the most general free energy up to second-order in derivatives. Following the procedure outlined in Section \ref{sec:hs-consti}, we are led to the linearized hydrostatic constitutive relations\footnote{The previous version of this paper contained in typo in the hydrostatic energy current in eq.~\eqref{eq:ideal-free-swave}, and a resultant typo in the speed of sound in eq~\eqref{eq:swave-sound-dispersion}.}
\begin{align}
\begin{split}
  \tilde\epsilon^\mu_\hs
  &= \epsilon\, u^\mu + (p-\tilde\mu q) \vec u^\mu
  + p_d\,v^\rho\tilde F_\rho{}^\mu 
  + \tilde\mu\nabla'_\nu J^{\mu\nu}_{\text{hs}}
  + \ldots, \\
  \tilde\pi^\mu_\hs
  &= \rho\,\vec u^\mu + \ldots , \\
  \tilde\tau^{\mu\nu}_\hs
  &= 
  p\, h^{\mu\nu}
  + \ldots, \\
  n_\mu J^\mu_\hs
  &= q + \ldots, \\
  J^{\mu\nu}_\hs
  &= p_d\, h^{\mu\nu} 
  - \half B_d\,\tr\,\tilde a\, h^{\mu\nu} 
  - G_d\,\tilde a^{\langle\mu\nu\rangle}
  + \ldots, \\ 
  X_\hs 
  &= \nabla'_\mu \lb q\,\vec u^\mu\rb
  + \ldots 
  - \nabla'_\mu\nabla'_\nu \lb
  p_d\, h^{\mu\nu} 
    - \half B_d\,\tr\,\tilde a\, h^{\mu\nu} 
  - G_d\,\tilde a^{\langle\mu\nu\rangle}
  \rb
  + \ldots,
\end{split}
\label{eq:ideal-free-swave}
\end{align}
where $p=p_0 + \frac{1}{2}\rho \vec{u}^2$ and
\begin{equation}
    \df p = s\df T + q\df\tmu 
    + \half\rho\,\df\vec u^2, \qquad 
    \epsilon = Ts + \tmu q - \rho\vec u^2.
\end{equation}
Note that we have only written down constitutive relations for the density part of $J^\mu$, since its spatial part is identically fixed by the dipole Ward identity. We have also written down ``constitutive relations'' for the Goldstone equation of motion $X$ in agreement with the adiabaticity equation, as we decided not to impose it identically as a definition of $\mu$. 

In flat spacetime and in the absence of background sources, the non-linear configuration equation for the scalar Goldstone is given by
\begin{equation}
    \dow_i\!\lb q\,u^i\rb
    =
    \dow_i\dow^i p_d
    - \dow_i\dow_j\!\lb 
    \lb B_d - \frac{2}{d}G_d \rb \delta^{ij}
    \dow_k\dow^k \phi
    + 2G_d \dow^i\dow^j \phi \rb
    + \ldots.
\end{equation}
This is solved by the generic hydrostatic configuration of the scalar Goldstone
\begin{equation}
    \langle \phi\rangle 
    = \phi_0 - \frac{1}{q_0} \pi^0_i x^i 
    - \half\xi_{ij} x^i x^j,
    \label{eq:swave-equilibrium-phi}
\end{equation}
which give rise to the same configurations for $\langle\phi_i\rangle$ as \eqref{eq:classical-config}. Note that, in contrast to p-waves, there is no equation in hydrostatic equilibrium enforcing a derivative suppression of the fluid velocity $u^i$. 

Similarly, we can also write down non-hydrostatic corrections to the constitutive relations following our discussion in Section \ref{sec:nhs-consti}. A representative but non-generic subset of these are
\begin{equation}
\begin{split}
    \tilde\epsilon^\mu_\nhs 
    &= T^2\kappa\, h^{\mu\nu}\delta_\scB n_\nu
    + \ldots, \\ 
    \tilde\tau^{\mu\nu}_\nhs 
    &= - T 
    \lb \zeta\, h^{\mu\nu}h^{\rho\sigma}
    + 2\eta\,h^{\mu\langle\rho}h^{\sigma\rangle\nu}
    \rb \delta_\scB h_{\rho\sigma}
    + \ldots, \\ 
    X_\nhs 
    &= - T\sigma_\phi\delta_\scB\phi
    + \ldots.
\end{split}
\end{equation}

\vspace{1em}
\noindent 
\textbf{Mode spectrum and order-mixing.}---We can use the hydrodynamic equations outlined above in the absence of background sources to obtain the linearized mode spectrum of s-wave dipole superfluids. In contrast to the p-wave phase, we  find a \emph{normal} sound mode in the longitudinal sector with linear dispersion relations
\begin{equation}
\begin{split}
    \omega 
    &= \pm 
    \sqrt{\frac{s^2\chi_{qq}}{\Xi \rho}} k
    + {\cal O}(k^2).
\end{split}
\label{eq:swave-sound-dispersion}
\end{equation}
 In the transverse sector, we similarly find the normal shear diffusion mode
\begin{equation}
    \omega = - i \frac{\eta}{\rho} k^2 + {\cal O}(k^3).
    \label{eq:swave-transverse-dispersion}
\end{equation}
In particular, we see that the shear mode is not subdiffusive for s-wave dipole superfluids. The above modes suggest the alternative derivative counting scheme we mentioned earlier with $z=1$. Note that both the normal modes above crucially depend on the kinetic mass density $\rho$. 

In addition to the modes above, we also find a pair of magnon-like second-sound modes in the longitudinal sector
\begin{align}
    \omega
    &= \pm 
    \sqrt{\frac{B_d + 2\frac{d-1}{d}G_d}
    {\chi_{qq}}} k^2
    + {\cal O}(k^4).
    \label{eq:swave-secondsound}
\end{align}
These are reminiscent of the second sound mode in ordinary U(1) superfluids, but has now been suppressed in $k$ due to the presence of dipole symmetry. We have gone beyond what is presented here and included many more derivative corrections to the constitutive relations. While we do not have a conclusive proof at this time, we comment that so far we have only found further corrections to this mode at ${\cal O}(k^4)$. In particular, in contrast to the magnon-like mode observed in the p-wave phase, the attenuation correction to this mode goes as $\omega \sim -ik^4$ and is purely subdiffusive. In this sense, this mode is ``purely fractonic'' in origin, and has no qualitative contamination from the non-fractonic hydrodynamic degrees of freedom.

The mode spectrum presented here is consistent with two derivative counting schemes. One is the $z=1$ scheme we suggested earlier in this Section, where $\omega \sim k$, so that the ``kinetic energy density'' $\rho$ is $\mathcal{O}(\partial^0)$, the viscosity $\eta$ is $\mathcal{O}(\partial^1)$, but the coefficients $B_d$ and $G_d$ are $\mathcal{O}(\partial^2)$. However, this scheme is in some tension with a more general hydrostatic analysis, which allows for equilibrium states~\eqref{eq:swave-equilibrium-phi} with superflow with $\phi \sim x^2$. 

The other scheme is that for the $p$-wave phase together with $\phi \sim \mathcal{O}(\partial^{-2})$. In this counting $\omega \sim k^2$, $B_d$ and $G_d$ are $\mathcal{O}(\partial^0)$, but $\eta$ and $\rho$ are $\mathcal{O}(\partial^2)$. While this counting is seemingly at odds with the existence of a sound mode, we note that (1) it is consistent with the existence of equilibrium states with superflow and (2) further is consistent with the linearized spectrum. However, with this counting the kinetic energy density $\rho$ is analogous to a dangerously irrelevant operator in quantum field theory, appearing in the denominator of the dispersion $\omega=\omega(k)$ and thereby changing the scalings of infrared dispersion relations.

Both of these countings are consistent with the linearized mode spectrum in the absence of superflow. We expect that only one is consistent in the presence of interactions, or equivalently in a superflow, and will report on the matter soon~\cite{swavearticle}. 

%-----------------------------------------------
\section{Discussion}
\label{S:discussion}
%-----------------------------------------------

In this work, we have systematically constructed the hydrodynamic description of translationally- and rotationally-invariant fluids with a spontaneously broken dipole symmetry. We have explained our algorithm in great detail already in the Introduction and Section~\ref{S:summary}. In this Discussion, we wrap up by drawing attention to a few interesting aspects of our analysis and point out a few natural questions and future directions suggested by our work.

First, thanks to the unconventional derivative counting appropriate to fluids with dipole symmetry, dipole superfluids have the unique feature that there is dissipative transport already at the leading order in derivatives. The thermal conductivity $\kappa$ is formally as important in the hydrodynamic equations as the ``perfect fluid'' part of the constitutive relations. This leading order dissipative transport is in the vector sector. There is dissipative transport in the scalar and tensor sectors as well, although these appear exclusively through derivative corrections to the leading order constitutive relations. In our analysis, we have classified all 47 transport coefficients consistent with the Second Law of thermodynamics up to first subleading derivative order. In doing so, we are able to deduce the full suite of inequality-type constraints on the transport coefficients enforced by the positivity of entropy production. As in relativistic hydrodynamics, we expect that out-of-equilibrium transport with even more derivatives is automatically consistent with the Second Law once the constraints presented here are satisfied.

In order to implement the Second Law constraints up to subleading order in derivatives, we built upon some existing techniques in the literature. We appropriately generalized the ``adiabaticity equation'' and the off-shell formulation of the Second Law of thermodynamics from~\cite{Loganayagam:2011mu} to fluids with dipole symmetry. We illustrated that the constitutive relations can be consistently classified into a hydrostatic part, obtained by variations of a hydrostatic effective action, and a non-hydrostatic part. In particular, using redefinitions of the hydrodynamic variables, we argued that it is possible to choose the non-hydrostatic corrections to only appear in $\epsilon^i$, $\tau^{ij}$, and $J^{ij}$. In doing so, we landed on a relatively simple parametrization of out-of-equilibrium transport given in~\eqref{eq:nonhs_subleading1} and showed that in this form it is straightforward to impose the positivity of entropy production.

Given our results, there are some natural directions to follow upon. A particularly interesting point to note is that hydrostatic equilibrium allows for the dipole flux $J^{ij}$ to have a nontrivial one-point function $\langle J^{ij} \rangle = p_d \delta^{ij}$ in the absence of dipole superflow. Indeed, the large $N$ models of~\cite{Jensen:2022iww} feature such a thermal one-point function. However, we expect that this is an indication that the thermodynamic state of the fluid is unstable and the system would prefer to transition to a state with nonzero superflow.

Furthermore, we have constructed a hydrodynamic description of p-wave dipole superfluids at the level of linearized constitutive relations. Our methods can be generalized to produce the full theory of nonlinear dipole superfluid dynamics, at the expense of keeping track of a dimension-dependent number of zeroth order scalars constructed from the zeroth order tensors $\tilde a_{\mu\nu}$, $\tilde F_{\mu\nu}$, and $F^n_{\mu\nu}$. Constructing such nonlinear hydrodynamics has the useful corollary of allowing us to describe transport in a state with superflow characterized by $\langle \partial_i \phi_j\rangle \neq 0$. These questions will be explored in more detail in~\cite{EricDipoleSuperflow}.

As another immediate follow-up to this work, we intend to construct a hydrodynamic description of dissipative s-wave dipole superfluids in an imminent publication~\cite{swavearticle}. The initial steps towards this construction were outlined in Section \ref{S:swave}. 
In particular, we found that the linearized spectrum and natural derivative-counting in this phase is quite different from its p-wave counterpart. It will be valuable to systematically revisit the constitutive relations of s-wave dipole superfluids and inspect the constraints imposed by the second law of thermodynamics.

We noted in our introduction that the dipole Goldstone field $\phi_i$ is UV-sensitive, in the sense that it is compact with a periodicity $\sim 2\pi/a$, with $a$ being the lattice spacing~\cite{Jensen:2012jh}. The U(1) density and flux $J^t$, $J^i$, the dipole flux $J^{ij}$, and the energy density and flux $\epsilon^t$, $\epsilon^i$ (in flat space) only depend on the derivatives of $\phi_i$, and are therefore UV-insensitive. On the other hand, the momentum density $\pi_i$ and the spatial stress tensor $\tau^{ij}$ depend directly on $\phi_i$, making these UV-sensitive. It would be interesting to explore what consequences does this sensitivity imply for the two-point functions of $\pi_i$ and $\tau^{ij}$, accounting for one-loop fluctuations of the dipole Goldstone.

In two spatial dimensions, the low-energy limit of elasticity theory is expected to be described by a dipole-symmetric phase, with disclination defects in the crystal playing the role of elementary charges~\cite{Pretko:2017kvd}; see also~\cite{Gromov:2017vir}. As such, we also expect our hydrodynamic description to describe a phase of liquid crystals with dislocations, seen as condensed disclination-antidisclination pairs. It would be interesting to see how this works out in detail and whether the low-energy description of these physical systems can indeed be reinterpreted as dipole-invariant field theories. More generally, it would be interesting to explore the relations between our results and transport in elastic media.

In nature, we expect any dipole symmetry to be emergent at low-energies, broken by irrelevant operators in the IR description. In a regime where the dipole symmetry is approximately preserved, such systems should still possess a hydrodynamic description at small temperatures, but with the explicit breaking of the dipole symmetry generating a small mass for the dipole Goldstone. It would interesting to investigate whether a hydrodynamic theory of a fluid with a light dipole Goldstone along the lines of~\cite{Armas:2021vku, Armas:2023tyx} can be turned into a diagnostic of these phases.

Lastly, in this work we have only considered the hydrodynamic description of systems with dipole symmetry. In the context of fractons, dipole symmetry is only a toy model for subsystem symmetries realized in lattice models with fractonic excitations. It would be interesting to explore whether subsystem-invariant theories also have phases which admit a hydrodynamic description like those considered in this work, and if so, how do these qualitatively differ from the ones with dipole symmetry.

%-----------------------------------------------
\subsection*{Acknowledgements}
%-----------------------------------------------

We are grateful to J.~Armas, P.~Glorioso, A.~Karch, P.~Kovtun, H.~T.~Lam, A.~Lucas, A.~Raz, and A.~Ritz for helpful discussions. We would especially like to thank J.~Armas for sharing a draft of~\cite{idealFractonHydro} prior to publication. The work of KJ, RL, and EM~was supported in part by the NSERC Discovery Grant program of Canada.
AJ is funded by the European Union’s Horizon 2020 research and innovation programme under the Marie Skłodowska-Curie grant agreement NonEqbSK No. 101027527. AJ is also partly supported by the Netherlands Organization for Scientific Research (NWO) and by the Dutch Institute for Emergent Phenomena (DIEP) cluster at the University of Amsterdam.

%-----------------------------------------------
\appendix
%-----------------------------------------------

%-----------------------------------------------
\section{Hydrostatic scalars}
\label{app:HS-scalars}
%-----------------------------------------------

In this appendix, we will enlist all hydrostatic scalars up to second order in derivatives that can appear in the free energy density ${\cal F}$ and characterize the hydrostatic constitutive relations. In particular, we will only enlist scalars that are at most quadratic in fluctuations. With these considerations, we find a total of seven hydrostatic scalars at zero-derivative order 
\begin{align}\begin{split}
    \mathcal{S}^{(0)} = 
    \left\{~~
    T, \quad \tmu, \quad 
    \half \tr\,\tilde{a}, \quad 
    \frac14 \tilde{a}^2, \quad 
    \frac14 \tilde{F}^2, \quad 
    \frac14 F_n^2,\quad 
    \half \tilde{F}\cdot F_n~~ \right\}.
\end{split}\end{align}
All transport coefficients in the theory can be seen as arbitrary functions of these zero-derivative scalars. The first three of these are linear in fluctuations, while the remaining four are quadratic. For a fully non-linear theory, there are other zero-derivative hydrostatic scalars one can construct combining chains of zero-derivative tensors, e.g. $\tilde F^{\mu\nu}\tilde F_{\nu\rho}\tilde F^{\rho\sigma}\tilde F_{\sigma\mu}$. 

Furthermore, for a parity-violating theory in even spatial dimensions, we can additionally construct zero derivative scalars as exterior products of $\tilde F_{\mu\nu}$ and $F^n_{\mu\nu}$, such as
\begin{align}
    \varepsilon^{\mu_1\mu_2\mu_2...\mu_{d}}
    \tilde{F}_{\mu_1\mu_2}\tilde{F}_{\mu_{3}\mu_4}
    \ldots 
    \tilde{F}_{\mu_{d-1}\mu_{d}},\qquad \ldots
\end{align}
where $\varepsilon^{\mu_1\mu_2\mu_2...\mu_{d}}$ is the spatial Levi-Civita symbol satisfying $n_{\mu_1} \varepsilon^{\mu_1\mu_2\mu_2...\mu_{d}} = 0$. This term is quadratic in fluctuations in $d=4$ and linear in $d=2$.
In odd spatial dimensions, we instead have one-derivative terms such as
\begin{align}
    \varepsilon^{\mu_1...\mu_{d}}
    \vec u_{\mu_1}
    \tilde{F}_{\mu_2\mu_3}...\tilde{F}_{\mu_{d-1}\mu_{d}}, \qquad 
    \varepsilon^{\mu_1...\mu_{d}}
    \dow_{\mu_1} {\cal S}_i^{(0)}
    \tilde{F}_{\mu_2\mu_3}...\tilde{F}_{\mu_{d-1}\mu_{d}}, \qquad\ldots
\end{align}
These are at least quadratic in fluctuations in $d=3$ and linear in $d=1$. We are focusing on parity-preserving fluids in this work, so we have not considered these parity-violating terms in the main text. Furthermore, due to our derivative counting scheme, there are no hydrostatic scalars at one-derivative order for a parity-preserving theory. So, the first corrections to the ideal order free energy appears at second order in derivatives.

We will now perform the counting of parity-preserving hydrostatic scalars at second order in derivatives and at most quadratic in fluctuations. First, we have one hydrostatic scalar that does not involve explicit derivatives
\begin{equation}
    \mathcal{S}^{(2)}_1\equiv 
    \half \vec u^2.
\end{equation}
Next, let us look at the hydrostatic scalars involving one explicit derivative. Hydrostatic identities $\delta_\scB(\ldots) = 0$ require that $u^\mu\dow_\mu{\cal S}^{(0)} = 0$. In addition to these, we have one more scalar hydrostatic identity relevant at second order in derivatives, $\nabla'_\mu\beta^\mu = 0$. We have 3 hydrostatic scalars of the kind $\vec u^\mu \dow_\mu S^{(0)}$, i.e.
\begin{equation}
    \mathcal{S}^{(2)}_2 \equiv \vec u^\mu \dow_\mu T, \qquad 
    \mathcal{S}^{(2)}_3 \equiv \vec u^\mu \dow_\mu\tmu, \qquad 
    \mathcal{S}^{(2)}_4 \equiv
    \half \vec u^\mu \dow_\mu\tr\,\tilde a.
\end{equation}
We have not counted scalars like $\bar u^\mu \dow_\mu\tilde a^2$ etc., because these are cubic in fluctuations. In principle, we also have a scalar like $\nabla'_\mu \vec{u}^\mu$. However, note that within an integral, for some transport coefficient $\lambda$, we find using integration by parts that
\begin{equation}
    \int \df^d x\sqrt{\gamma}\,\lambda\, \nabla'_\mu \vec u^\mu 
    = - \int \df^d x\sqrt{\gamma}\,\vec u^\mu \dow_\mu \lambda 
    = - \int \df^d x\sqrt{\gamma}\,
    \sum_I\frac{\dow\lambda}{\dow{\cal S}^{(0)}_I}
    \vec u^\mu \dow_\mu{\cal S}^{(0)}_I.
    \label{eq:der-argument}
\end{equation}
Therefore, it is not an independent hydrostatic scalar for the purposes of the hydrostatic partition function. We will use similar arguments to kill other ``total-derivative'' scalars in our hydrostatic counting. Similarly, we have scalars like $\vec u_\mu \nabla'_\nu S^{\mu\nu}$, where $S^{\mu\nu}$ is a placeholder for zero-derivative tensors $\tilde a^{\mu\nu}$, $\tilde F^{\mu\nu}$, and $F_n^{\mu\nu}$. Using the same procedure as above, these can be replaced with $S^{\mu\nu}\nabla_\mu \vec u_\nu$, resulting in 3 scalars
\begin{align}
    \mathcal{S}^{(2)}_{5}
    \equiv \tilde{a}^{\mu\nu}\nabla_\mu \vec u_\nu, \qquad
    \mathcal{S}^{(2)}_{6}
    \equiv \tilde{F}^{\mu\nu}\nabla_\mu \vec u_\nu, \qquad
    \mathcal{S}^{(2)}_{7} 
    \equiv F_n^{\mu\nu}\nabla_\mu \vec u_\nu. 
\end{align}

Next, let us consider hydrostatic scalars with two explicit derivatives. There are scalars of the kind $\nabla_\mu{\cal S}^{(0)}\nabla^\mu {\cal S}^{(0)}$, where we have defined $\nabla^\mu = h^{\mu\nu}\nabla_\nu$. Up to quadratic order in fluctuations, this yields 6 possibilities 
\begin{equation}
\begin{gathered}
    \mathcal{S}^{(2)}_{8} 
    = \half \nabla_\mu T\nabla^\mu T, \qquad 
    \mathcal{S}^{(2)}_{9}
    = \nabla_\mu T\nabla^\mu \tmu, \qquad 
    \mathcal{S}^{(2)}_{10} 
    = \half\nabla_\mu T\nabla^\mu \tr\,\tilde a, \\
    \mathcal{S}^{(2)}_{11} = \half\nabla_\mu\tmu\nabla^\mu\tmu, \qquad 
    \mathcal{S}^{(2)}_{12} = \half \nabla_\mu\tmu\nabla^\mu\tr\,\tilde a, \qquad 
    \mathcal{S}^{(2)}_{13} 
    = \frac18 \nabla_\mu\tr\,\tilde a\,\nabla^\mu \tr\,\tilde a.
\end{gathered}
\end{equation}
Using the integration by parts argument outlined above, we can skip the total-derivative hydrostatic scalars of the kind $h^{\mu\nu}\nabla_\mu\nabla_\nu{\cal S}^{(0)}$.  Likewise, we can skip scalars of the kind $\nabla_\mu\nabla_\nu S^{\mu\nu}$ in favor of $\nabla_\mu S^{\mu\nu}\nabla_\nu {\cal S}^{(0)}$, which in turn can be skipped in favor of $S^{\mu\nu}\nabla_\mu\nabla_\nu {\cal S}^{(0)}$. This results in three more hydrostatic scalars
\begin{align}
    \mathcal{S}^{(2)}_{14}
    \equiv
    \half \tilde{a}^{\mu\nu}\nabla_\mu\nabla_\nu T, \qquad
    \mathcal{S}^{(2)}_{15} \equiv \half \tilde{a}^{\mu\nu}\nabla_\mu\nabla_\nu\tmu, \qquad
    \mathcal{S}^{(2)}_{16} \equiv
    \frac14 \tilde{a}^{\mu\nu}\nabla_\mu\nabla_\nu \tr\,\tilde{a}.
\end{align}
Note that terms like $\tilde F^{\mu\nu}\nabla_\mu\nabla_\nu{\cal S}^{(0)} = \half \tilde F^{\mu\nu} F^n_{\mu\nu} v^\lambda\dow_\lambda {\cal S}^{(0)} = -\half \tilde F^{\mu\nu} F^n_{\mu\nu} \vec u^\lambda\dow_\lambda {\cal S}^{(0)}$ are quartic in fluctuations and hence are not important for our purposes. In addition to these, there are 4 hydrostatic scalars of the kind $\nabla_\mu S_{\nu\rho}\nabla^\mu S^{\nu\rho}$, i.e.
\begin{equation}
\begin{gathered}
    \mathcal{S}^{(2)}_{17} 
    \equiv
    \half \nabla_\mu\tilde{a}_{\nu\rho}
    \nabla^\mu\tilde{a}^{\nu\rho}, \qquad 
    \mathcal{S}^{(2)}_{18} 
    \equiv
    \half\nabla_\mu\tilde{F}_{\nu\rho}
    \nabla^\mu\tilde{F}^{\nu\rho}, \qquad 
    \mathcal{S}^{(2)}_{19} 
    \equiv 
    \half \nabla_\mu F^n_{\nu\rho}\nabla^\mu F_n^{\nu\rho}, \\ 
    \mathcal{S}_{20}^{(2)}\equiv
    \nabla_\mu F^n_{\nu\rho}\nabla^\mu \tilde{F}^{\nu\rho},
\end{gathered}
\end{equation}
and 6 of the kind $\nabla_\mu S_{\nu\rho}\nabla^\nu S^{\mu\rho}$, i.e.
\begin{align}\begin{gathered}
    \mathcal{S}^{(2)}_{21} 
    \equiv
    \half  \nabla_\mu
    \tilde{a}_{\nu\rho}\nabla^\nu
    \tilde{a}^{\mu\rho}, \qquad
    \mathcal{S}^{(2)}_{22} 
    \equiv \half\nabla_\mu\tilde{F}_{\nu\rho}
    \nabla^\nu\tilde{F}^{\mu\rho}, \qquad
    \mathcal{S}_{23}^{(2)}
    \equiv 
    \half\nabla_\mu F^n_{\nu\rho} \nabla^\nu F_n^{\mu\rho}, \\
    \mathcal{S}^{(2)}_{24} \equiv\nabla_\mu\tilde{a}_{\nu\rho}\nabla^\nu\tilde{F}^{\mu\rho},
    \qquad
    \mathcal{S}_{25}^{(2)} 
    \equiv \nabla_\mu \tilde{a}_{\nu\rho}\nabla^\nu F_n^{\mu\rho}, \qquad 
    \mathcal{S}_{26}^{(2)}\equiv\nabla_\mu F^n_{\nu\rho}\nabla^\nu \tilde{F}^{\mu\rho}.
\end{gathered}\end{align}
We have skipped scalars of the kind $S_{\mu\nu}\nabla_\rho\nabla^\rho S^{\mu\nu}$, $S_{\mu\nu}\nabla_\rho\nabla^\mu S^{\rho\nu}$, and $\nabla^\rho S_{\rho\nu}\nabla_\mu S^{\mu\nu}$ using integration by parts. Finally, there are two scalars constructed out of the Riemann tensor
\begin{align}\begin{split} 
    \mathcal{S}^{(2)}_{27} 
    \equiv R^{\lambda}_{\;\;\mu\lambda\nu}h^{\mu\nu}, \qquad 
    \mathcal{S}^{(2)}_{28}
    \equiv R^{\lambda}_{\;\;\mu\lambda\nu}\tilde a^{\mu\nu}.
\end{split}\end{align}

This exhausts the list of potential hydrostatic scalars. In summary, at second order in derivatives, there are 28 hydrostatic scalars which must be added to the ideal order pressure to capture linearized hydrodynamics. Interestingly, with the exception of the Ricci scalar $\mathcal{S}^{(2)}_{27}$, all other two-derivative hydrostatic scalars in our counting are quadratic in fluctuations. This means that we can take the respective transport coefficients to be constants for our purposes.

%-----------------------------------------------
\section{Computation of response functions}
\label{app:responsefunctions}
%-----------------------------------------------

In this appendix, we will outline an analytic path to computing hydrodynamic response functions of our model. We will break this problem into two steps. We recall that the conservation equations of the system can be written in terms of dipole-invariant objects in \eqref{eq:dipole-invariant-Wardidentity}. Therefore, we will first use the dipole-invariant energy, momentum, and charge conservation equations in \eqref{eq:dipole-invariant-Wardidentity} to determine $\delta T$, $\delta u^i$, and $\delta\tmu$ in terms of the spacetime background fields $\delta n_t$, $\delta n_i$, $\delta g_{ij}$, and $\delta v^i$, and the dipole-invariant versions of the gauge fields $\delta\tilde A_t$, $\delta \tilde A_i$, and $\delta \tilde a_{ij}$. Pretending for the moment that $\delta\tilde A_t$, $\delta \tilde A_i$, and $\delta \tilde a_{ij}$ are background sources, we can use these solutions to compute the response functions ${\tilde G}^R_{\tilde O\tilde O}$ between dipole-invariant objects through
\begin{equation}
    \begin{split}
        \delta \tilde s
        &= \begin{pmatrix}
            {- \delta n_t} \\
            {- \delta n_i} \\
            {- \delta v^i} \\
            \half \delta g_{ij} \\
            \delta \tilde A_t \\
            \delta \tilde A_i \\ 
            \half \delta \tilde a_{ij}
        \end{pmatrix}, \qquad
        \langle {\tilde O} \rangle_s
        = \sqrt{\gamma}\,
        \begin{pmatrix}
            \tilde\epsilon^t + \tilde\pi_k v^k \\
            \tilde\epsilon^i - J^{ik} \tilde F_{tk} \\
            \tilde\pi_i - \tilde\tau_{ik}n^k \\
            \tilde\tau^{ij} - 2\tilde A^{(i}_{~k}J^{j)k} \\
            J^t \\
            J^i \\
            J^{ij}
        \end{pmatrix}.
    \end{split}
    \label{eq:invariant-operators}
\end{equation}
The answers thus obtained will, of course, not be the physical response functions because we still need to solve for $\delta\phi_i$. Nonetheless, the dipole-invariant response functions are more analytically tractable and can be used as building blocks to construct the full physical response functions. The second part of the problem will be to solve the dipole conservation equation in \eqref{eq:dipole-invariant-Wardidentity} for $\delta\phi_i$ in terms of the physical background fields. These can be used to convert the dipole-invariant response functions into their physical counterparts and also obtain the hydrodynamic mode spectrum.

Let us imagine that we have obtained the dipole-invariant response functions $\tilde G^R_{\tilde O\tilde O}$. We will like to relate this to the physical response functions that will actually be observable in an experiment. Firstly, let us note the transformation between dipole-invariant and physical background fields
\begin{equation}
\begin{split}
    \delta\tilde A_{t}
    &= \delta A_t - \frac{1}{q}\pi_i^0 \delta v^i, \\
    \delta\tilde A_{i}
    &= \delta A_i + \delta\phi_i, \\
    \delta\tilde a_{ij}
    &= \delta a_{ij} + 2ik_{(i} \delta\phi_{j)}
    - \frac{i}{q} \pi^k_0
    \lb 2k_{(i}\delta g_{j)k} - k_k \delta g_{ij} \rb.
    \label{eq:tildeA-var}
\end{split}
\end{equation}
Note that the equilibrium value of $\tilde A_i$ is $\pi_i^0/q$, while that of $\tilde A_t$ and $\tilde a_{ij}$ is zero. Using these together with \eqref{eq:physical-operators} and \eqref{eq:invariant-operators}, we can obtain the dictionary between physical and invariant operators up to linear order in fluctuations, i.e.
\begin{equation}
\begin{split}
    \langle\epsilon^t\rangle_s
    &= \langle{\tilde{\epsilon}}^t\rangle_s 
    + \pi_i^0 \delta v^i, \\
    \langle\epsilon^i\rangle_s
    &= \langle{\tilde{\epsilon}}^i\rangle_s 
    - i\omega p_d \delta^{ij}
    \delta\langle\phi_j\rangle_s \nn\\
    &\qquad 
    - \frac{p_d}{q} \lb ik_m\pi^m_0 \delta^i_k
    - ik_k\pi_0^i \rb \delta v^k
    + i\omega \frac{p_d}{q} \lb 2\delta^{i(k}\pi_0^{l)}
    - \delta^{kl} \pi^i_0 \rb \half \delta g_{kl}
    , \nn
\end{split}
\end{equation}
\begin{equation}
\begin{split}
    \langle\pi_i\rangle_s
    &= \langle\tilde{\pi}_i\rangle_s 
    + \frac1q \pi_i^0 \langle J^t \rangle_s 
    + q\,\delta\langle\phi_i\rangle_s \nn\\
    &\qquad 
    + \pi_i^0 \delta n_t 
    + \frac{p_d}{q} \lb 
    i k_m \pi_0^m \delta_i^k 
    - ik_i \pi_0^k\rb \delta n_k
    , \\
    \langle{\cal\tau}^{ij}\rangle_s 
    &= \langle{\tilde{\tau}}^{ij}\rangle_s 
    + \frac{2}{q} \pi^{(i}_0 ik_k \langle J^{j)k} \rangle
    - \frac1q \pi_0^k ik_k \langle J^{ij} \rangle
    + p_d \lb 2i k^{(i} \delta^{j)k} 
    - \delta^{ij} ik^k \rb \langle\delta\phi_k \rangle_s \nn\\ 
    &\qquad 
    + \frac{p_d}{q}  \lb 2\pi^{(i}_0 \delta^{j)k}
    - \delta^{ij} \pi_0^k \rb i\omega \delta n_k \nn\\
    &\qquad 
    + \frac{p_d}{q} \lb 
        4\pi^{(i}_0\delta^{j)(k}ik^{l)}
    - 4ik^{(i}\delta^{j)(k} \pi_0^{l)}
    - 2ik^{(i} \pi^{j)}_0 \delta^{kl}
    + 2\delta^{ij} i k^{(k} \pi_0^{l)}
    \rb \half \delta g_{kl}.
\end{split}
\end{equation}
These transformation rules can be summarized as
\begin{equation}
\begin{split}
    {\cal R}^\dagger \cdot \delta s
    &= \delta \tilde s
    - {\cal X}^i \langle\delta\phi_i\rangle_s, \\
    \langle O\rangle_s 
    &= 
    {\cal R}\cdot\langle\tilde O\rangle_s
    + \cC^i\delta\langle\phi_i\rangle_s
    - {\cal Y}\cdot \delta s.
    \label{eq:compact-trans}
\end{split}
\end{equation}
Here ``$\dagger$'' denotes the transpose conjugate operation and we have identified the transformation coefficients as
\begin{align}
    {\cal R}
    &= 
    \begin{pmatrix}
        {\mathbb 1}_{4\times 4} & \rvline &
        \begin{matrix}
        0 & 0 & 0 \\ 
        0 & 0 & 0 \\
        1/q\, \pi_i^0 & 0 & 0 \\
        0 & 0 & 
        1/q\lb 2\pi^{(i}_0 \delta^{j)}_{(k} ik_{l)}
        - ik_m\pi_0^m \delta^i_{(k}\delta^j_{l)} \rb 
        \end{matrix} \\
        \hline 
        {\mathbb 0}_{3\times 4} & \rvline & {\mathbb 1}_{3\times 3}
    \end{pmatrix}, \nn
\end{align}
\begin{align}
    \cC
    &= 
    \begin{pmatrix}
        \begin{matrix}
        0 & 0 & \pi_k^0 & 0 \\ 
        0 & 0 & - p_d/q
        \begin{pmatrix}
            ik_m\pi^m_0 \delta^i_k \\
            - \pi_0^i ik_k  
        \end{pmatrix}
        & -i\omega p_d/q
        \begin{pmatrix}
            2\delta^{i(k}\pi_0^{l)} \\
            - \delta^{kl} \pi^i_0
        \end{pmatrix} \\ 
        \pi_i^0 & 
        p_d/q 
        \begin{pmatrix}
            ik_m \pi_0^m \delta_i^k \\
            - ik_i \pi_0^k
        \end{pmatrix}
        & 0 & 0 \\ 
        0  & i\omega p_d/q
        \begin{pmatrix}
            2\pi^{(i}_0 \delta^{j)k} \\ 
            - \delta^{ij} \pi_0^k    
        \end{pmatrix}
        & 0 
        & p_d/q
        \begin{pmatrix}
            4ik^{(i}\delta^{j)(k} \pi_0^{l)}
            - 4\pi^{(i}_0\delta^{j)(k}ik^{l)} \\
            - 2\delta^{ij} i k^{(k} \pi_0^{l)}    
            + 2ik^{(i} \pi^{j)}_0 \delta^{kl}
        \end{pmatrix}
        \end{matrix} & \rvline 
        & {\mathbb 0}_{4\times 3} \\
        \hline
        {\mathbb 0}_{3\times 4}
        & \rvline & {\mathbb 0}_{3\times 3}
    \end{pmatrix}, \nn
\end{align}
\begin{align}
    {\cal Y}^m
    &= 
    \begin{pmatrix}
        0 \\
        {-i\omega p_d\delta^{im}} \\
        {q\delta^m_i} \\
        {ip_d} \lb 2\delta^{m(i}k^{j)}
        - \delta^{ij}k^m \rb \\
        \hline
        {\mathbb 0}_{3\times 1}
    \end{pmatrix}, \qquad 
    {\cal X}^m
    = 
    \begin{pmatrix}
        {\mathbb 0}_{4\times 1} \\ 
        \hline 
        0 \\ 
        \delta^{m}_i \\
        ik_{(i}\delta_{j)}^m
    \end{pmatrix}.
\end{align}
The free row indices are denoted by $i,j$ while the free column indices by $k,l$. Note that ${\cal C}^\dagger = {\cal C}$. Note also that for an equilibrium state with zero momentum density, i.e. $\pi^i_0 = 0$, these simplify to ${\cal R} = {\mathbb 1}$ and ${\cal C} = 0$, while ${\cal X}^i$ and ${\cal Y}^i$ are independent of $\pi^i_0$.

Using \eqref{eq:compact-trans} and noting that $\langle\tilde O \rangle_s$ only depends on $\delta \tilde s$, we can write down the relation between physical and dipole-invariant response functions
\begin{equation}
    G^R_{OO}
    = {\cal R} \cdot 
    \tilde G^R_{\tilde O\tilde O} \cdot 
    {\cal R}^\dagger
    + {\cal R}\cdot \lb 
    \tilde G^R_{\tilde O\tilde O} \cdot
    {\cal X}^i - {\cal Y}^i \rb \otimes 
    \frac{\delta\langle\phi_i\rangle_s}{\delta s}
    + {\cal C}.
    \label{eq:G-trans-partial}
\end{equation}
All we need now to complete the mapping are the solutions for $\langle\phi_i\rangle$ using the dipole conservation Ward identity. At linearized order in momentum space, it is given as
\begin{equation}
\begin{split}
    \langle J^i\rangle_s
    - ik_j \langle J^{ij} \rangle_s
    &=
    q\, \delta v^i
    + i\omega p_d \delta^{ik} \delta n_k
    + p_d \lb \delta^{i(k} ik^{l)} 
    - \delta^{kl} ik^i
    \rb \delta g_{kl} \nn\\
    \implies 
    {\cal X}^{\dagger i} \cdot \langle {\tilde O} \rangle_s
    &=
    - {\cal Y}^{\dagger i}\cdot \delta\tilde s.
\end{split}
\end{equation}
Since both $J^{ij}$ and $J^i$ are dipole-invariant operators, we can express this in terms of dipole-invariant response functions as
\begin{equation}
\begin{split}
    \lb {\cal X}^{\dagger i} \cdot \tilde G^R_{\tilde O\tilde O}
    - {\cal Y}^{\dagger i} \rb \cdot 
    \delta \tilde s
    &= 0.
\end{split}
\label{eq:EOM-W}
\end{equation}
Using \eqref{eq:compact-trans} into \eqref{eq:EOM-W}, we can read out the solution for $\langle\phi_i\rangle_s$ in terms of the background fields as
\begin{equation}
\begin{split}
    \langle \delta\phi_i \rangle_s
    = - M_{ij}\lb {\cal X}^{\dagger j} \cdot 
    \tilde G^R_{\tilde O\tilde O}
    - {\cal Y}^{\dagger j} \rb \cdot {\cal R}^\dagger \cdot \delta s,
\end{split}
\label{eq:phi-sol-W}
\end{equation}
where the coefficient matrix $M_{ij}$ is defined as the inverse of
\begin{equation}
\begin{split}
    (M^{-1})^{ij}
    &= {\cal X}^{\dagger i} \cdot
    \tilde G^R_{\tilde O\tilde O} 
    \cdot {\cal X}^j.
\end{split}
\end{equation}
Plugging this back into \eqref{eq:G-trans-partial}, we find
\begin{equation}
    G^R_{OO}
    = {\cal R}\cdot \bigg[ 
    \tilde G^R_{\tilde O\tilde O} 
    - \lb \tilde G^R_{\tilde O\tilde O} \cdot
    {\cal X}^i - {\cal Y}^i \rb 
     M_{ij}\lb {\cal X}'^{\dagger j} \cdot
    \tilde G^R_{\tilde O\tilde O}
    - {\cal Y}^{\dagger j} \rb \bigg] \cdot {\cal R}^\dagger
    + {\cal C}.
    \label{eq:G-trans}
\end{equation}
Note that the terms inside the square brackets are independent of $\pi^i_0$, which are nothing but the physical correlation functions at zero momentum density. The physical correlation functions at nonzero momentum density can be obtained by performing a global dipole transformation and eventually result in an operator rotation ${\cal R}$ and a contact term ${\cal C}$. 

%-----------------------------------------------
\subsection*{Onsager relations}
%-----------------------------------------------

When the underlying microscopic description has some discrete time-reversal symmetry, the dipole-invariant response functions are also required to satisfy the same Onsager relations as the physical ones in \eqref{eq:onsager}, i.e.
\begin{equation}
    \Big[\tilde G^{R}_{\tilde O\tilde O}(\omega,\vec k)\Big]^\rmT 
    = \Theta \cdot
    G^{R}_{\tilde O\tilde O}(\omega,-\vec k)
    \cdot \Theta,
    \label{eq:onsager-invariant}
\end{equation}
together with the Onsager constraints for the coefficient matrices
\begin{equation}
\begin{gathered}
    \Theta\cdot{\cal X}^\dagger\big|_{k\to -k} = -{\cal X}, \qquad
    \Theta\cdot{\cal Y}^\dagger\big|_{k\to -k} = -{\cal Y}, \\ 
    \Theta\cdot{\cal R}^\dagger\cdot\Theta\big|_{k\to -k}
    = {\cal R}^\rmT, \qquad 
    \Theta\cdot{\cal C}^\dagger\cdot\Theta\big|_{k\to -k}
    = {\cal C}^\rmT, \\ 
    M_{ij}\big|_{k\to -k} = M_{ji}. 
\end{gathered}    
\end{equation}
required for consistency of the transformation rules \eqref{eq:G-trans}. 

%-----------------------------------------------
\section{Transformation properties of conservation equations}
%-----------------------------------------------

In this appendix, we give some details regarding the equations of motion. Let us express the conservation equations for off-shell configurations of the dipole Goldstone field $\phi_\mu$ as
\begin{align}
\begin{split}
    \mathsf{E}_{\epsilon}
    &= \nabla_\mu' \epsilon^\mu
    + v^\mu f_\mu 
    + (\tau^{\mu\nu} + \tau^{\mu\nu}_d)
    h_{\lambda\nu}\nabla_\mu v^\lambda, \\ 
    \mathsf{E}^\nu_\pi
    &= 
    \nabla_\mu'\lb v^\mu \pi^\nu 
    + \tau^{\mu\nu} + \tau^{\mu\nu}_d \rb
    - h^{\nu\mu}f_\mu 
    + \pi_\mu h^{\nu\lambda}\nabla_\lambda v^\mu, \\ 
    \mathsf{E}_q
    &= \nabla_\mu'J^\mu, \\
    \mathsf{E}_d^\nu 
    &= \nabla_\mu' J^{\mu\nu}  - h^\nu_\mu J^\mu
    + X^\nu,
\end{split}
\end{align}
where $f_\mu$ and $\tau^{\mu\nu}_d$ are defined in \eqref{eq:f-X}. Using \eqref{eq:dipole-var-currents}, we note that $f_\mu$ and $\tau^{\mu\nu}_d$ transform under dipole transformations as
\begin{align}\begin{split}
    f_\mu 
    &\to f_\mu 
    + 
    \lb
    R^\lambda_{~\mu\rho\nu} 
    + F^n_{\nu\rho} \nabla_\mu v^\lambda
    \rb \psi^\rho J^{\nu}_{~\lambda}  \\
    &\qquad
    - J^\nu h_{\mu\lambda} \nabla_\nu\psi^{\lambda} 
    - (J^\nu n_\nu) \psi^\rho \half \lie_v h_{\mu\rho} 
    + \nabla'_{\nu} \lb\psi_\mu X^\nu \rb
    \\
    &\qquad 
    - n_{\mu}  \nabla'_{\nu} \lb  
    \psi^\rho \nabla_\rho v^\lambda J^{\nu}_{~\lambda}  \rb
    + n_{\mu} \lb \nabla_\nu' J^{\nu\rho} + X^\rho \rb \psi^\sigma \half \lie_v h_{\rho\sigma}, \\
    \tau_d^{\mu\nu} 
    &\to \tau_d^{\mu\nu}  - J^{\rho\nu} \nabla_\rho \psi^\mu.
\end{split}\end{align}
The conservation equations $\mathsf{E}_q$ and $\mathsf{E}^\mu_d$ are manifestly invariant under dipole transformations. For the other two, after some non-trivial algebra, we can show that they transform as
\begin{align}
\begin{split}
    \mathsf{E}_{\epsilon}
    &\to \mathsf{E}_{\epsilon},  \\
    \mathsf{E}^\nu_\pi
    &\to \mathsf{E}^\nu_\pi
    - \psi^\nu \mathsf{E}_q
    - \nabla_\mu'\lb \psi^{\nu} \mathsf{E}_d^\mu \rb.
\end{split}
\label{eq:E-relations}
\end{align}
In deriving these, we have used the curvature identities
\begin{equation}
\begin{split}
    [\nabla_\mu,\nabla_\nu] V^\lambda 
    &= R^\lambda_{~\rho\mu\nu} V^\rho 
    - F^n_{\mu\nu} v^\rho\nabla_\rho V^\lambda, \\ 
    [\nabla_\mu,\nabla_\nu] V_\lambda 
    &= - R^\rho_{~\lambda\mu\nu} V_\rho 
    - F^n_{\mu\nu} v^\rho\nabla_\rho V_\lambda, \\ 
    3R^\lambda_{~[\rho\mu\nu]}
    &= 3 F^n_{[\mu\nu} \nabla_{\rho]} v^\lambda, \\ 
    2h_{\lambda(\rho} R^\lambda_{~\sigma)\mu\nu}
    &= 2n_\rho\nabla_{[\mu}\lie_v h_{\nu]\sigma}
    + 2n_\sigma\nabla_{[\mu} \lie_v h_{\nu]\rho}.
\end{split}
\end{equation}
Therefore the full system of conservation equations is invariant under dipole transformations.

As we motivated in the main text, the dipole-invariant combinations of energy current, momentum density, and stress tensor themselves satisfy a set of conservation equations
\begin{align}
\begin{split}
    \mathsf{E}_{\tilde\epsilon}
    &= \nabla_\mu' \tilde\epsilon^\mu
    + v^\mu \tilde f_\mu 
    + (\tilde\tau^{\mu\nu} + \tilde\tau^{\mu\nu}_d)
    h_{\lambda\nu}\nabla_\mu v^\lambda, \\ 
    \mathsf{E}^\nu_{\tilde\pi}
    &= 
    \nabla_\mu'\lb v^\mu \tilde\pi^\nu 
    + \tilde\tau^{\mu\nu} + \tilde\tau^{\mu\nu}_d \rb
    - h^{\nu\mu} \tilde f_\mu 
    + \tilde\pi_\mu h^{\nu\lambda}\nabla_\lambda v^\mu,
\end{split}
\end{align}
where $\tilde f_\mu$ and $\tilde\tau^{\mu\nu}_d$ are defined in \eqref{eq:tf-X}. The dipole-invariant quantities are obtained by performing a dipole transformation along $\phi_\mu$. Therefore, using \eqref{eq:E-relations} we can read out the relations
\begin{align}
\begin{split}
    \mathsf{E}_{\tilde\epsilon}
    &= \mathsf{E}_{\epsilon},  \\
    \mathsf{E}^\nu_{\tilde\pi}
    &= \mathsf{E}^\nu_\pi
    - \phi^\nu \mathsf{E}_q
    - \nabla_\mu'\lb \phi^{\nu} \mathsf{E}_d^\mu \rb.
\end{split}
\end{align}

We can express the off-shell Second Law of thermodynamics in \eqref{eq:off-shell-2nd} in terms of the quantities outlined above, i.e.
\begin{align}
    \nabla'_\mu s^\mu 
    &- \beta^\rho n_\rho \mathsf{E}_\epsilon
    + \beta^\rho h_{\rho\nu} \mathsf{E}^\nu_\pi
    + \lb\Lambda_\beta  + \beta^\lambda A_\lambda\rb \mathsf{E}_q
    + \lb\psi_\nu^\beta + h_{\nu\lambda}\beta^\rho A^\lambda_{~\rho}\rb
    \mathsf{E}^\nu_d
    = \Delta \geq 0.
\end{align}
It can be checked that the Second Law statement is invariant under dipole transformations, provided that we also shift the entropy current as
\begin{equation}
    s^\mu \to s^\mu + \beta^\nu \psi_\nu \mathsf{E}^\mu_d.
\end{equation}
Note that this transformation vanishes on-shell, so the physical entropy is still invariant. We can also note the transformation properties of the free energy current defined in \eqref{eq:free-E-current}, i.e.
\begin{align}\begin{split}
    N^\mu 
    &\to N^\mu 
    - \lb 2\psi^{(\rho} J^{\sigma)\mu} 
    - \psi^{\mu} J^{\rho\sigma}
    \rb \half \lie_\beta h_{\rho\sigma}
    + \frac{1}{\sqrt{\gamma}}
    \dow_\nu\!\lb\sqrt{\gamma}\,
    2\psi^{[\nu} J^{\mu]}_{~\rho}\beta^\rho\rb.
\end{split}\end{align}
We have used the fact that $1/\sqrt{\gamma}\, \dow_\nu (\sqrt{\gamma}\, X^{\mu\nu}) = \nabla'_\nu X^{\mu\nu} + v^\mu \half F^n_{\rho\sigma}X^{\rho\sigma} $. The second term in the transformation is irrelevant because its gradient vanishes and does not contribute to the adiabaticity equation. However, the free energy current still transforms non-trivially under dipole transformations even for on-shell configurations, and the transformation law is similar to the conserved energy current $\epsilon^\mu$ as given in \eqref{eq:dipole-var-currents}.

Correspondingly, we can denote the Second Law statement in the dipole-invariant basis using the entropy current
\begin{equation}
    \tilde s^\mu = s^\mu + \beta^\nu \phi_\nu \mathsf{E}^\mu_d,
\end{equation}
leading to
\begin{align}
    \nabla'_\mu \tilde s^\mu 
    &- \beta^\rho n_\rho \mathsf{E}_{\tilde\epsilon}
    + \beta^\rho h_{\rho\nu} \mathsf{E}^\nu_{\tilde\pi}
    + \lb \Lambda_\beta  + \beta^\lambda \tilde A_\lambda\rb \mathsf{E}_q
    + \lb\tilde\psi_\nu^\beta
    + h_{\nu\lambda}\beta^\rho \tilde A^\lambda_{~\rho}\rb
    \mathsf{E}^\nu_d
    = \Delta \geq 0,
\end{align}
where $\tilde\psi_\mu = \psi_\mu - h_\mu^\nu \lie_\beta\phi_\nu$.

%-----------------------------------------------
\section{Comparison with previous works}
\label{app:comparison}
%-----------------------------------------------

Our results for p-wave superfluids have some overlap with previous works of \cite{Glodkowski:2022xje, Glorioso:2023chm}, and in this Appendix we would like to emphasize the commonalities and distinctions between our works. 

We have argued at length in this work that the global dipole symmetry of a physical system must be spontaneously broken in the low-energy description for there to exist non-trivial charge transport, leading to a vector Goldstone $\phi_i$. In the absence of dipole superfluid, i.e. when $\langle \partial_i \phi_j \rangle = 0$, we can identify the one-point function of momentum density with that of the dipole Goldstone, i.e. $\langle \pi_i \rangle = \langle \phi_i \rangle/q_0$. In \cite{Glodkowski:2022xje}, this identification is posited even out-of-equilibrium, though, as we show, this is not true on curved backgrounds, or when $\tilde{\pi}_i\neq 0$, for instance in the presence of a kinetic mass density $\rho$. By utilizing the dipole-invariant construction of hydrodynamics, we are able to relax this assumption and explicitly identify the role of the Goldstone field in the momentum response beyond equilibrium thermodynamics. A tangible feature of this is the appearance of the kinetic mass density $\rho$ in momentum response function $G^R_{\pi_\perp\pi_\perp}$.

As we emphasize in the main text, a consistent formulation of p-wave dipole superfluid hydrodynamics involves a novel derivative counting scheme in the presence of background sources. Fundamental to this counting is the requirement that the hydrodynamic fields $\tmu$ and $T$ scale as $O(\partial^0)$. The existence of superflow indicates the Goldstone $\phi_i$ scales as $O(\partial^{-1})$, from which it follows via the homogeneity of hydrodynamic equations that $A_i\sim O(\partial^{-1})$ and ${u}^i\sim O(\partial)$. Note that while ${u}^i$ is a hydrodynamic field, the dipole Ward identity requires that it vanishes in equilibrium, as we demonstrate in \eqref{eq:equilibrium_u0}. In fact, this counting scheme is implicitly present in  the work of \cite{Glodkowski:2022xje} as well, as illustrated via the inclusion of thermal conduction at the same derivative order as the ideal order hydrostatic terms in the dipole-invariant energy flux $\tilde{\epsilon}^i$. 

Next, the authors of \cite{Glodkowski:2022xje} noticed that the use of a \emph{shifted energy current} and effective chemical potential led to a more familiar expression for the first law and allowed for more straightforward Second Law constraints. In fact, these are none other than the dipole-invariant versions of the energy flux $\tilde{\epsilon}^i$ and our dipole-invariant definition of the chemical potential. That such a rewriting is useful then becomes transparent as the hydrodynamic theory written in terms of dipole-invariant fields is effectively that of a charged boost-agnostic fluid without dipole symmetry. 

Conveniently, we find that with a certain consistent truncation, our results match with those of \cite{Glodkowski:2022xje}. For the sake of matching, we will work in flat spacetime background. To compare, consider the ideal constitutive relations for the U(1) and dipole currents from \eqref{eq:ideal_flat_constitutive}. The linearized Goldstone equations of motion then imply eq.~\eqref{eq:equilibrium_u0}. So, denoting the notation of \cite{Glodkowski:2022xje} with a hat, we identify 
\begin{align}
    \hat F^{ij} 
    = - (J^{ij} -\chi_m\tilde{F}^{ij}), \quad 
    \hat n = q, \quad 
    \hat \mu = \tilde\mu, \quad 
    \hat V^i = u^i, \quad 
    \hat p_i = q\phi_i, \qquad \hat{V}_{ij} = - \partial_{(i}\phi_{j)}~,
\end{align}
 and because $\hat{V}_{[ij]} = 0$ in \cite{Glodkowski:2022xje}, to match, we then must set $\chi_m= 0$. This is the equivalent of setting $\hat{F}^{ij} = -J^{ij}$, an equality obtained from the ideal Second Law in their work. Continuing, we see that
\begin{align}
    \df p 
    = s \df T 
    + \hat n \df \hat\mu 
    - \lb 
    p_d \delta_{ij}-G_d a_{\langle i j \rangle}
    \rb \df \hat{V}_{ij},
\end{align}
so that 
\begin{align}
    \delta F_{ij} = \delta_{ij}\left(
    B_d \delta \hat V^k_{~k}
    + \chi_{sd}\delta T 
    + \chi_{qd}\delta \hat\mu\right)
    + 2G_d\delta \hat V_{\langle ij\rangle}.
\end{align}
which tells us to identify $\hat f_{\|}=B_d$ and $\hat f_\perp = 2G_d$, as well as set $\chi_{sd} = \chi_{qd}=0$. These are reasonable if in our model, we choose $p_d = - \half B_d \,\tr\, \tilde{a}$ which then vanishes in equilibrium; see \eqref{eq:thermodynamicderivatives}. We can also see that $\hat{\bar{f}} = \hat n_0^{-1}(\hat f_{\|}
+\frac{d-1}{d}\hat f_\perp) = B_d'/q_0$ in our notation.

Having made these identifications, we can consider constitutive relations. In the absence of background sources our hydrostatic dipole-invariant energy current is
\begin{align}
    \tilde{\epsilon}^i_{\text{ideal}} = (\epsilon+p){u}^i + (p_d \delta^{ij}
    - G_d\tilde{a}^{\langle ij \rangle})
    \tilde{F}_{tj} \to (\epsilon+p)\hat{V}^i -\hat{F}^{ij}\partial_t\phi_j,
\end{align}
 agreeing with \cite{Glodkowski:2022xje}. Importantly, to match the two expressions, we must include certain non-linear terms in $\tilde{\epsilon}^i$. Including non-hydrostatic corrections, in flat spacetime, we find 
 \begin{align}
     \tilde{\epsilon}^i = \tilde{\epsilon}^i_{\text{ideal}} 
     + T^2\kappa \partial^i\frac1T.
 \end{align}
This allows us to identify $\hat\alpha = T^2\kappa$.

Turning to higher order non-hydrostatic terms, whereas we predict 7 non-trivial non-dissipative terms and \cite{Glodkowski:2022xje} predicts 6, both our works predict 12 non-trivial transport coefficients which contribute to entropy production. In fact, once we have integrated by parts, the expressions for entropy production are quite similar, being expressible in terms of $\partial_{i}\partial_j T$, $\partial_i\partial_j \hat\mu$, and $\partial_i \hat V_j$, though the relation between transport coefficients is somewhat obscured. The extra coefficient we find is $\bar{\Omega}_\epsilon$ which explicitly contains a time derivative. It seems reasonable that using equations of motion, this can be replaced with spatial derivatives, but whether or not this term can be eliminated altogether at second order was not evident to us.

After imposing time reversal invariance, we are left with only 6 dissipative coefficients, as opposed to 12, and we retain 6 non-dissipative coefficients at second order (now, $\bar{\Omega}_\epsilon = 0$). Hence, our total counts after Onsager's relations are imposed are the same, but our accounting of entropy production differs slightly. This is partially related to our choice of hydrodynamic frame, where we impose $h_{\mu\nu}J^\mu_{\nhs} = 0$. Nevertheless, this distinction does not affect the leading behavior of the hydrodynamic modes in $k$. Given the matching discussed above, where we turn off many coefficients, we find the same expressions for the longitudinal modes to $O(k^2)$ and the transverse mode to $O(k^4)$.

In summary, the work of \cite{Glodkowski:2022xje}, can be seen as a consistent truncation of the work presented here, subject to the caveat that the relationship between the momentum $\pi_i$ and the Goldstone $\phi_i$ is required to hold out-of-equilibrium. In this respect, our method can be used to find response functions for the model presented in that work, and can be used as a diagnostic to identify relevant contributions to the hydrodynamics. For instance, when $\rho, p_d \neq 0$, then 
\begin{align}
    \lim_{\omega \to 0} \frac{-1}{\omega^2}\,\text{Re} \,G^R_{\epsilon^i\epsilon^j}(\omega) = \frac{\rho p_d^2}{q^2},
\end{align}
whereas for \cite{Glodkowski:2022xje} would find a vanishing result. Likewise, in the limit that $q\to 0$, we would recover the response of a neutral fluid
\begin{align}
    G_{\pi_\perp \pi_\perp}^R = -\rho + \frac{\omega \rho^2}{\rho\omega + i\eta k^2}.
\end{align}

We would also like to mention the work of \cite{Glorioso:2023chm} which found similar results to our work and to the results of \cite{Glodkowski:2022xje} in the context of p-wave dipole superfluids. The work considers dissipation up to $O(\partial^2)$ in the limit that energy is not conserved, as well as hydrodynamics with energy conservation up to $O(\partial)$ via inclusion of the thermal conductivity term. Overall, the schematic structure of the hydrodynamic modes is similar to what we find, but because of the non-trivial mixing between energy, charge, and the dipole current, we found that it was necessary to consider all currents simultaneously within a consistent derivative scheme. We nevertheless note a curious observation that the subdiffusive constant $D_\perp$ is oblivous to explicit derivative corrections appearing in the non-hydrostatic sector in the sense that the only relevant dissipative parameter is $\eta$, but not terms like $\eta_d$ which appear at the same order in the tensorial sector. In fact, the subdiffusion constant matches the result of \cite{Glorioso:2023chm} given in the absence of energy conservation, suggesting that the effects of energy conservation are subleading to the coupling of stress to dipole fluctuations in the transverse sector, leading to a universal propagation behavior. This is connected to the fact that the dissipative vector terms appear $O(\partial^{-2})$ enhanced relative to the scalar and tensor terms and the energy flux only sees the vector corrections.

\bibliographystyle{JHEP}
\bibliography{refs2}

\providecommand{\href}[2]{#2}\begingroup\raggedright\begin{thebibliography}{10}

\bibitem{Chamon:2004lew}
C.~Chamon, \emph{{Quantum Glassiness}},
  \href{https://doi.org/10.1103/physrevlett.94.040402}{\emph{Phys. Rev. Lett.}
  {\bfseries 94} (2005) 040402}
  [\href{https://arxiv.org/abs/cond-mat/0404182}{{\ttfamily
  cond-mat/0404182}}].

\bibitem{2011AnPhy.326..839B}
S.~{Bravyi}, B.~{Leemhuis} and B.M.~{Terhal}, \emph{{Topological order in an
  exactly solvable 3D spin model}},
  \href{https://doi.org/10.1016/j.aop.2010.11.002}{\emph{Annals of Physics}
  {\bfseries 326} (2011) 839}
  [\href{https://arxiv.org/abs/1006.4871}{{\ttfamily 1006.4871}}].

\bibitem{Haah:2011drr}
J.~Haah, \emph{{Local stabilizer codes in three dimensions without string
  logical operators}},
  \href{https://doi.org/10.1103/physreva.83.042330}{\emph{Phys. Rev. A}
  {\bfseries 83} (2011) 042330}
  [\href{https://arxiv.org/abs/1101.1962}{{\ttfamily 1101.1962}}].

\bibitem{Vijay:2015mka}
S.~Vijay, J.~Haah and L.~Fu, \emph{{A New Kind of Topological Quantum Order: A
  Dimensional Hierarchy of Quasiparticles Built from Stationary Excitations}},
  \href{https://doi.org/10.1103/PhysRevB.92.235136}{\emph{Phys. Rev. B}
  {\bfseries 92} (2015) 235136}
  [\href{https://arxiv.org/abs/1505.02576}{{\ttfamily 1505.02576}}].

\bibitem{Nandkishore:2018sel}
R.M.~Nandkishore and M.~Hermele, \emph{{Fractons}},
  \href{https://doi.org/10.1146/annurev-conmatphys-031218-013604}{\emph{Ann.
  Rev. Condensed Matter Phys.} {\bfseries 10} (2019) 295}
  [\href{https://arxiv.org/abs/1803.11196}{{\ttfamily 1803.11196}}].

\bibitem{Seiberg:2020bhn}
N.~Seiberg and S.-H.~Shao, \emph{{Exotic Symmetries, Duality, and Fractons in
  2+1-Dimensional Quantum Field Theory}},
  \href{https://doi.org/10.21468/SciPostPhys.10.2.027}{\emph{SciPost Phys.}
  {\bfseries 10} (2021) 027}
  [\href{https://arxiv.org/abs/2003.10466}{{\ttfamily 2003.10466}}].

\bibitem{Vijay:2016phm}
S.~Vijay, J.~Haah and L.~Fu, \emph{{Fracton Topological Order, Generalized
  Lattice Gauge Theory and Duality}},
  \href{https://doi.org/10.1103/PhysRevB.94.235157}{\emph{Phys. Rev. B}
  {\bfseries 94} (2016) 235157}
  [\href{https://arxiv.org/abs/1603.04442}{{\ttfamily 1603.04442}}].

\bibitem{Williamson:2016jiq}
D.J.~Williamson, \emph{{Fractal symmetries: Ungauging the cubic code}},
  \href{https://doi.org/10.1103/PhysRevB.94.155128}{\emph{Phys. Rev. B}
  {\bfseries 94} (2016) 155128}
  [\href{https://arxiv.org/abs/1603.05182}{{\ttfamily 1603.05182}}].

\bibitem{You:2018oai}
Y.~You, T.~Devakul, F.J.~Burnell and S.L.~Sondhi, \emph{{Subsystem symmetry
  protected topological order}},
  \href{https://doi.org/10.1103/PhysRevB.98.035112}{\emph{Phys. Rev. B}
  {\bfseries 98} (2018) 035112}
  [\href{https://arxiv.org/abs/1803.02369}{{\ttfamily 1803.02369}}].

\bibitem{Slagle:2017wrc}
K.~Slagle and Y.B.~Kim, \emph{{Quantum Field Theory of X-Cube Fracton
  Topological Order and Robust Degeneracy from Geometry}},
  \href{https://doi.org/10.1103/PhysRevB.96.195139}{\emph{Phys. Rev. B}
  {\bfseries 96} (2017) 195139}
  [\href{https://arxiv.org/abs/1708.04619}{{\ttfamily 1708.04619}}].

\bibitem{Seiberg:2020wsg}
N.~Seiberg and S.-H.~Shao, \emph{{Exotic $U(1)$ Symmetries, Duality, and
  Fractons in 3+1-Dimensional Quantum Field Theory}},
  \href{https://doi.org/10.21468/SciPostPhys.9.4.046}{\emph{SciPost Phys.}
  {\bfseries 9} (2020) 046} [\href{https://arxiv.org/abs/2004.00015}{{\ttfamily
  2004.00015}}].

\bibitem{Pretko:2018jbi}
M.~Pretko, \emph{{The Fracton Gauge Principle}},
  \href{https://doi.org/10.1103/PhysRevB.98.115134}{\emph{Phys. Rev. B}
  {\bfseries 98} (2018) 115134}
  [\href{https://arxiv.org/abs/1807.11479}{{\ttfamily 1807.11479}}].

\bibitem{Pretko:2017kvd}
M.~Pretko and L.~Radzihovsky, \emph{{Fracton-Elasticity Duality}},
  \href{https://doi.org/10.1103/PhysRevLett.120.195301}{\emph{Phys. Rev. Lett.}
  {\bfseries 120} (2018) 195301}
  [\href{https://arxiv.org/abs/1711.11044}{{\ttfamily 1711.11044}}].

\bibitem{Doshi:2020jso}
D.~Doshi and A.~Gromov, \emph{{Vortices and Fractons}},
  \href{https://arxiv.org/abs/2005.03015}{{\ttfamily 2005.03015}}.

\bibitem{tilted}
E.~Guardado-Sanchez, A.~Morningstar, B.M.~Spar, P.T.~Brown, D.A.~Huse and
  W.S.~Bakr, \emph{Subdiffusion and heat transport in a tilted two-dimensional
  fermi-hubbard system},
  \href{https://doi.org/10.1103/PhysRevX.10.011042}{\emph{Phys. Rev. X}
  {\bfseries 10} (2020) 011042}.

\bibitem{Jensen:2022iww}
K.~Jensen and A.~Raz, \emph{{Large $N$ fractons}},
  \href{https://arxiv.org/abs/2205.01132}{{\ttfamily 2205.01132}}.

\bibitem{Pretko:2020cko}
M.~Pretko, X.~Chen and Y.~You, \emph{{Fracton Phases of Matter}},
  \href{https://doi.org/10.1142/S0217751X20300033}{\emph{Int. J. Mod. Phys. A}
  {\bfseries 35} (2020) 2030003}
  [\href{https://arxiv.org/abs/2001.01722}{{\ttfamily 2001.01722}}].

\bibitem{subsystemWIP}
K.~Jensen and A.~Raz, \emph{{Forthcoming}}, .

\bibitem{GromovMultipole}
A.~Gromov, \emph{Towards classification of fracton phases: The multipole
  algebra}, \href{https://doi.org/10.1103/PhysRevX.9.031035}{\emph{Phys. Rev.
  X} {\bfseries 9} (2019) 031035}.

\bibitem{Distler:2021qzc}
J.~Distler, A.~Karch and A.~Raz, \emph{{Spontaneously broken subsystem
  symmetries}}, \href{https://doi.org/10.1007/JHEP03(2022)016}{\emph{JHEP}
  {\bfseries 03} (2022) 016}
  [\href{https://arxiv.org/abs/2110.12611}{{\ttfamily 2110.12611}}].

\bibitem{Gorantla:2022eem}
P.~Gorantla, H.T.~Lam, N.~Seiberg and S.-H.~Shao, \emph{{Global dipole
  symmetry, compact Lifshitz theory, tensor gauge theory, and fractons}},
  \href{https://doi.org/10.1103/PhysRevB.106.045112}{\emph{Phys. Rev. B}
  {\bfseries 106} (2022) 045112}
  [\href{https://arxiv.org/abs/2201.10589}{{\ttfamily 2201.10589}}].

\bibitem{Gorantla:2022ssr}
P.~Gorantla, H.T.~Lam, N.~Seiberg and S.-H.~Shao, \emph{{2+1d Compact Lifshitz
  Theory, Tensor Gauge Theory, and Fractons}},
  \href{https://arxiv.org/abs/2209.10030}{{\ttfamily 2209.10030}}.

\bibitem{Bhattacharyya:2007vjd}
S.~Bhattacharyya, V.E.~Hubeny, S.~Minwalla and M.~Rangamani, \emph{{Nonlinear
  Fluid Dynamics from Gravity}},
  \href{https://doi.org/10.1088/1126-6708/2008/02/045}{\emph{JHEP} {\bfseries
  02} (2008) 045} [\href{https://arxiv.org/abs/0712.2456}{{\ttfamily
  0712.2456}}].

\bibitem{Banerjee:2008th}
N.~Banerjee, J.~Bhattacharya, S.~Bhattacharyya, S.~Dutta, R.~Loganayagam and
  P.~Surowka, \emph{{Hydrodynamics from charged black branes}},
  \href{https://doi.org/10.1007/JHEP01(2011)094}{\emph{JHEP} {\bfseries 01}
  (2011) 094} [\href{https://arxiv.org/abs/0809.2596}{{\ttfamily 0809.2596}}].

\bibitem{Crossley:2015evo}
M.~Crossley, P.~Glorioso and H.~Liu, \emph{{Effective field theory of
  dissipative fluids}},
  \href{https://doi.org/10.1007/JHEP09(2017)095}{\emph{JHEP} {\bfseries 09}
  (2017) 095} [\href{https://arxiv.org/abs/1511.03646}{{\ttfamily
  1511.03646}}].

\bibitem{Haehl:2018lcu}
F.M.~Haehl, R.~Loganayagam and M.~Rangamani, \emph{{Effective Action for
  Relativistic Hydrodynamics: Fluctuations, Dissipation, and Entropy Inflow}},
  \href{https://doi.org/10.1007/JHEP10(2018)194}{\emph{JHEP} {\bfseries 10}
  (2018) 194} [\href{https://arxiv.org/abs/1803.11155}{{\ttfamily
  1803.11155}}].

\bibitem{Jensen:2017kzi}
K.~Jensen, N.~Pinzani-Fokeeva and A.~Yarom, \emph{{Dissipative hydrodynamics in
  superspace}}, \href{https://doi.org/10.1007/JHEP09(2018)127}{\emph{JHEP}
  {\bfseries 09} (2018) 127}
  [\href{https://arxiv.org/abs/1701.07436}{{\ttfamily 1701.07436}}].

\bibitem{Jain:2020vgc}
A.~Jain, \emph{{Effective field theory for non-relativistic hydrodynamics}},
  \href{https://doi.org/10.1007/JHEP10(2020)208}{\emph{JHEP} {\bfseries 10}
  (2020) 208} [\href{https://arxiv.org/abs/2008.03994}{{\ttfamily
  2008.03994}}].

\bibitem{Armas:2020mpr}
J.~Armas and A.~Jain, \emph{{Effective field theory for hydrodynamics without
  boosts}}, \href{https://doi.org/10.21468/SciPostPhys.11.3.054}{\emph{SciPost
  Phys.} {\bfseries 11} (2021) 054}
  [\href{https://arxiv.org/abs/2010.15782}{{\ttfamily 2010.15782}}].

\bibitem{swavearticle}
A.~Jain, K.~Jensen, R.~Liu and E.~Mefford, \emph{S-wave dipole superfluid
  hydrodynamics}, .

\bibitem{Gromov:2020yoc}
A.~Gromov, A.~Lucas and R.M.~Nandkishore, \emph{{Fracton hydrodynamics}},
  \href{https://doi.org/10.1103/PhysRevResearch.2.033124}{\emph{Phys. Rev.
  Res.} {\bfseries 2} (2020) 033124}
  [\href{https://arxiv.org/abs/2003.09429}{{\ttfamily 2003.09429}}].

\bibitem{Iaconis:2020zhc}
J.~Iaconis, A.~Lucas and R.~Nandkishore, \emph{{Multipole conservation laws and
  subdiffusion in any dimension}},
  \href{https://doi.org/10.1103/PhysRevE.103.022142}{\emph{Phys. Rev. E}
  {\bfseries 103} (2021) 022142}
  [\href{https://arxiv.org/abs/2009.06507}{{\ttfamily 2009.06507}}].

\bibitem{Glorioso:2021bif}
P.~Glorioso, J.~Guo, J.F.~Rodriguez-Nieva and A.~Lucas, \emph{{Breakdown of
  hydrodynamics below four dimensions in a fracton fluid}},
  \href{https://arxiv.org/abs/2105.13365}{{\ttfamily 2105.13365}}.

\bibitem{Grosvenor:2021rrt}
K.T.~Grosvenor, C.~Hoyos, F.~Pe\~na Ben\'\i{}tez and P.~Sur\'owka,
  \emph{{Hydrodynamics of ideal fracton fluids}},
  \href{https://doi.org/10.1103/PhysRevResearch.3.043186}{\emph{Phys. Rev.
  Res.} {\bfseries 3} (2021) 043186}
  [\href{https://arxiv.org/abs/2105.01084}{{\ttfamily 2105.01084}}].

\bibitem{Glodkowski:2022xje}
A.~G\l{}\'odkowski, F.~Pe\~na Ben\'\i{}tez and P.~Sur\'owka,
  \emph{{Hydrodynamics of dipole-conserving fluids}},
  \href{https://arxiv.org/abs/2212.06848}{{\ttfamily 2212.06848}}.

\bibitem{Glorioso:2023chm}
P.~Glorioso, X.~Huang, J.~Guo, J.~Rodriguez-Nieva and A.~Lucas,
  \emph{{Goldstone bosons and fluctuating hydrodynamics with dipole and
  momentum conservation}},  \href{https://arxiv.org/abs/2301.02680}{{\ttfamily
  2301.02680}}.

\bibitem{idealFractonHydro}
J.~Armas and E.~Have, \emph{{Ideal fracton superfluids}}, .

\bibitem{Jain:2021ibh}
A.~Jain and K.~Jensen, \emph{{Fractons in curved space}},
  \href{https://doi.org/10.21468/SciPostPhys.12.4.142}{\emph{SciPost Phys.}
  {\bfseries 12} (2022) 142}
  [\href{https://arxiv.org/abs/2111.03973}{{\ttfamily 2111.03973}}].

\bibitem{Jensen:2014ama}
K.~Jensen, \emph{{Aspects of hot Galilean field theory}},
  \href{https://doi.org/10.1007/JHEP04(2015)123}{\emph{JHEP} {\bfseries 04}
  (2015) 123} [\href{https://arxiv.org/abs/1411.7024}{{\ttfamily 1411.7024}}].

\bibitem{landaubook}
L.~Landau and E.~Lifshitz, \emph{Fluid Mechanics (Second Edition)}, Pergamon,
  second edition~ed. (1987),
  \href{https://doi.org/https://doi.org/10.1016/B978-0-08-033933-7.50006-4}{https://doi.org/10.1016/B978-0-08-033933-7.50006-4}.

\bibitem{Kovtun:2012rj}
P.~Kovtun, \emph{{Lectures on hydrodynamic fluctuations in relativistic
  theories}}, \href{https://doi.org/10.1088/1751-8113/45/47/473001}{\emph{J.
  Phys. A} {\bfseries 45} (2012) 473001}
  [\href{https://arxiv.org/abs/1205.5040}{{\ttfamily 1205.5040}}].

\bibitem{Banerjee:2012iz}
N.~Banerjee, J.~Bhattacharya, S.~Bhattacharyya, S.~Jain, S.~Minwalla and
  T.~Sharma, \emph{{Constraints on Fluid Dynamics from Equilibrium Partition
  Functions}}, \href{https://doi.org/10.1007/JHEP09(2012)046}{\emph{JHEP}
  {\bfseries 09} (2012) 046} [\href{https://arxiv.org/abs/1203.3544}{{\ttfamily
  1203.3544}}].

\bibitem{Jensen:2012jh}
K.~Jensen, M.~Kaminski, P.~Kovtun, R.~Meyer, A.~Ritz and A.~Yarom,
  \emph{{Towards hydrodynamics without an entropy current}},
  \href{https://doi.org/10.1103/PhysRevLett.109.101601}{\emph{Phys. Rev. Lett.}
  {\bfseries 109} (2012) 101601}
  [\href{https://arxiv.org/abs/1203.3556}{{\ttfamily 1203.3556}}].

\bibitem{Bhattacharyya:2012xi}
S.~Bhattacharyya, S.~Jain, S.~Minwalla and T.~Sharma, \emph{{Constraints on
  Superfluid Hydrodynamics from Equilibrium Partition Functions}},
  \href{https://doi.org/10.1007/JHEP01(2013)040}{\emph{JHEP} {\bfseries 01}
  (2013) 040} [\href{https://arxiv.org/abs/1206.6106}{{\ttfamily 1206.6106}}].

\bibitem{Glorioso:2016gsa}
P.~Glorioso and H.~Liu, \emph{{The second law of thermodynamics from symmetry
  and unitarity}},  \href{https://arxiv.org/abs/1612.07705}{{\ttfamily
  1612.07705}}.

\bibitem{Haehl:2018uqv}
F.M.~Haehl, R.~Loganayagam and M.~Rangamani, \emph{{Inflow Mechanism for
  Hydrodynamic Entropy}},
  \href{https://doi.org/10.1103/PhysRevLett.121.051602}{\emph{Phys. Rev. Lett.}
  {\bfseries 121} (2018) 051602}
  [\href{https://arxiv.org/abs/1803.08490}{{\ttfamily 1803.08490}}].

\bibitem{Jensen:2018hhx}
K.~Jensen, R.~Marjieh, N.~Pinzani-Fokeeva and A.~Yarom, \emph{{An entropy
  current in superspace}},
  \href{https://doi.org/10.1007/JHEP01(2019)061}{\emph{JHEP} {\bfseries 01}
  (2019) 061} [\href{https://arxiv.org/abs/1803.07070}{{\ttfamily
  1803.07070}}].

\bibitem{Bhattacharyya:2013lha}
S.~Bhattacharyya, \emph{{Entropy current and equilibrium partition function in
  fluid dynamics}}, \href{https://doi.org/10.1007/JHEP08(2014)165}{\emph{JHEP}
  {\bfseries 08} (2014) 165} [\href{https://arxiv.org/abs/1312.0220}{{\ttfamily
  1312.0220}}].

\bibitem{Hernandez:2017mch}
J.~Hernandez and P.~Kovtun, \emph{{Relativistic magnetohydrodynamics}},
  \href{https://doi.org/10.1007/JHEP05(2017)001}{\emph{JHEP} {\bfseries 05}
  (2017) 001} [\href{https://arxiv.org/abs/1703.08757}{{\ttfamily
  1703.08757}}].

\bibitem{Armas:2018atq}
J.~Armas and A.~Jain, \emph{{Magnetohydrodynamics as superfluidity}},
  \href{https://doi.org/10.1103/PhysRevLett.122.141603}{\emph{Phys. Rev. Lett.}
  {\bfseries 122} (2019) 141603}
  [\href{https://arxiv.org/abs/1808.01939}{{\ttfamily 1808.01939}}].

\bibitem{Armas:2018zbe}
J.~Armas and A.~Jain, \emph{{One-form superfluids \& magnetohydrodynamics}},
  \href{https://doi.org/10.1007/JHEP01(2020)041}{\emph{JHEP} {\bfseries 01}
  (2020) 041} [\href{https://arxiv.org/abs/1811.04913}{{\ttfamily
  1811.04913}}].

\bibitem{Bidussi:2021nmp}
L.~Bidussi, J.~Hartong, E.~Have, J.~Musaeus and S.~Prohazka, \emph{{Fractons,
  dipole symmetries and curved spacetime}},
  \href{https://doi.org/10.21468/SciPostPhys.12.6.205}{\emph{SciPost Phys.}
  {\bfseries 12} (2022) 205}
  [\href{https://arxiv.org/abs/2111.03668}{{\ttfamily 2111.03668}}].

\bibitem{Slagle:2018kqf}
K.~Slagle, A.~Prem and M.~Pretko, \emph{{Symmetric Tensor Gauge Theories on
  Curved Spaces}},
  \href{https://doi.org/10.1016/j.aop.2019.167910}{\emph{Annals Phys.}
  {\bfseries 410} (2019) 167910}
  [\href{https://arxiv.org/abs/1807.00827}{{\ttfamily 1807.00827}}].

\bibitem{Novak:2019wqg}
I.~Novak, J.~Sonner and B.~Withers, \emph{{Hydrodynamics without boosts}},
  \href{https://doi.org/10.1007/JHEP07(2020)165}{\emph{JHEP} {\bfseries 07}
  (2020) 165} [\href{https://arxiv.org/abs/1911.02578}{{\ttfamily
  1911.02578}}].

\bibitem{deBoer:2020xlc}
J.~de~Boer, J.~Hartong, E.~Have, N.A.~Obers and W.~Sybesma, \emph{{Non-Boost
  Invariant Fluid Dynamics}},
  \href{https://doi.org/10.21468/SciPostPhys.9.2.018}{\emph{SciPost Phys.}
  {\bfseries 9} (2020) 018} [\href{https://arxiv.org/abs/2004.10759}{{\ttfamily
  2004.10759}}].

\bibitem{Gouteraux:2022kpo}
B.~Gout\'eraux, E.~Mefford and F.~Sottovia, \emph{{Thermodynamic origin of the
  Landau instability of superfluids}},
  \href{https://arxiv.org/abs/2212.10410}{{\ttfamily 2212.10410}}.

\bibitem{EricDipoleSuperflow}
E.~Mefford, \emph{Dipole superflow hydrodynamics}, .

\bibitem{Jensen:2013kka}
K.~Jensen, R.~Loganayagam and A.~Yarom, \emph{{Anomaly inflow and thermal
  equilibrium}}, \href{https://doi.org/10.1007/JHEP05(2014)134}{\emph{JHEP}
  {\bfseries 05} (2014) 134} [\href{https://arxiv.org/abs/1310.7024}{{\ttfamily
  1310.7024}}].

\bibitem{Banerjee:2015uta}
N.~Banerjee, S.~Dutta and A.~Jain, \emph{{Equilibrium partition function for
  nonrelativistic fluids}},
  \href{https://doi.org/10.1103/PhysRevD.92.081701}{\emph{Phys. Rev. D}
  {\bfseries 92} (2015) 081701}
  [\href{https://arxiv.org/abs/1505.05677}{{\ttfamily 1505.05677}}].

\bibitem{Armas:2019sbe}
J.~Armas and A.~Jain, \emph{{Viscoelastic hydrodynamics and holography}},
  \href{https://doi.org/10.1007/JHEP01(2020)126}{\emph{JHEP} {\bfseries 01}
  (2020) 126} [\href{https://arxiv.org/abs/1908.01175}{{\ttfamily
  1908.01175}}].

\bibitem{Armas:2020bmo}
J.~Armas and A.~Jain, \emph{{Hydrodynamics for charge density waves and their
  holographic duals}},
  \href{https://doi.org/10.1103/PhysRevD.101.121901}{\emph{Phys. Rev. D}
  {\bfseries 101} (2020) 121901}
  [\href{https://arxiv.org/abs/2001.07357}{{\ttfamily 2001.07357}}].

\bibitem{Gromov:2017vir}
A.~Gromov, \emph{{Chiral Topological Elasticity and Fracton Order}},
  \href{https://doi.org/10.1103/PhysRevLett.122.076403}{\emph{Phys. Rev. Lett.}
  {\bfseries 122} (2019) 076403}
  [\href{https://arxiv.org/abs/1712.06600}{{\ttfamily 1712.06600}}].

\bibitem{Grosvenor:2021hkn}
K.T.~Grosvenor, C.~Hoyos, F.~Pe\~na Benitez and P.~Sur\'owka,
  \emph{{Space-Dependent Symmetries and Fractons}},
  \href{https://doi.org/10.3389/fphy.2021.792621}{\emph{Front. in Phys.}
  {\bfseries 9} (2022) 792621}
  [\href{https://arxiv.org/abs/2112.00531}{{\ttfamily 2112.00531}}].

\bibitem{Haehl:2015uoc}
F.M.~Haehl, R.~Loganayagam and M.~Rangamani, \emph{{Topological sigma models \&
  dissipative hydrodynamics}},
  \href{https://doi.org/10.1007/JHEP04(2016)039}{\emph{JHEP} {\bfseries 04}
  (2016) 039} [\href{https://arxiv.org/abs/1511.07809}{{\ttfamily
  1511.07809}}].

\bibitem{Jensen:2014aia}
K.~Jensen, \emph{{On the coupling of Galilean-invariant field theories to
  curved spacetime}},
  \href{https://doi.org/10.21468/SciPostPhys.5.1.011}{\emph{SciPost Phys.}
  {\bfseries 5} (2018) 011} [\href{https://arxiv.org/abs/1408.6855}{{\ttfamily
  1408.6855}}].

\bibitem{Geracie:2015xfa}
M.~Geracie, K.~Prabhu and M.M.~Roberts, \emph{{Fields and fluids on curved
  non-relativistic spacetimes}},
  \href{https://doi.org/10.1007/JHEP08(2015)042}{\emph{JHEP} {\bfseries 08}
  (2015) 042} [\href{https://arxiv.org/abs/1503.02680}{{\ttfamily
  1503.02680}}].

\bibitem{Jain:2016rlz}
A.~Jain, \emph{{Theory of non-Abelian superfluid dynamics}},
  \href{https://doi.org/10.1103/PhysRevD.95.121701}{\emph{Phys. Rev. D}
  {\bfseries 95} (2017) 121701}
  [\href{https://arxiv.org/abs/1610.05797}{{\ttfamily 1610.05797}}].

\bibitem{Liu1972MethodOL}
I.-S.~Liu, \emph{Method of lagrange multipliers for exploitation of the entropy
  principle}, {\emph{Archive for Rational Mechanics and Analysis} {\bfseries
  46} (1972) 131}.

\bibitem{Loganayagam:2011mu}
R.~Loganayagam, \emph{{Anomaly Induced Transport in Arbitrary Dimensions}},
  \href{https://arxiv.org/abs/1106.0277}{{\ttfamily 1106.0277}}.

\bibitem{Haehl:2015pja}
F.M.~Haehl, R.~Loganayagam and M.~Rangamani, \emph{{Adiabatic hydrodynamics:
  The eightfold way to dissipation}},
  \href{https://doi.org/10.1007/JHEP05(2015)060}{\emph{JHEP} {\bfseries 05}
  (2015) 060} [\href{https://arxiv.org/abs/1502.00636}{{\ttfamily
  1502.00636}}].

\bibitem{Jain:2018jxj}
A.~Jain, \emph{{A universal framework for hydrodynamics}}, Ph.D. thesis, Durham
  U., CPT, 6, 2018.

\bibitem{Kovtun:2019hdm}
P.~Kovtun, \emph{{First-order relativistic hydrodynamics is stable}},
  \href{https://doi.org/10.1007/JHEP10(2019)034}{\emph{JHEP} {\bfseries 10}
  (2019) 034} [\href{https://arxiv.org/abs/1907.08191}{{\ttfamily
  1907.08191}}].

\bibitem{Hiscock:1983zz}
W.A.~Hiscock and L.~Lindblom, \emph{{Stability and causality in dissipative
  relativistic fluids}},
  \href{https://doi.org/10.1016/0003-4916(83)90288-9}{\emph{Annals Phys.}
  {\bfseries 151} (1983) 466}.

\bibitem{Bemfica:2020zjp}
F.S.~Bemfica, M.M.~Disconzi and J.~Noronha, \emph{{First-Order
  General-Relativistic Viscous Fluid Dynamics}},
  \href{https://doi.org/10.1103/PhysRevX.12.021044}{\emph{Phys. Rev. X}
  {\bfseries 12} (2022) 021044}
  [\href{https://arxiv.org/abs/2009.11388}{{\ttfamily 2009.11388}}].

\bibitem{Bhattacharyya:2014bha}
S.~Bhattacharyya, \emph{{Entropy Current from Partition Function: One
  Example}}, \href{https://doi.org/10.1007/JHEP07(2014)139}{\emph{JHEP}
  {\bfseries 07} (2014) 139} [\href{https://arxiv.org/abs/1403.7639}{{\ttfamily
  1403.7639}}].

\bibitem{Iaconis:2019hab}
J.~Iaconis, S.~Vijay and R.~Nandkishore, \emph{{Anomalous Subdiffusion from
  Subsystem Symmetries}},
  \href{https://doi.org/10.1103/PhysRevB.100.214301}{\emph{Phys. Rev. B}
  {\bfseries 100} (2019) 214301}
  [\href{https://arxiv.org/abs/1907.10629}{{\ttfamily 1907.10629}}].

\bibitem{1963AnPhy..24..419K}
L.P.~{Kadanoff} and P.C.~{Martin}, \emph{{Hydrodynamic equations and
  correlation functions}},
  \href{https://doi.org/10.1016/0003-4916(63)90078-2}{\emph{Annals of Physics}
  {\bfseries 24} (1963) 419}.

\bibitem{Bhattacharya:2011eea}
J.~Bhattacharya, S.~Bhattacharyya and S.~Minwalla, \emph{{Dissipative
  Superfluid dynamics from gravity}},
  \href{https://doi.org/10.1007/JHEP04(2011)125}{\emph{JHEP} {\bfseries 04}
  (2011) 125} [\href{https://arxiv.org/abs/1101.3332}{{\ttfamily 1101.3332}}].

\bibitem{Armas:2021vku}
J.~Armas, A.~Jain and R.~Lier, \emph{{Approximate symmetries,
  pseudo-Goldstones, and the second law of thermodynamics}},
  \href{https://arxiv.org/abs/2112.14373}{{\ttfamily 2112.14373}}.

\bibitem{Armas:2023tyx}
J.~Armas and A.~Jain, \emph{{Approximate higher-form symmetries, topological
  defects, and dynamical phase transitions}},
  \href{https://arxiv.org/abs/2301.09628}{{\ttfamily 2301.09628}}.

\end{thebibliography}\endgroup

\end{document}